\documentclass[%
 reprint,
superscriptaddress,
 amsmath,amssymb,
 aps,
]{revtex4-1}
\usepackage[utf8]{inputenc}
\usepackage{floatflt}
\usepackage{subfigure}
\usepackage[section]{placeins}
\usepackage{graphicx}
\usepackage{dcolumn}
\usepackage{bm}
\usepackage{verbatim}
\usepackage{mathtools}
\usepackage{multirow}
\usepackage{bigdelim} 
\usepackage{xcolor}

\begin{document}

\newcommand{\sgn}{\operatorname{sgn}}

\preprint{APS/123-QED}

\title{Parasitic conduction channels in topological insulator thin films
} 

\author{Sven Just}
\altaffiliation[Present address: ]{Leibniz-Institut für Festk\"{o}rper und Werkstoffforschung Dresden, 01069 Dresden, Germany}
\affiliation{Peter Gr\"{u}nberg Institut (PGI-3), Forschungszentrum J\"{u}lich, 52425 J\"{u}lich, Germany}
\affiliation{J\"ulich Aachen Research Alliance (JARA), Fundamentals of Future Information Technology, 52425 J\"{u}lich, Germany}
\affiliation{Experimental Physics VI A, RWTH Aachen University, Otto-Blumenthal-Stra\ss e, 52074 Aachen, Germany}

\author{Felix Lüpke}
\altaffiliation[Present address: ]{Oak Ridge National Laboratory, TN, USA}
\affiliation{Peter Gr\"{u}nberg Institut (PGI-3), Forschungszentrum J\"{u}lich, 52425 J\"{u}lich, Germany}
\affiliation{J\"ulich Aachen Research Alliance (JARA), Fundamentals of Future Information Technology, 52425 J\"{u}lich, Germany}

\author{Vasily Cherepanov}
\affiliation{Peter Gr\"{u}nberg Institut (PGI-3), Forschungszentrum J\"{u}lich, 52425 J\"{u}lich, Germany}
\affiliation{J\"ulich Aachen Research Alliance (JARA), Fundamentals of Future Information Technology, 52425 J\"{u}lich, Germany}

\author{F. Stefan Tautz}
\affiliation{Peter Gr\"{u}nberg Institut (PGI-3), Forschungszentrum J\"{u}lich, 52425 J\"{u}lich, Germany}
\affiliation{J\"ulich Aachen Research Alliance (JARA), Fundamentals of Future Information Technology, 52425 J\"{u}lich, Germany}
\affiliation{Experimental Physics VI A, RWTH Aachen University, Otto-Blumenthal-Stra\ss e, 52074 Aachen, Germany}

\author{Bert Voigtl\"{a}nder}
\email[Corresponding author: ]{b.voigtlaender@fz-juelich.de}
\affiliation{Peter Gr\"{u}nberg Institut (PGI-3), Forschungszentrum J\"{u}lich, 52425 J\"{u}lich, Germany}
\affiliation{J\"ulich Aachen Research Alliance (JARA), Fundamentals of Future Information Technology, 52425 J\"{u}lich, Germany}
\affiliation{Experimental Physics VI A, RWTH Aachen University, Otto-Blumenthal-Stra\ss e, 52074 Aachen, Germany}

\date{\today}



\newcommand{\Si}{Si(111)--(1$\times$1) }
\newcommand{\Sii}{Si(111)--(7$\times$7) }
\newcommand{\TeSi}{Te/Si(111)--(1$\times$1) }
\newcommand{\BiSi}{Bi/Si(111)--($\sqrt3\times\sqrt3$) }

\begin{abstract}

Thin films of topological insulators (TI) usually exhibit multiple parallel conduction channels for the transport of electrical current. Beside the topologically protected surface states (TSS), parallel channels may exist, namely the interior of the not-ideally insulating TI film, the interface layer to the substrate, and the substrate itself. To be able to take advantage of the auspicious transport properties of the TSS, the influence of the parasitic parallel channels on the total current transport has to be minimized. Because the conductivity of the interior (bulk) of the thin TI film is difficult to access by measurements, we propose here an approach for calculating the mobile charge carrier concentration in the TI film. To this end, we calculate the near-surface band bending using parameters obtained experimentally from surface-sensitive measurements, namely (gate-dependent) four-point resistance measurements and angle-resolved photoelectron spectroscopy (ARPES). While in most cases another parameter in the calculations, i.e. the concentration of unintentional dopants inside the thin TI film, is unknown, it turns out that in the thin-film limit the band bending is largely independent of the dopant concentration in the film. Thus, a well-founded estimate of the total mobile charge carrier concentration and the conductivity of the interior of the thin TI film proves possible. Since the interface and substrate conductivities can be measured by a four-probe conductance measurement prior to the deposition of the TI film, the total contribution of all parasitic channels, and therefore also the contribution of the vitally important TSS, can be determined reliably.
\end{abstract}

\maketitle

\section{Introduction}

Topological insulators are candidates for future electronic devices and might be used for low-power spintronics or quantum computing, due to the special properties of their topologically protected surface states (TSS), such as spin-momentum locking and prohibited direct backscattering \cite{Hsieh_2009,Roushan_2009}. In recent years, the compound materials Bi$_2$Se$_3$, Bi$_2$Te$_3$ and Sb$_2$Te$_3$, which belong to the class of van-der-Waals bonded chalcogenides, have emerged as promising TI for applications at room temperature, essentially because of their pronounced band gap \cite{Hsieh_2009,Zhang_2009}. However, for taking advantage of the topological properties of the TI in any transport device, the electrical current has to be transmitted predominantly through the TSS. But for TI bulk crystals the parallel bulk conductance plays a significant role. 
\begin{figure}[b!]
\centering
\includegraphics[width=0.47\textwidth]{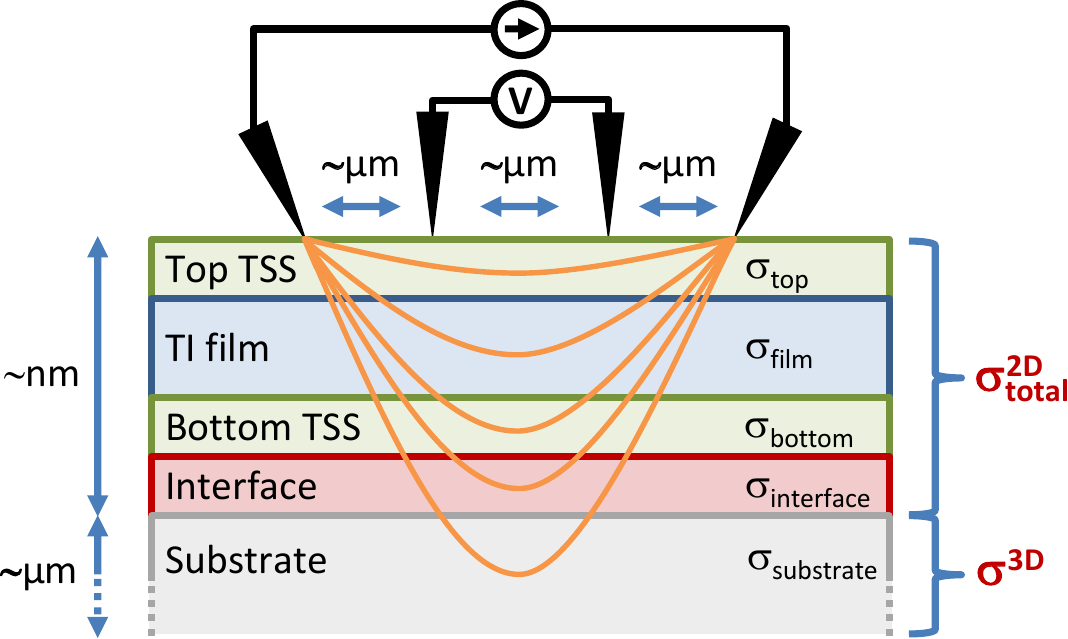}
\caption{(Color online) Multiple parallel conduction channels in a topological insulator thin film. The current transport can occur through the top and bottom TSS channels, but also through the interior of the TI film, through the interface layer between film and substrate as well as through the substrate itself. Position-dependent four-probe measurements on the surface can only differentiate between the total 2D conductivity, i.e. the sum of all parallel channels in the thin film, and the 3D substrate conductivity. Note that different from the schematic the film thickness ($\sim$ nm) is much smaller than the usual distance between the tips ($\sim$ $\mu$m). }
\label{fig1}
\vspace{-0ex}
\end{figure}
The transport through this parasitic channel may even surpass the current transport through the TSS, as has been observed  recently \cite{Barreto_2014,Ren_2010}. 
While the influence of the bulk can be suppressed if TI materials are grown as thin films on sufficiently low-conducting substrates, also in this case multiple parallel conduction channels that participate in the total current transport may be present. This is visualized in Fig.\,\ref{fig1}. Beside the TSS channels, the current can also flow through the interior of the not-ideally insulating TI film, through the potentially highly conductive interface layer between the TI film and the substrate, and through the substrate itself. Thus, it is an important task to design future devices based on thin TI films in a manner that minimizes the fraction of the current through the parasitic parallel transport channels. 

Clearly, to achieve this goal the conductivities of the parasitic channels in actual MBE-grown thin TI films have to be determined first. At first glance, obvious experiments to this end are distance-dependent four-probe measurements on the surface to disentangle two-dimensional (2D) from three-dimensional (3D) conductance channels \cite{Just_2015,Miccoli_2015,Voigtlaender_2018}. However, the thickness of the grown TI films (typically $10$ nm) poses a challenge here. Specifically, the small vertical length scale makes it impossible to separate the bulk conductance of the TI film as a 3D conductance from the 2D conductance of the TSS, since for this purpose the probes would have to be spaced closer than one quarter of the film thickness. But this is  not possible, because this spacing would approach the tip radius \cite{Just_2015,Miccoli_2015,Voigtlaender_2018,Wells1,Just_2017}. For relevant MBE-grown TI films, position-dependent four-probe measurements are thus only suitable to separate the total 2D conductivity $\sigma^{\mathrm{2D}}_{\mathrm{total}}$ of the thin TI film, which comprises contributions from the top and bottom TSS, the interior of the thin film and the interface channels (Fig. \ref{fig1}), from the 3D substrate conductivity \cite{Miccoli_2015,Wells1,Durand,Just_2015}. However, since  the conductivity of the substrate channel can easily be minimized by choosing an appropriately low-doped substrate as template for the grown TI thin film, separating out its conductivity is not the main issue. 

%

It is thus evident that a different approach has to be taken to decompose what in practical distance-dependent four-tip measurements appears to be an undifferentiated 2D conductivity, into the intrinsic TSS on the one hand and the parasitic thin-film and interface contributions on the other. In this paper, we address this problem with a two-pronged strategy, dealing separately with the interface and the thin-film channels. 
Regarding the interface conductivity, we observe that in the common case of van-der-Waals epitaxy the interaction between the TI film and the substrate is small. As a consequence, a surface reconstruction of the substrate, established prior to growing the TI, often survives the subsequent film growth, as evidenced by transmission electron microscopy \cite{Luepke_2017}. This suggests that the sought-after interface conductivity can in fact be measured as the surface conductivity of the relevant surface reconstruction without the TI film. Such measurements can be routinely performed with a distance-dependent four-probe experiment \cite{Just_2015,Miccoli_2015,Voigtlaender_2018}. In appendix A, we report surface conductivity measurements for several surface reconstructions on Si(111) that are relevant for the growth of (Bi$_{1-x}$Sb$_x$)$_2$Te$_3$. They reveal that the surface conductivity strongly depends on the type of surface reconstruction (Table \ref{tab1}). One may therefore expect that this pronounced variation translates to the interface conductivity, meaning that the pertinent surface reconstruction needs to be taken into account when disentangling the parasitic conduction channels from the TSS.  

The interior of the TI film is more difficult to separate out experimentally, simply because it is sandwiched between the top and bottom TSS channels. We therefore resort to a calculation of its conductivity from experimentally accessible parameters.  Generically, the TI film conductivity is determined by the charge carrier concentration and the charge carrier mobility. In this paper, we regard the charge carrier mobility as a given material parameter of the TI and concentrate on the determination of the charge carrier concentration, which in a thin film is principally determined by the dopant concentration and the band bending. 
Because the concentration of (unintentional) dopants stemming from the growth process is not known, the charge carrier concentration and conductivity of the film cannot be calculated directly. However, we will show here that in the thin-film limit a variation of the dopant concentration does not influence the mobile charge carrier concentration in a significant way. This is in stark contrast to the situation of a half-infinite bulk crystal, where the doping through dopants  in the material would lead to a strong shift of the band edges with respect to the Fermi energy inside the bulk, and therefore also to a strong effect on the mobile charge carrier concentration. In a thin film, the shift of the bands with respect to the Fermi energy and the near-surface band-bending is much smaller. The reason is the long screening length compared to the film thickness \cite{Kim_2012,Yang_2015,ViolBarbosa_2013}. The problem at hand therefore boils down to a calculation of the band bending in the thin film, from which the total charge carrier concentration in the film can be determined by integration, irrespective of the (unknown) dopant concentration.  

In order to determine the band bending, we use two distinct levels of approximation. In the symmetric approximation, we assume that the TI has identical properties at its surface to the vacuum and its interface to the substrate. In this case, the position of the surface Fermi level relative to the Dirac point, which can be measured at the top TSS by angle-resolved photoemission (ARPES), is enough to calculate the band bending. In contrast, in the asymmetric approximation the surface Fermi level and the Fermi level in the thin film at the interface to the substrate are allowed to differ from each other. In this case, ARPES alone is not enough to determine the band bending in the film. However, if supplemented by  gate-dependent four-probe transport measurements on the surface of the TI, the combined experimental information is sufficient to determine the band bending, the carrier concentration and thus the film conductivity.


\begin{table}
\begin{ruledtabular} 
\centering 
\begin{tabular}{ l l }
&\\[-1.5ex]	
Surface reconstruction & Surface conductivity $\sigma_S$\\[1ex]
\hline\\[-1ex]
Te/Si(111)-(7$\times$7) & $(8.3\,\pm\,0.5) \cdot 10^{-6}\,\mathrm{S}/\square$\\ 
Te/Si(111)-(1$\times$1) & $(2.6\,\pm\,0.5) \cdot 10^{-7}\,\mathrm{S}/\square$ \cite{Luepke_2017}\\
Bi/Si(111)-($\sqrt{3}\times\sqrt{3}$) 1 ML & $(1.4\,\pm\,0.1) \cdot 10^{-4}\,\mathrm{S}/\square$ \cite{Just_2015}\\
\end{tabular}
\caption{Interface conductivities of different passivated and reconstructed interfaces of Si(111) used as a substrate for van-der-Waals epitaxy of thin TI films.}
\label{tab1}
\end{ruledtabular} 
\end{table}

\section{Band bending in topological insulators}
In principle, the bulk of a TI should be insulating, if the Fermi energy is located in the bulk band gap with only the Dirac cone of the TSS crossing it. But  unintentional doping during the growth process may cause a considerable bulk conductivity \cite{Barreto_2014,Ren_2010}. This in turn may result in an unwanted, substantial current flowing through the interior of the TI film rather than through the TSS. Fortunately, it is principally possible to influence the bulk conductivity of the TI by growing the material as a thin film with a large surface-to-volume ratio, realized for instance by film thicknesses in the range $10$ to $100$\,nm \cite{Lanius_2016}.  
In any thin film, there is an influence of the surface and interface states on the film's bulk conductivity, because charge may be transferred between the surface states and the interior of the film, resulting in near-surface and near-interface band bendings and corresponding space-charge regions, and a concomitant reduction or increase of the concentration of mobile charge carriers in the film. 

For trivial, non-topological surface and interface states with their often large density of states, the pinning of the Fermi level at the surface and interface states usually plays a decisive role for the band bending \cite{Lueth}. In contrast, there exists no Fermi level pinning by topological surface states, because the density of states in their Dirac cones is comparatively small. As a consequence, the filling levels of the TSS (i.e., the Dirac cone) on the energy axis are expected to \textit{change} with each charge that is transferred between TSS and film interior. This warrants the re-examination of the common phenomenon of bend bending in thin films for the special situation of a TI material.

In this paper, we model the TI thin film as a narrow-bandgap semiconductor without Fermi level pinning. We calculate the near-surface band bending induced by the top and bottom TSS in a semi-classical approach, which includes both classical electrostatics and quantization effects that arise from the vertical confinement of electrons in the thin film with thicknesses in the range of several nm. Specifically, we start by solving Poisson's equation, which relates the band curvature to the space charge density, under appropriate boundary conditions. The resultant band bending potential is used in Schr\"{o}dinger's equation to calculate the quantized eigenenergies which give rise to multiple subbands inside the conduction and valence bands. Since these subbands result in a modification of the effective density of states of valence and conduction bands, Poisson's equation must be solved a second time. Note that an exact solution would require a fully self-consistent solution of Poisson's and Schr\"{o}dinger's equations. Here, we truncate this self-consistency cylce after one and a half iterations, deriving our final band bending and mobile charge carrier concentrations from the second solution of Poisson's equation. Initially, we apply this scheme to the simple case of a semi-infinite bulk crystal with a single surface to vacuum. This serves as a reference for thin film calculations which involve a surface to vacuum as well as an interface to the film substrate. Here, we first assume symmetric boundary conditions, i.e. equal charge carrier densities in the top and bottom TSS. Finally, the more general case of asymmetric boundary conditions in a thin film is presented. Further details of the calculation are described below and in the supplemental material \cite{suppl}.  

Throughout this paper, the following notation is used: The chemical potential (Fermi energy $E_{F}$) is constant across the TI film. The $z$-dependent position of the valence band edge is denoted as $E_{V}(z)$. Relative to the valence band edge, the Fermi levels at the top surface, in the bulk of the TI, and at the bottom interface to the substrate are designated as $E^{\mathrm{top}}_F \equiv E_{F} - E^{\mathrm{top}}_{V}$, $E^{\mathrm{bulk}}_F \equiv E_{F}-E^{\mathrm{bulk}}_{V}$, and $E^{\mathrm{bottom}}_F \equiv E_{F}-E^{\mathrm{bottom}}_{V}$, respectively. 

\subsection{Near-surface band bending in a semi-infinite bulk crystal}
\label{sec:bulkbandbending}
\subsubsection{Formalism}

\begin{figure*}[t!]
	\centering
        \includegraphics[height=0.25\textwidth]{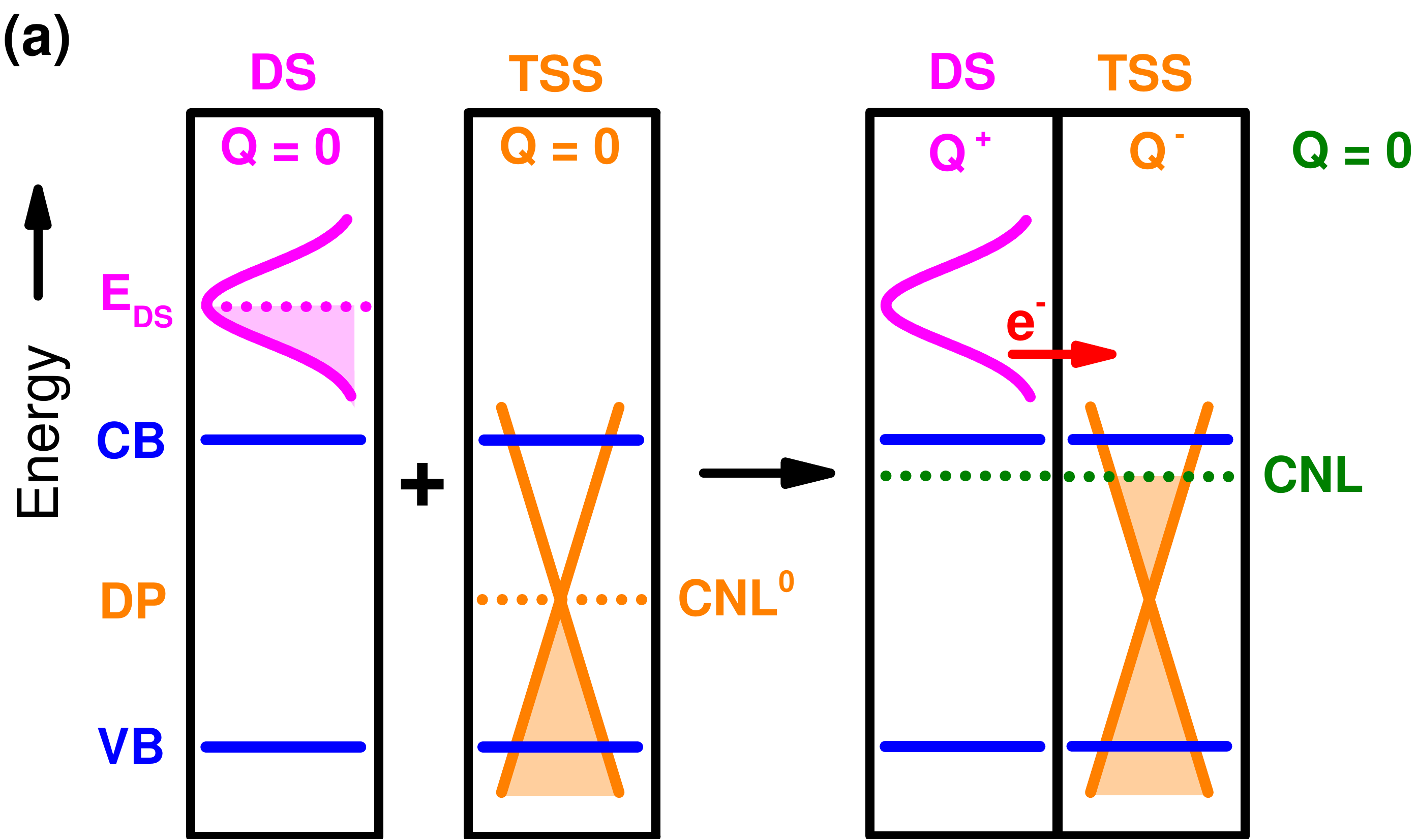}
        \includegraphics[height=0.25\textwidth]{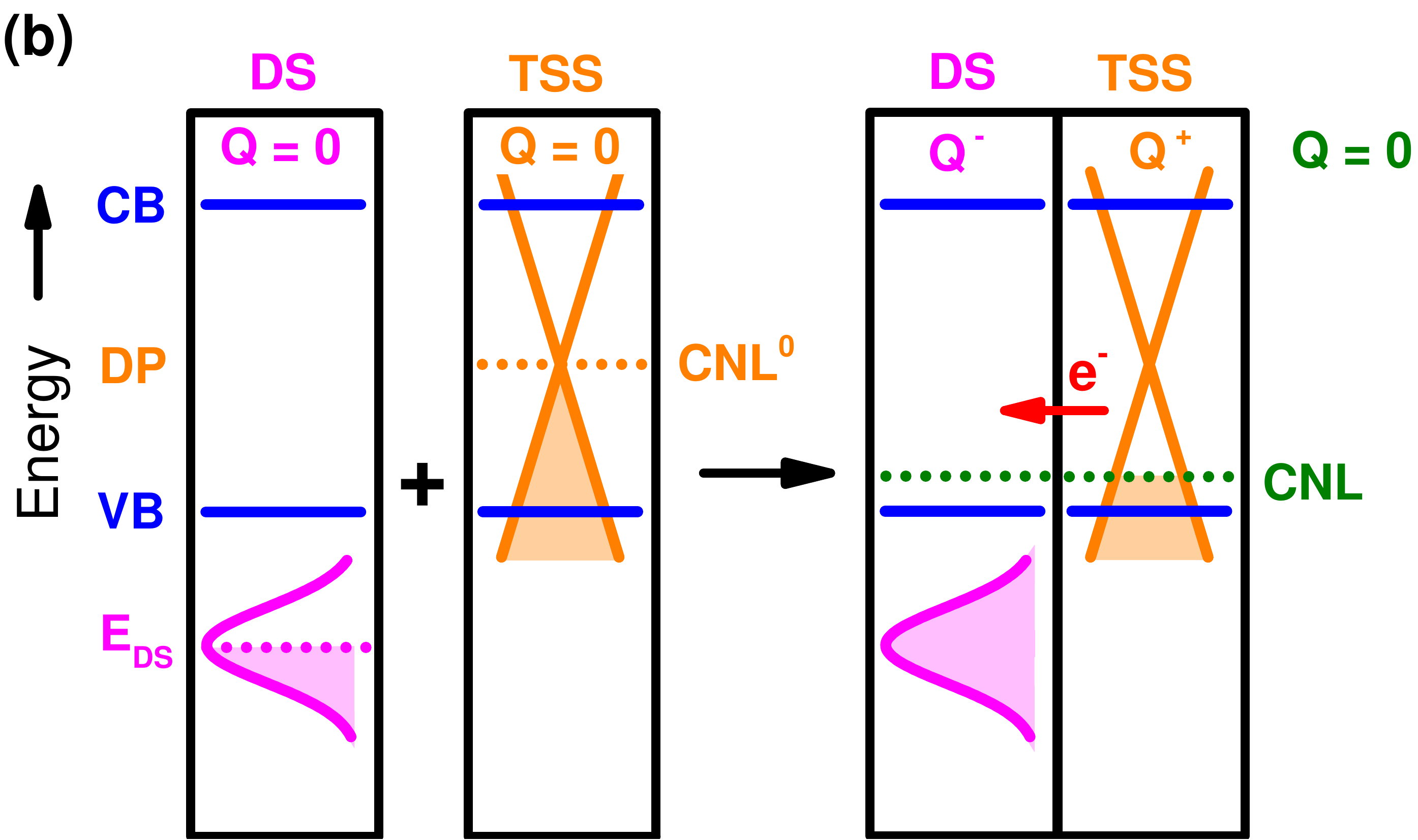}
	\caption{(Color online) Principle of the formation of the CNL level of the TI surface for the presence of additional charged trivial defect states (DS). Both the DS (magenta) located around $E_{\mathrm{DS}}$ and the TSS (Dirac cone, orange) would be filled up to their individual charge neutrality levels (dotted lines), if they were artificially separated from each other (left parts of (a) and (b)). However, if the DS and TSS are combined (right parts), a charge transfer occurs, resulting in a common aligned CNL (green line). This final CNL is influenced by the position of the DS ($E_{\mathrm{DS}}$) either above (a) or below (b) the Dirac point (DP).} 
\label{supfig0}
\vspace{-0cm}
\end{figure*}

Poisson's equation  $d^2 \Phi /dz^2= -\rho(z)/\epsilon_0\epsilon_r$ relates the space charge density $\rho(z)$, measured in units of charge per area, to the spatial curvature of the bands in the space charge region. In writing down this equation, we assume that parallel to the surface there is no spatial dependence of the bands. The potential $\Phi(z)$ is given by $q\Phi(z) = E_{F} - E_{\mathrm{intrinsic}}(z)$, with $q$ being the elementary charge and $E_{\mathrm{intrinsic}}(z)$ the intrinsic level, while the band bending is defined as $V(z) \equiv \Phi(z) - \Phi_b$, where $\Phi_b$ is the potential in the bulk, i.e. $\Phi_b \equiv \Phi(z \rightarrow \infty)$. In our calculations of band bending, we use specific material parameters of the TI as input, in particular the band gap and the effective masses of electrons and holes. Moreover, we use a parabolic approximation for the bulk band edges in order to calculate the effective densities of states $N^{\mathrm{eff}}_C$ and $N^{\mathrm{eff}}_V$ in the valence and conduction bands that determine the mobile charge carrier densities in the (bulk) TI \cite{suppl}. Finally, we assume the non-degeneracy of the TI, such that the Fermi distribution can be approximated by the Boltzmann distribution.  
Introducing the dimensionless potentials $u(z) = q\Phi(z)/k_B T$ and $v(z) = qV(z)/k_B T$, where $T$ is the temperature and $k_B$ the Boltzmann constant, Poisson's equation becomes \cite{Lueth,Many}
\begin{align}
	\frac{d^2v}{dz^2} & = -\frac{q^2}{\epsilon_0 \epsilon_r k_B T} \left(n_b - p_b + p_b e^{-v} - n_b e^v\right). \label{pois} 
\end{align}
$n_b$ and $p_b$ denote the electron and hole densities in the bulk, respectively, which are proportional to $e^{u_b}$. The first two terms on the left correspond to the static charge caused by the dopant atoms in the material and the last two terms on the right specify how the mobile charge carrier densities in the space charge region are modified by the band bending $v(z)$. Eq. (\ref{pois}) can be expressed as 
\begin{align}
	\frac{d^2v}{dz^2} & = \frac{1}{L^2} \left(\frac{\sinh(u_b+v)}{\cosh(u_b)} - \tanh(u_b) \right), \label{pois2}
\end{align}
with the effective Debye length 
\begin{align}
L=\sqrt{\frac{\epsilon_0 \epsilon_r k_B T}{q^2(n_b+p_b)}}.
\label{eq:Debye}
\end{align}
Multiplying both sides of Eq. (\ref{pois2}) with $2\frac{dv}{dz}$ and using $2\frac{dv}{dz} \frac{d^2v}{dz^2}=\frac{d}{dz}(\frac{dv}{dz} \frac{dv}{dz})$, it can be integrated to yield, after taking the square root,
\begin{align}
	\frac{dv}{dz} & = \sgn(-v) \frac{\sqrt{2}}{L}\sqrt{\frac{\cosh(u_b+v)}{\cosh(u_b)}-v \, \tanh(u_b)+c}. \label{ersteAbl}
\end{align}
The constant $c$ must be determined from the boundary conditions. For a semi-infinite bulk crystal the boundary conditions are $\frac{dv}{dz}=0$ for $z\rightarrow \infty$ and $v(0)=v_{\mathrm{top}}=(E^{\mathrm{top}}_{F}-E^{\mathrm{bulk}}_{F})/k_B T$. Inserting these boundary conditions into Eq. (\ref{ersteAbl}) and expressing the hyperbolic functions by exponentials leads to the solution \cite{Lueth,Many}
\begin{align}
	z(v) & = A \int_{v_{\mathrm{top}}}^{v(z)} \hspace{-1.5ex} \sqrt{\frac{e^{u_b} + e^{-u_b}}{e^{u_b}(e^v - v - 1) + e^{-u_b}(e^{-v} + v - 1)}}\, dv  \label{bb}
\end{align}
with the prefactor $A= \sgn(-v) \frac{L}{\sqrt{ 2}}$. 
Eq. (\ref{bb}) can be integrated numerically and inverted to determine the band bending $v(z)$. Since the latter is influenced decisively by the boundary condition, i.e the surface position of the Fermi level relative to the band edges ($E^{\mathrm{top}}_{F}$), we now turn to a discussion of the factors which determine this parameter.  

\subsubsection{Topological surface states}
The Fermi energy at the surface of a conventional semiconductor is often pinned at a fixed position $E_F^{\mathrm{top}}$ relative to the band edges. This is a consequence of a high density of states of the semiconductor surface states. In contrast, the surface density of states of the intrinsic TSS in TIs is relatively low, which is the outcome of the specific linear dispersion of the TSS, i.e., the Dirac cone $E(k) = \hbar v_F k$ \cite{Novoselov_2005}. As a result, there is no pinning of $E^{\mathrm{top}}_{F}$ by the intrinsic TSS at the surface of a TI. 

The charge carrier density of Dirac electrons in the TSS, generated by charge transfer from the film,  can be written as a function of energy $E$ up to which the cone is filled as \cite{Luepke_2017_2} 
\begin{align}
	n_{\mathrm{TSS}}(E) & = \frac{1}{4 \pi \hbar^2 v_F^2} \left(E^2 - E_0^2 \right)\,\mathrm{.} 
\end{align}
Here, $E_0$ represents an initial filling level that may be caused by surface doping due to surface defects or adsorbates (see below). In the absence of surface defects, $E_0=0$. 
The filling level of the Dirac cone in a TI, i.e. the surface Fermi energy $E_F^\mathrm{top}$, can be determined by surface-sensitive ARPES measurements and then used as a known parameter in the band bending calculations. Since there is no pinning, $E^{\mathrm{top}}_{F}$ is subject to several external factors, in particular the bulk dopant concentration, the density of adsorbates on the surface, and the presence of a gate voltage. The influence of these parameters will be discussed in more detail below. 

\subsubsection{Non-topological, trivial defect states}
The charge neutrality level (CNL) of a surface state is the position of the Fermi level at which the surface is uncharged. In the case of a TI the CNL of the intrinsic TSS coincides with the Dirac point. However, the presence of additional trivial defect states (DS) at the surface of the TI, for example surface vacancies or additional adsorbates, may have a strong influence on the CNL of the TSS. Specifically, the surface doping caused by such trivial defects may modify the filling level of the TSS, thus compensating the additional charges of the DS in order to fulfill surface neutrality. Therefore, the CNL of the combined TSS/DS system (considered as isolated from the bulk) is generally not located at the Dirac point. 
A schematic that illustrates the shift of the surface CNL from the intrinsic value of the TSS, i.e. the Dirac point, to a new value that is determined by surface charge neutrality between the DS and the TSS is displayed in Fig. \ref{supfig0}. The density of states of the trivial DS (magenta) is centered around $E_{\mathrm{DS}}$. Two different cases are shown in panels a and b. Initially (before electrical connection with the TSS), the DS is neutral and filled up to its charge neutrality level $E_{\mathrm{DS}}$ (dotted magenta line). The isolated Dirac cone of the TSS (orange) is also neutral and filled up to the Dirac point which represents the initial charge neutrality level CNL$^0$ of the TI surface (dotted orange line). If the trivial DS and the TSS are connected, a charge transfer occurs, yielding a common value for the CNL of the TI surface (dotted green line). For a DS above the conduction band (CB), as shown in Fig. \ref{supfig0}(a), electrons flow into the TSS, increasing the filling level of the Dirac cone and reducing the filling level of the DS.  As a result, the DS becomes positively charged and the TSS negatively charged, but in total the neutrality condition is maintained. Similarly, for a DS positioned below the valence band (VB), Fig. \ref{supfig0}(b), charge transfer occurs in the reverse direction. Electrons from the TSS flow into the DS, which results in a negative DS and a positive TSS with a common CNL below the Dirac point. In the examples in Fig. \ref{supfig0}, the DS is either completely filled or completely depleted by the charge transfer to or from the TSS, respectively. However, depending on the properties of the DS, i.e. its position $E_{\mathrm{DS}}$ and width, it is also conceivable that the DS remains partially filled after charge transfer. 

In conclusion, the CNL of the TI surface and thus the filling level of the TSS is influenced by the presence of any additional trivial DS. Both a change of the filling level of the DS and a shift of $E_{\mathrm{DS}}$ provoke a corresponding shift of the surface CNL. This overall surface CNL becomes relevant when the charge exchange between the surface and the bulk of the TI is enabled and influences the resultant near-surface band bending.  

\subsubsection{Charge transfer between TSS and bulk}
Generally speaking, band bending is governed by the condition of charge neutrality: the total charge (per unit area) in surface states and the total mobile charge per unit area in the space charge layer must be equal and of opposite sign. This charge neutrality involves charge transfer between the surface and the space charge layer. To understand this charge transfer conceptually, we consider a situation in which surface and bulk are initially disconnected \cite{Heinzel_2010}. In this situation, the filling level of the TI surface (comprising TSS and DS) corresponds to its charge neutrality level (CNL) at which the separated TI surface itself is neutral (only in the absence of any DS, this CNL coincides with the Dirac point). In the separated bulk, neutrality implies the absence of any band bending below the surface, thus the charge neutrality level corresponds to the bulk Fermi level. If surface and bulk are connected to each other, the separate CNL of bulk and surface will in general not be aligned to each other. As a consequence, charge will flow from the component with the higher CNL to the component with the lower one. In the process, surface and space charge layer will become charged and the bands in the near-surface region will bend, but overall charge neutrality is conserved. The CNL of the TI surface (TSS plus DS) can thus be described as the initial filling level up to which the (neutral) TI surface is filled before equilibration with the bulk. As such, it is not a quantity that can be measured. 

\begin{figure*}[t!]
\centering
\includegraphics[height=0.214\textwidth]{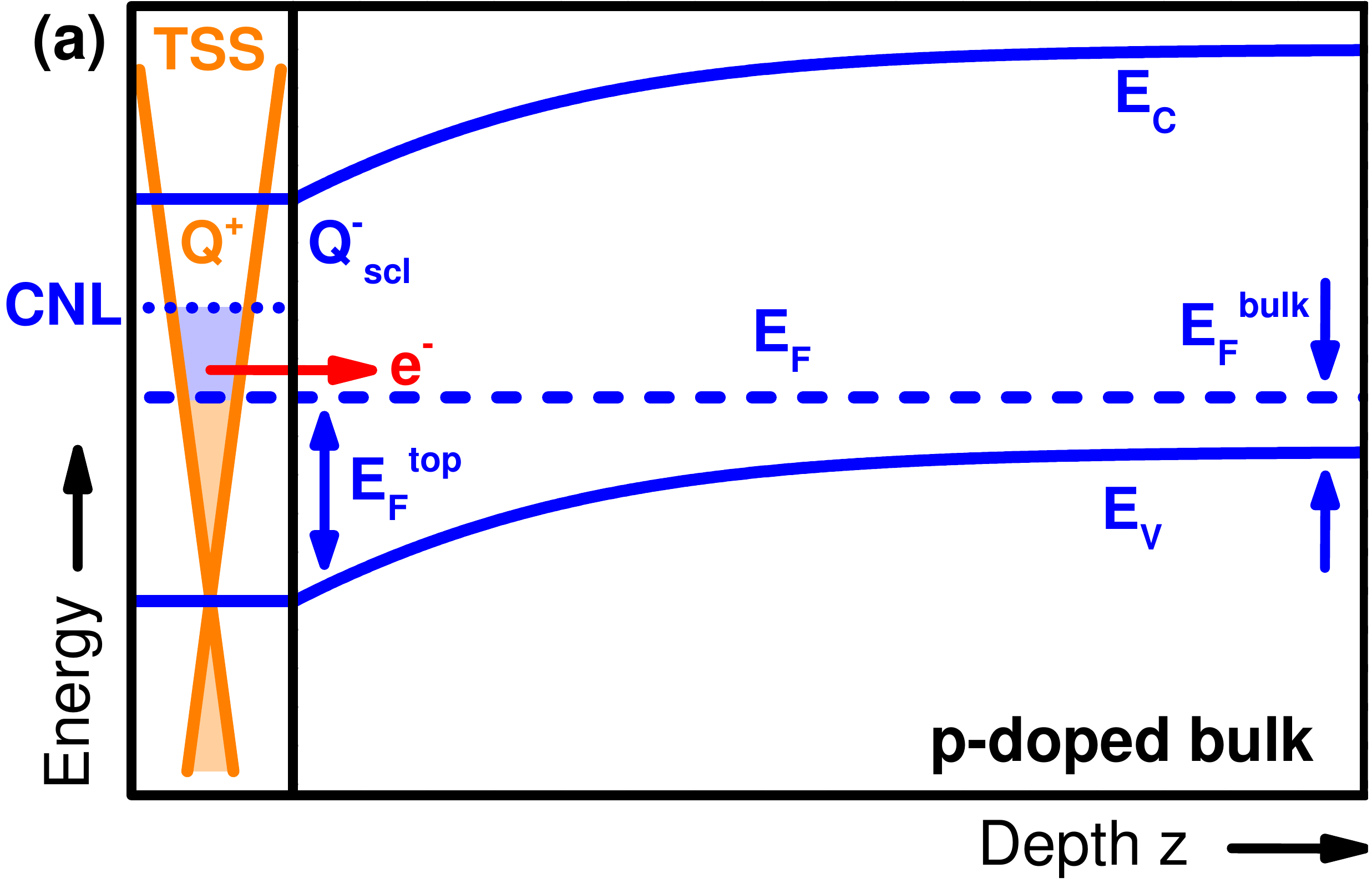}
\hspace{-0.20cm}
\includegraphics[height=0.214\textwidth]{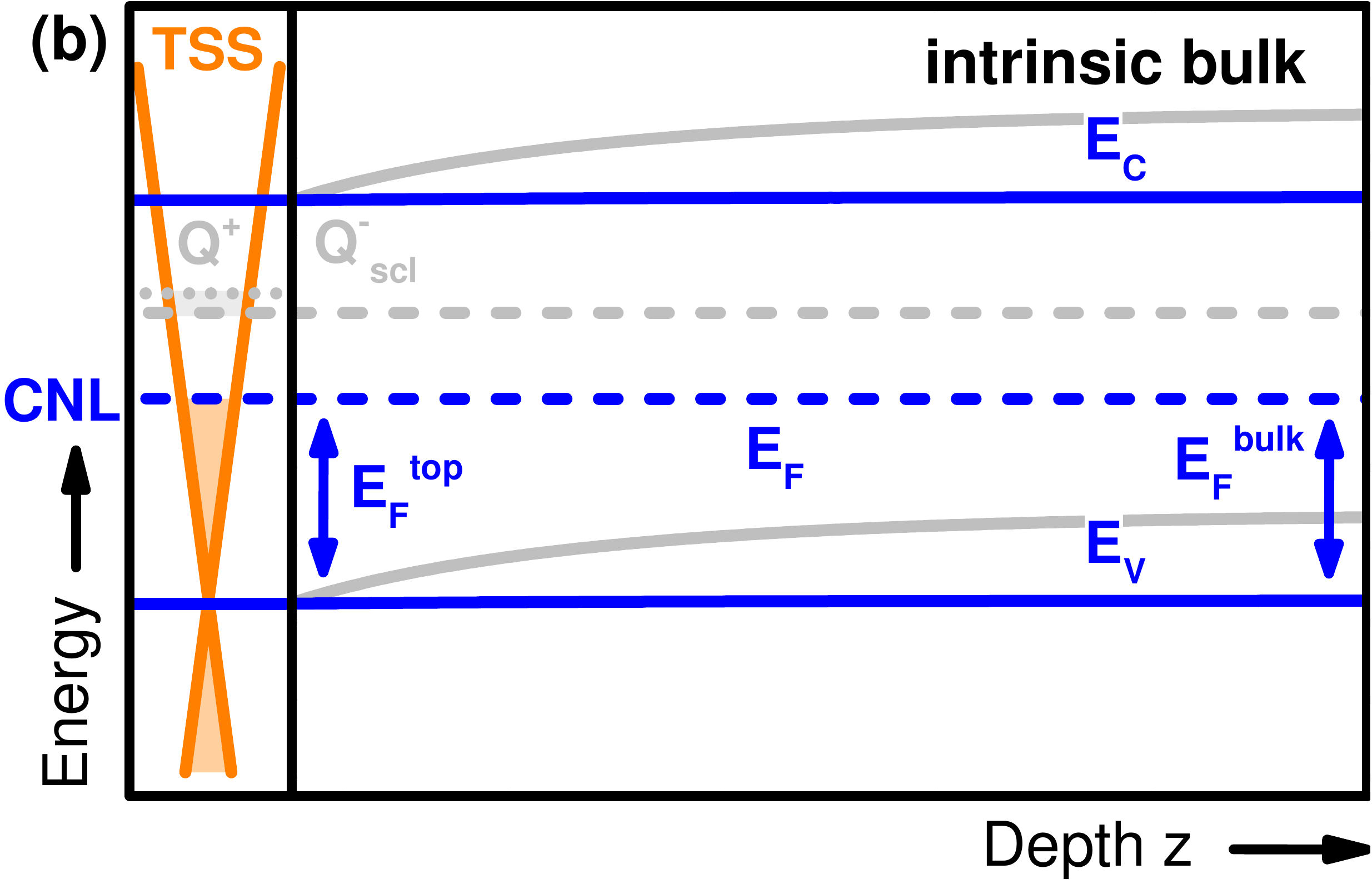}
\hspace{-0.20cm}
\includegraphics[height=0.214\textwidth]{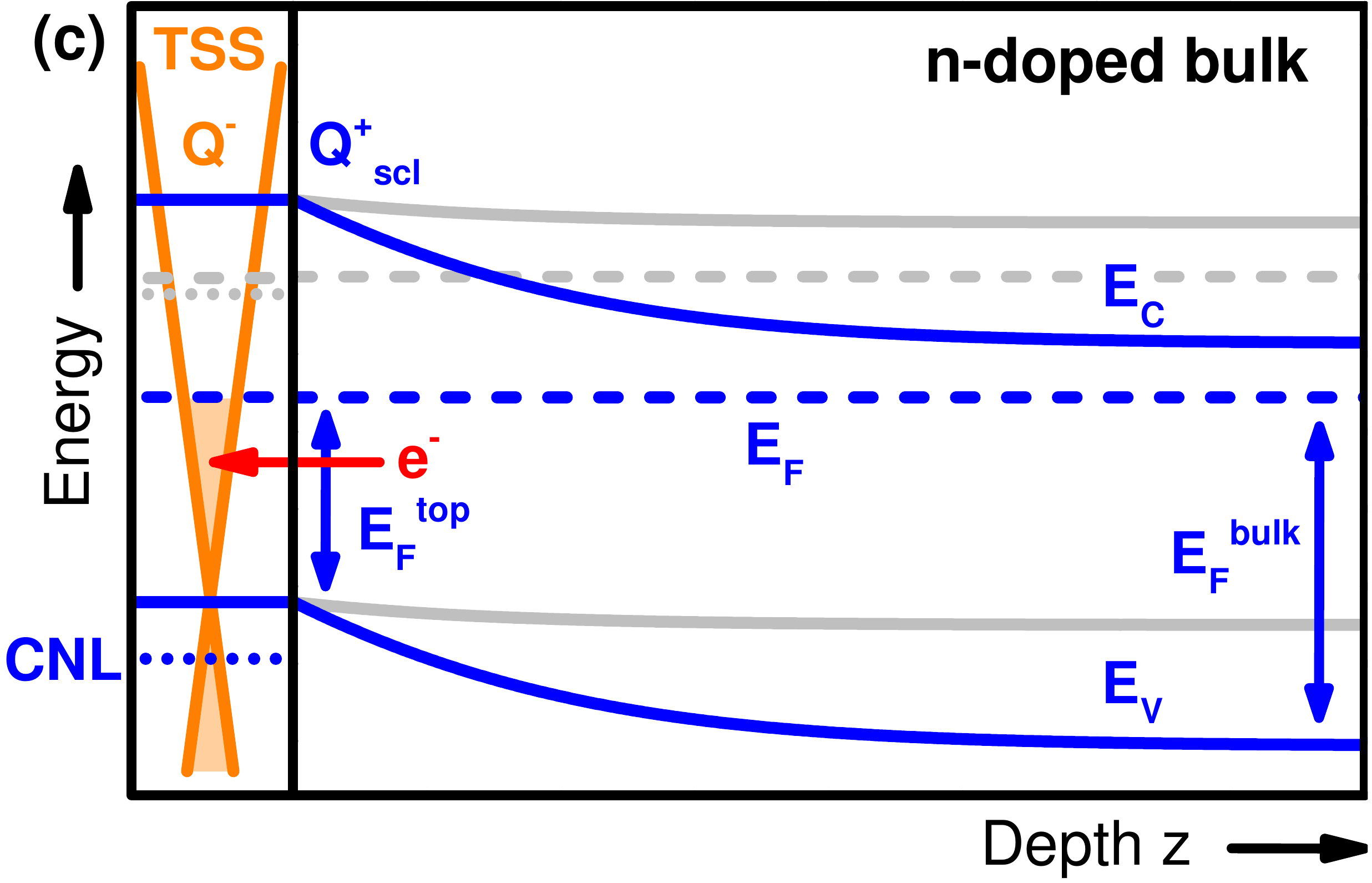}
	\caption{(Color online) Principle of band bending in extended TI bulk crystals for three different bulk dopant concentrations, i.e. (a) p-doped ($E_F^{\mathrm{bulk}}=35\,\mathrm{meV}$, $p_{\mathrm{bulk}}=4\times 10^{17}\,\mathrm{cm}^{-3}$), (b) intrinsic ($E_F^{\mathrm{bulk}}=130\,\mathrm{meV}$, $n_{\mathrm{bulk}}=p_{\mathrm{bulk}}=1\times 10^{16}\,\mathrm{cm}^{-3}$), and (c) n-doped ($E_F^{\mathrm{bulk}}=225\,\mathrm{meV}$, $n_{\mathrm{bulk}}=4\times 10^{17}\,\mathrm{cm}^{-3}$), in combination with fixed values for either the surface Fermi energy $E^{\mathrm{top}}_{F}$ (blue) or the charge neutrality level CNL (gray) of the TI surface (TSS plus DS). On the left of each panel, the partially filled (highlighted orange area) Dirac cone of the TSS is shown, and on the right the calculated conduction and valence bands are plotted as function of depth $z$ into the crystal. The Fermi energy $E_F$ (dashed line, constant) and both the surface Fermi energy $E^{\mathrm{top}}_{F}$ and the bulk Fermi energy $E^{\mathrm{bulk}}_{F}$ (blue arrows) are indicated. For the blue diagrams $E^{\mathrm{top}}_{F}$ is fixed to a midgap value of $130\,\mathrm{meV}$, resulting in different band bendings and CNL positions (dotted lines) from (a) to (c). In contrast, the gray diagrams (b,c) exhibit a fixed CNL at $230\,\mathrm{meV}$ (same as in (a)), so that in this case $E^{\mathrm{top}}_{F}$ varies from (a) to (c) and the shape of band bending differs from the blue-colored case. Further calculation parameters are listed in Tab. \ref{tab2}.}
\label{fig2ii}
\vspace{-0ex}
\end{figure*}

\subsubsection{Relevant parameters for overall band bending}
For an extended bulk crystal, the surface CNL, the surface Fermi energy $E^{\mathrm{top}}_{F}$, and the bulk Fermi energy $E^{\mathrm{bulk}}_{F}$ (which represents the bulk dopant concentration) all influence the charge that finally resides in the TSS and space charge layer, and thus the band bending. The three parameters are interdependent: if two of them are given, the third one is fixed by the charge neutrality condition. It is clear that the surface CNL, and thus also the initial filling level of the TSS, is a conceptual quantity that describes a neutral TI surface (comprising both TSS and DS) which is isolated from the bulk of the TI crystal; it is therefore not measurable. However, the surface Fermi energy $E^{\mathrm{top}}_{F}$ and thus the final filling level of the Dirac cone can be directly observed by ARPES measurements. In contrast, the exact value for the bulk Fermi energy $E^{\mathrm{bulk}}_{F}$ is not known, as the exact density of dopants in the bulk TI material is often unspecified, so that $E^{\mathrm{bulk}}_{F}$ has to remain as a free parameter in the calculations.
%
  
\subsubsection{Example calculations}
In order to explore the interplay of the three key parameters and their impact on the band bending in a semi-infinite bulk TI crystal, Fig. \ref{fig2ii} outlines band diagrams for three different bulk dopant concentrations, i.e. nondegenerate p-doped (Fig. \ref{fig2ii}(a)), intrinsic (Fig. \ref{fig2ii}(b)) and nondegenerate n-doped (Fig. \ref{fig2ii}(c)). For the calculated diagrams, a specific value for either $E^{\mathrm{top}}_{F}$ (blue) or the surface CNL (gray) is used -- the respective other parameter results from the condition of charge neutrality.  
For all three blue diagrams in Fig. \ref{fig2ii}(a)-(c), the value of $E^{\mathrm{top}}_{F}$ is fixed to the midgap position. Such a situation was experimentally realized in the quaternary BiSbTeSe system \cite{Arakane_2012}, for example.

For the p-doped TI in Fig. \ref{fig2ii}(a), a downward band bending with a negatively charged space charge region is obtained. The TSS is positively charged, as indicated by the calculated CNL above $E^{\mathrm{top}}_{F}$. The filling level of the Dirac cone has thus been reduced and electrons have flown into the space charge region (highlighted blue area). In case of an even stronger downward band bending, as obtained for example by a degenerately doped bulk, a non-topological two-dimensional electron gas (2DEG) with quantized states is expected near the surface. This has been observed experimentally \cite{Bianchi_2010,King_2011,Benia_2011}. In this limit, the description by Poisson's equation has to be replaced by the Schr\"{o}dinger-Poisson approach even for an extended bulk crystal. 

In Fig. \ref{fig2ii}(b), the bulk Fermi energy $E^{\mathrm{bulk}}_{F}$ is in the midgap position, representing the intrinsic character of the bulk of the TI. In conjunction with the assumed midgap position of $E^{\mathrm{top}}_{F}$ this results in flat bands. In this case the calculated CNL is equal to $E^{\mathrm{top}}_F$, indicating that no charge transfer has occurred. Note that the CNL in panel (b) is different from the one in obtained in panel (a), because its value must depend on $E^{\mathrm{bulk}}_{F}$ if a constant midgap position of $E^{\mathrm{top}}_{F}$ is assumed. 

For the n-doped TI in Fig. \ref{fig2ii}(c), an upward band bending is obtained, if a midgap position of $E^{\mathrm{top}}_{F}$ is assumed (blue lines). The calculated CNL is now far below the $E^{\mathrm{top}}_{F}$, indicating a charge transfer of electrons from the bulk into the Dirac cone. 

If on the other hand the CNL is assumed to be constant (instead of $E^{\mathrm{top}}_{F}$ as discussed above), the resulting band bendings for the same three bulk dopant concentrations as considered before turn out to be completely different. This is demonstrated by the gray band diagrams in Fig. \ref{fig2ii}(b)-(c). Physically, a constant value for the CNL corresponds to a fixed surface configuration with a specific density of surface defects or adsorbates. Specifically, in Fig. \ref{fig2ii}(b) the flat bands  for the midgap position of $E^{\mathrm{bulk}}_{F}$ give way to a downward band bending, because now $E^{\mathrm{top}}_{F}$, being dependent on both $E^{\mathrm{bulk}}_{F}$ and CNL, is not positioned midgap any more. Similarly, in Fig. \ref{fig2ii}(c)  the same position of the CNL as for the gray bands in panel (b) yields a much weaker upward band bending, with a different position of $E^{\mathrm{top}}_{F}$(dashed gray line). 
Two conclusions can be drawn from the above examples: (1) Comparing blue versus gray bands in each panel Fig. \ref{fig2ii}(a) to (c), we observe that a variation of the CNL due to surface defect states directly influences $E^{\mathrm{top}}_{F}$, if the bulk dopant concentration remains constant. Therefore, if the surface defect concentration is increased over time, a shifting $E^{\mathrm{top}}_{F}$ is expected. This  effect of surface degradation due to long-time storage of TIs in vacuum or exposure to air has indeed been observed experimentally by ARPES measurements \cite{Bianchi_2010,Kong_2011,Analytis_2010}. (2) Comparing the blue bands in Fig. \ref{fig2ii}(a) and  gray bands in Fig. \ref{fig2ii}(b)-(c) among each other shows that bulk dopant concentration, which varies from panel (a) to (c), has a strong influence on $E^{\mathrm{top}}_{F}$ and therefore also on the band bending. Thus, for an extended semi-infinite bulk crystal both the surface Fermi energy $E^{\mathrm{top}}_{F}$ and the bulk dopant concentration, represented by $E^{\mathrm{bulk}}_{F}$, must be known to perform an exact calculation of the near-surface band bending (Note that this makes the knowledge of difficult-to-determine CNL superfluous, because the latter is at any rate determined by $E^{\mathrm{top}}_{F}$ and $E^{\mathrm{bulk}}_{F}$). Remarkably, this is in contrast to the situation for thin TI films, for which the band bending can be calculated reasonably well \textit{without} the knowledge of $E^{\mathrm{bulk}}_{F}$, as we will show in the next section. 

\subsection{Symmetric band bending in a thin film}

%
\subsubsection{Top and bottom topological surface states}
In the case of a thin TI film, there exist two TSS, one each at the top and bottom surfaces of the film (note that the bottom surface of the film corresponds to its interface to the substrate). The bottom TSS is not directly accessible by surface sensitive methods and therefore difficult to investigate. Thus, if no further information is available, it is a reasonable first approximation to assume that the properties, in particular the filling levels, of the bottom TSS are identical to the top TSS. We consider this symmetric approximation, in which only information about the top TSS is needed and in which the boundary conditions on the top and bottom surfaces are the same, in the present section.  

For a thin film of thickness $d$ with two surfaces and $E^{\mathrm{top}}_{F}=E^{\mathrm{bottom}}_{F}$, the problem is symmetric with respect to $z_0 = d/2$, and the appropriate boundary conditions are $\frac{dv}{dz} = 0$ for $z_0 = d/2$ and $v(0)=v(d)=v_{\mathrm{top}}=(E^{\mathrm{top}}_{F} - E^{\mathrm{bulk}}_{F})/k_B T$.  The solution of Eq. (\ref{ersteAbl}) for these boundary conditions, which evidently must lead to a symmetric band bending, can be expressed by 
\begin{widetext}
\begin{align}
	z(v) & = \sgn(-v) \frac{L}{\sqrt{2}} \int_{v_{\mathrm{top}}}^{v(z)} \sqrt{\frac{e^{u_b} + e^{-u_b}}{e^{u_b}\left(e^v - e^{v\left(\frac{d}{2}\right)} - v + v\left(\frac{d}{2}\right)\right) + e^{-u_b}\left(e^{-v} - e^{-v\left(\frac{d}{2}\right)} + v - v\left(\frac{d}{2}\right)\right)}}\, dv ,\,\, \mbox{for}\;0\le z\le\frac{d}{2}
	\label{eq:symmetric}
\end{align}
\end{widetext}
This equation has to be calculated iteratively, because the potential $v\left(\frac{d}{2}\right)$ is not known a priori. In the first iteration step  $v\left(\frac{d}{2}\right)$ is determined by numerical inversion, and then the equation is solved for the remaining values $v(z)$ in the interval $0\le z<\frac{d}{2}$. 
The case of symmteric band bending in a thin film has been considered before in the framework of the Schottky approximation \cite{Brahlek_2015,Brahlek_2014}. However, the latter approximation is only valid for depletion layers with a strong band bending $|e V_{\mathrm{top}}|\,\gg\,k_B T$, whence all free charge carriers are transferred from the interior of the film into the TSS, resulting in a completely depleted and insulating material in the film. But since both direction and quantity of the charge transfer are influenced by several parameters, as illustrated in section \ref{sec:bulkbandbending} in detail, depletion layers with small band bending as well as accumulation layers are conceivable, for which a description within the Schottky approximation is insufficient.  

\subsubsection{Quantization}
Due to finite thickness of the film, the quantization arising from the confinement of electrons in the direction perpendicular to the film must be taken into account. As stated above, this requires the solution of coupled Poisson's and Schr\"{o}dinger's equations.   For weak band bending, which turns out to be the relevant situation for the thin TI films studied here (see below), we can approximate the potential as a square potential with infinite barriers when solving the Schr\"{o}dinger equation. Further details of the calculation can be found in the supplemental material \cite{suppl}. It turns out that the calculated mobile charge carrier density in the thin TI film, which is discussed in the following, is reduced by a factor of $2$ to $2.5$ compared to the purely classical approach in which only Poisson's equation is solved.  

\begin{figure*}[t!]
\centering
\includegraphics[height=0.231\textwidth]{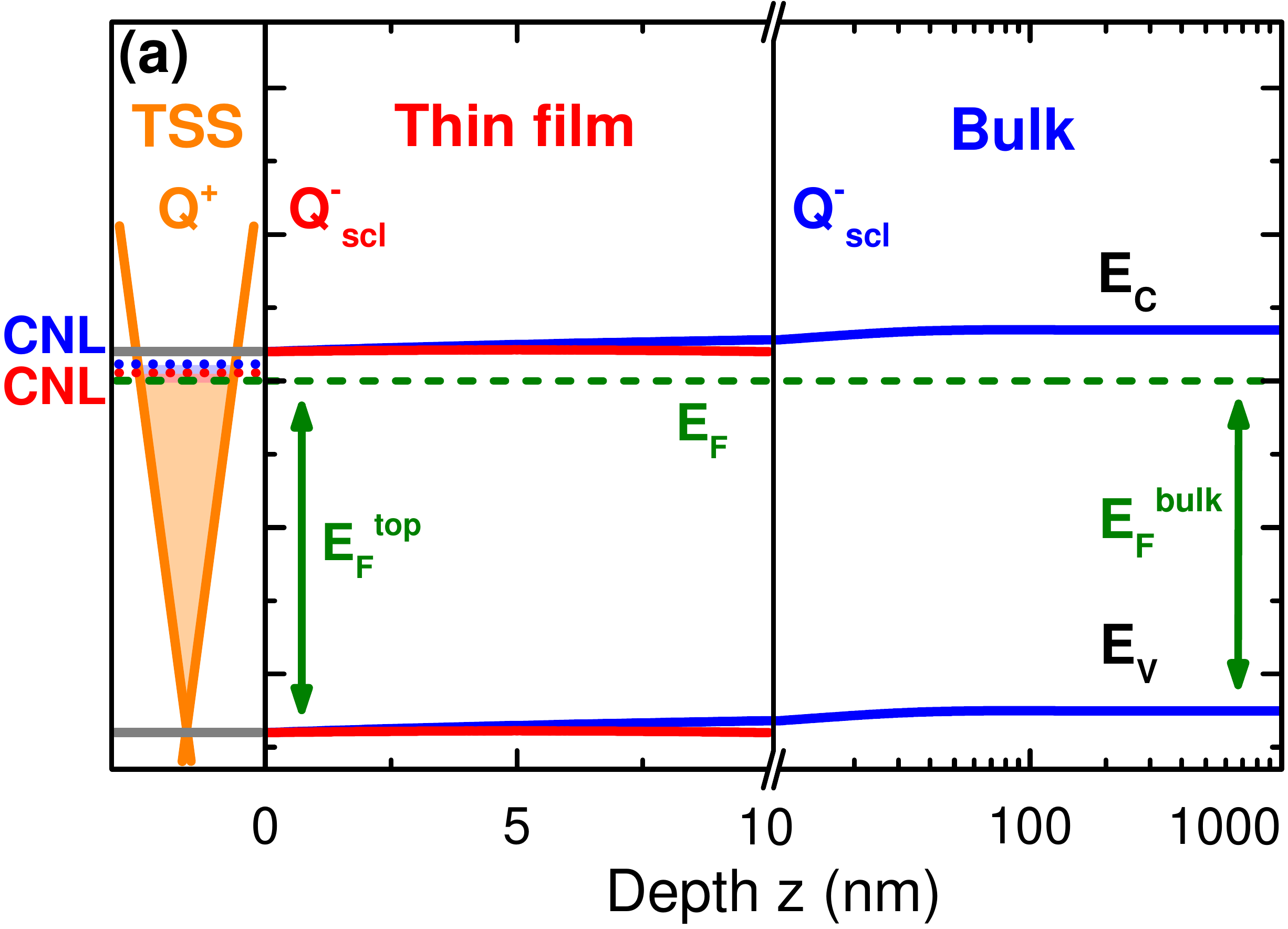}
\hspace{-0.175cm}
\includegraphics[height=0.231\textwidth]{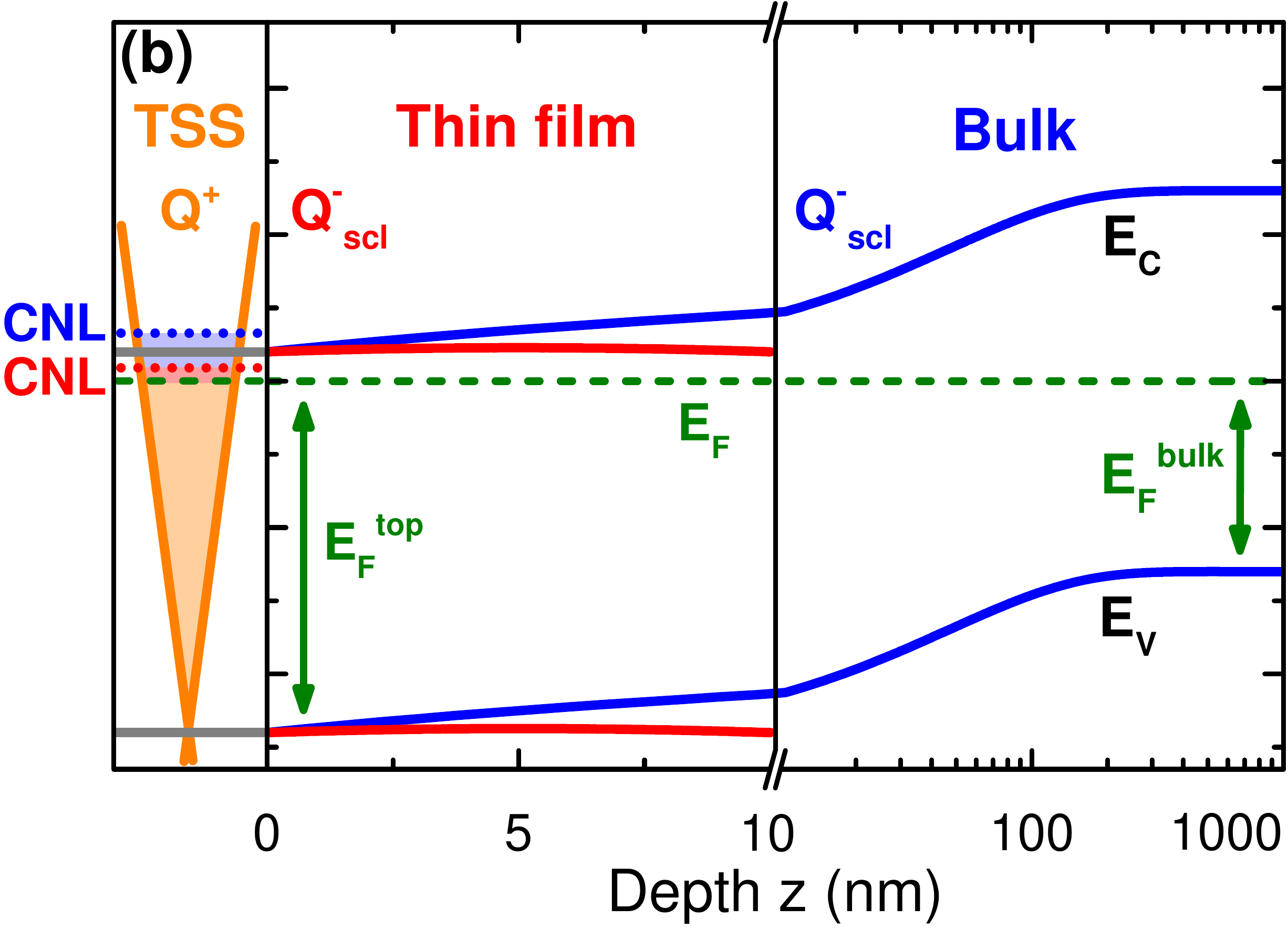}
\hspace{-0.175cm}
\includegraphics[height=0.231\textwidth]{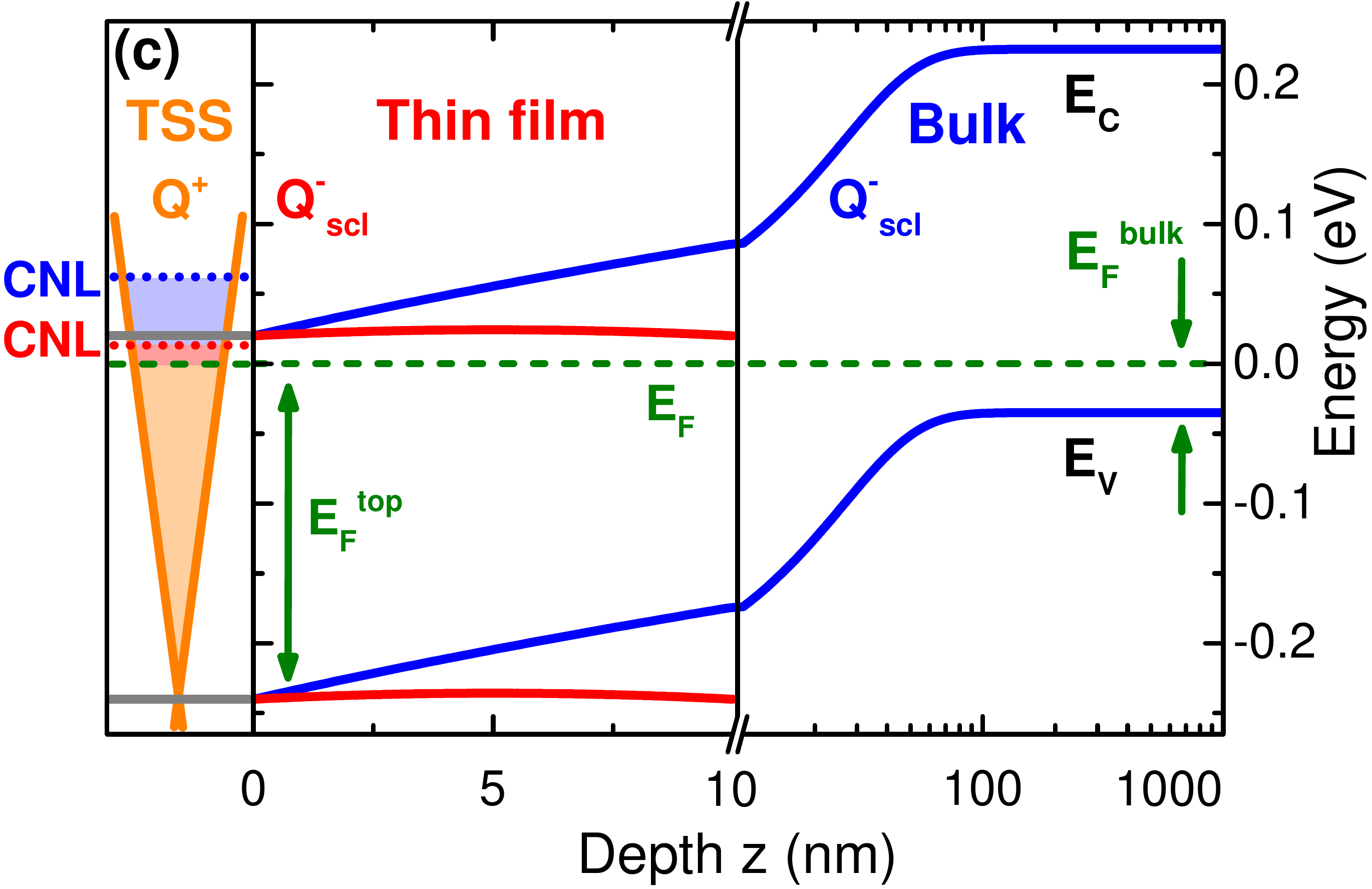}
	\caption{(Color online) Calculated band diagrams using the symmetric approximation for a $10\,\mathrm{nm}$ BiSbTe$_3$ thin film for three different dopant concentrations, i.e. (a) n-doped ($E_F^{\mathrm{bulk}}=225\,\mathrm{meV}$, $n_{\mathrm{bulk}}=4\times 10^{17}\,\mathrm{cm}^{-3}$), (b) intrinsic ($E_F^{\mathrm{bulk}}=130\,\mathrm{meV}$, $n_{\mathrm{bulk}}=p_{\mathrm{bulk}}=1\times 10^{16}\,\mathrm{cm}^{-3}$), and (c) p-doped ($E_F^{\mathrm{bulk}}=35\,\mathrm{meV}$, $p_{\mathrm{bulk}}=4\times 10^{17}\,\mathrm{cm}^{-3}$). For all diagrams, the surface Fermi energy $E^{\mathrm{top}}_{F}$ is set to 20 meV below the conduction band edge resulting from ARPES measurements \cite{Luepke_2017_2}. Further parameters of the calculation are listed in Tab. \ref{tab2}. In the central parts of the panels (a) - (c), the conduction and valence bands of the thin film are shown as function of depth $z$ from the surface (red lines), while in the left parts of (a) - (c) the partially filled Dirac cone (orange) of the top TSS is depicted. The blue lines show the band bending in a corresponding extended bulk crystal exhibiting the same dopant concentration as the thin film. 
	The Fermi energy (dashed green line) and both $E^{\mathrm{bulk}}_{F}$ and $E^{\mathrm{top}}_{F}$ (arrows) are indicated. 
	In contrast to the bulk case, the band positions in the thin film are not influenced significantly by the dopant concentration.    
The CNL (colored dotted lines) is different for the thin film (red) and the bulk (blue), but in both cases it is positioned above $E^{\mathrm{top}}_{F}$, giving rise to a downward band bending due to a transfer of negative charges from the TSS to the film or bulk.}
\label{fig3}
\vspace{-0ex}
\end{figure*}

\subsubsection{Influence of the film thickness}
In addition to the three parameters which jointly determine the band bending of a semi-infinite bulk TI crystal (see above), a fourth parameter becomes relevant in the thin-film limit --- the film thickness $d$.  It influences the amount of charge which can be transferred between the surface and the TI film. As the film thickness decreases, the total charge that can be transferred into the space charge region becomes increasingly limited, with the result of a less pronounced band bending, because charge neutrality imposes the same limit on the surface charges (TSS/DS) on both surfaces. Again, all four parameters $E^{\mathrm{top}}_{F}$, $E^{\mathrm{bulk}}_{F}$, CNL and $d$ are interdependent, such that one parameter is given by the other three. 

In order to illustrate the effect of limited charge transfer by means of an example, we display in Fig. \ref{fig3} the calculated band bending at room temperature for the TI material BiSbTe$_3$ (considered in more detail in section \ref{bisbte}) as function of depth $z$ for three different bulk dopant concentrations, i.e. n-doped (Fig. \ref{fig3}(a)), intrinsic (Fig. \ref{fig3}(b)) and p-doped (Fig. \ref{fig3}(c)). The red band diagrams belong to a 10 nm thin film, while the blue bands represent the corresponding extended bulk crystal with the same bulk dopant concentration and position of the surface Fermi energy $E^{\mathrm{top}}_{F}$. For the calculation, the band gap is set to 260 $\mathrm{meV}$ and a fixed $E^{\mathrm{top}}_F = 240\,\mathrm{meV}$ has been chosen. Moreover, the Dirac point coincides with the edge of the valence band and the effective mass is set to $m^* = 0.15\,m_e$, in agreement with ARPES measurements \cite{Luepke_2017_2}. All  parameters of the calculation are summarized in Tab. \ref{tab2}. 

In Fig. \ref{fig3} we observe that for the bulk crystal (blue curves) the dopant concentration, represented in the calculations by the value of the bulk Fermi energy $E^{\mathrm{bulk}}_{F}$, strongly influences the actual $E^{\mathrm{bulk}}_{F}$, and results in a strong increase of the near-surface band bending in Fig. \ref{fig3}(b) (intrinsic material) and \ref{fig3}(c) (p-doped material), since $E^{\mathrm{top}}_{F}$ is fixed close to the conduction band by construction. In contrast, in the thin-film limit (red curves) the bending of the bands remains weak and largely independent of the dopant concentration, even for different dopant types (Fig. \ref{fig3}(a) n-doped vs. \ref{fig3}(c) p-doped). Thus, in the thin film the valence band position relative to the Fermi level can differ strongly from its value in a bulk crystal with the same dopant concentration (Fig. \ref{fig3}(c)).    
This at first glance surprising behavior of the film can be rationalized by the behavior of the CNL of the TSS, which turns out to be very different for the thin film (red dotted lines in Fig. \ref{fig3}) and the bulk (blue dotted lines): On the one hand, charge transfer from the TSS into the TI (only this direction occurs in Fig. \ref{fig3}) is strongly suppressed for the thin film -- the red (thin film) CNL appears always close to $E_{F}$ at the surface, while the blue (bulk) CNL is consistently located above the red one, for Fig. \ref{fig3} (c) actually substantially above the red. On the other hand, even the comparatively little charge that is transferred from the TSS into the thin film is sufficient to change the whole film from intrinsic to n-type (Fig. \ref{fig3} (b)) or indeed from p-type to n-type (Fig. \ref{fig3} (c)). 
At this point, we must distinguish between the dopant and doping concentrations in the TI material. The former describes the concentration of defects in the material, determined by the growth conditions, while the latter specifies the concentration of mobile charge carriers in the material. In an extended crystal the concentration of dopants directly controls the concentration of mobile carriers, i.e. the doping, but for a thin TI film with an additional source of charges, the TSS, this is not true. For example, in Fig. \ref{fig3} (b) the dopant concentration in the film corresponds to an intrinsic bulk TI material, but the additionally transferred charges from the TSS are sufficient to nearly fully n-dope the film (Fig. \ref{fig3}(b)), or even completely saturate all acceptors in the p-type film material of Fig. \ref{fig3}(c) and still result in essentially the same $n$-doping as in Fig. \ref{fig3} (b). As there is a small increase in charge transfer from Fig. \ref{fig3} (a) to (c), also the band bending increases slightly, but it is still very weak compared to the bulk case (blue). 

\begin{table}[b]
\begin{ruledtabular}
\centering
\begin{tabular}{ l l }
&\\[-1.5ex]
Parameter & Value\\[1ex]
\hline\\[-1ex]
$m^*$ & $0.15\,m_e$ \\   
	$v_{\mathrm{Fermi}}$ & $5.6 \times 10^{-5}\,\mathrm{ms}^{-1}$ \\
	$T$ & $300\,\mathrm{K}$\\
	$E_{\mathrm{Dirac}}$ & $0\,\mathrm{eV}$ (at valence band)\\
	$E_{\mathrm{gap}}$ & $0.26\,\mathrm{eV}$\\[0.5ex]
	$E^{\mathrm{top}}_{F}$ ($V_{\mathrm{gate}} = 0\,\mathrm{V}$) & $0.24\,\mathrm{eV}$\\[0.5ex]
	$n^{\mathrm{top}}_{\mathrm{TSS}}$ ($V_{\mathrm{gate}} = 0\,\mathrm{V}$) & $4 \times 10^{12}\,\mathrm{cm}^{-2}$\\
\end{tabular}
	\caption{Different fixed parameters used as input for the band banding calculations for the TI system BiSbTe$_3$ as reported in \cite{Luepke_2017_2}.}
\label{tab2}
\end{ruledtabular}
\end{table}

In conclusion, Fig. \ref{fig3} illustrates a significant difference compared to the case of the semi-infinite bulk crystal in the section \ref{sec:bulkbandbending}: For a 10 nm thin film the band bending across the complete film is largely independent of the dopant concentration in the film; in fact, the bands remain nearly flat at the surface position $E^{\mathrm{top}}_{F}$ for widely varying bulk dopant levels. Thus, the position of the Fermi energy inside the film deviates strongly from the bulk Fermi energy $E^{\mathrm{bulk}}_{F}$ of a corresponding semi-infinite bulk crystal with the same dopant concentration. Notably, this allows the approximation of the total mobile charge carrier concentration in the thin film from information gained from surface-sensitive measurements, even if the dopant concentration inside the film material remains unknown. 

\subsubsection{Screening}
The weak band bending in the thin TI film is linked to the long screening length $L$ compared to the small film thickness $d$. This is immediately obvious from Eq. (\ref{pois2}), which shows that the curvature of the band bending potential $v$ is inversely proportional to the square of $L$. A small curvature of course also limits the value of $v$ that can be reached over the thickness $d\ll L$ of the film. 
If the film thickness $d$ approaches the screening length $L$, the total charge transfer between TSS and film becomes larger, resulting in an increasingly stronger band bending. For $d>2L$, the space charge region in the film becomes the same as in an extended bulk crystal. 

According to Eq. (\ref{eq:Debye}), the screening length $L$ also depends on the carrier densities $n_b+p_b$. The larger $n_b+p_b$, the shorter is $L$. Therefore, the near independence of the band bending on the bulk dopant concentration in the thin film is only found for not too high dopant levels in the TI material. Furthermore, a long screening length can also be caused by a large dielectric constant. Many TI materials have such large dielectric constants. For example, for BiSbTe$_3$ a value of $\epsilon_r \approx 100$ is reported \cite{Kim_2012,Yang_2015,ViolBarbosa_2013}. 

\subsubsection{Degenerate doping}
In Fig. \ref{fig3}, only non-degenerate doping levels in the TI have been considered, all of which apparently lead to a downward band bending. Of course, it is in principle also possible that the dopant concentration is sufficiently high for the TI material to be degenerately n-doped, with $E_{F}^\mathrm{bulk}$ above the conduction band edge. According to our calculations, electrons from the film then flow into the TSS. But because of the limited number of available charge carriers, the Fermi energy in the film drops below the conduction band and in the end nearly coincides with $E_{F}^\mathrm{top}$, resulting in a weak upward band bending. Thus, again the finite film thickness $d\ll L$ leads to an effective suppression of band bending in the film in the case of a slight degenerate doping.  

Only for very strong degenerate doping the screening length can become so small that a stronger band bending is obtained in the thin TI film leading to a pronounced dependency of the charge carrier density in the film on the film dopant concentration. 
However, the case of such a strong degenerately doped TI film, where the Fermi energy is located deep inside the bulk bands, is undesirable, as in this case the interior of the film will become the dominant parasitic conduction channel due to the high DOS of the bulk states. State of the art growth of ternary or even quaternary materials avoids this undesirable case of degenerate doping.   
Furthermore, it is clear that for degenerately doped materials the Boltzmann approximation is not valid any more. Instead, all calculations must employ the Fermi-Dirac distribution. Indeed, such calculations are possible, but the equations are more complex.   

Thus, all calculations presented in this paper are performed for the more relevant case of non-degenerate doping. We note that already a surface Fermi level $E^{\mathrm{top}}_{F}$ only 20 meV below the conduction band edge, which is assumed in Fig. \ref{fig3}, stretches the validity of the Boltzmann approximation. In this case, the deviation between the Fermi-Dirac and Boltzmann distributions is up to 50 \%, but as the Boltzmann distribution is larger compared to the Fermi-Dirac distribution, the calculated band bending will be overestimated. The true band bending will be even smaller. 

\subsubsection{The example of BiSbTe$_3$} \label{bisbte}
Here, we focus on the specific TI material (Bi$_{1-x}$Sb$_{x}$)$_2$Te$_3$ with $x=0.5$, i.e. BiSbTe$_3$, which was studied extensively in ref. \cite{Luepke_2017_2,Kellner_2015}. Fig. \ref{fig3} that we used above to illustrate generic properties of band bending in thin TI films, was already calculated for this case. The assumption of a fixed $E^{\mathrm{top}}_F$ in Fig. \ref{fig3} (a) to (c) corresponds to the situation in which a definite value of $E^{\mathrm{top}}_F$ is measured by ARPES on an MBE-grown BiSbTe$_3$ film; the different values of the surface CNL in Fig. \ref{fig3} (a) to (c) would then indicate different surface concentrations of adsorbates and/or surface defects. From this perspective, each band diagram in Fig. \ref{fig3} corresponds to a distinct initial state which after charge transfer between the surface and the film interior results in the final state observed by a particular ARPES measurement (i.e. in the present case $E^{\mathrm{top}}_F$ = 240 meV \cite{Luepke_2017_2}). Our aim is to identify the most likely initial state and therefore also the corresponding band bending scenario.

In BiSbTe$_3$ we observe $E^{\mathrm{top}}_F$  to be very close to the conduction band edge, as measured by ARPES \cite{Luepke_2017_2,Kellner_2015}. As demonstrated in Fig. \ref{fig3}, this results in \textit{weak} downward band bending for p-, intrinsic and n-type dopant concentrations. 
On the other hand, if the film material is strongly or even degenerately n-doped, a result can also be a weak upward bending (see section above). 
Generally, such a situation with upward band bending, obtained for a $E^{\mathrm{top}}_F$ positioned further away from the conduction band edge, has been discussed recently within the Schottky approximation \cite{Brahlek_2015,Brahlek_2014}.
However, a high dopant concentration in the bulk would be in contradiction with the aim to grow films of ternary TI material systems with a minimal conductivity of the interior of the film gap \cite{Lanius_2016}. For this reason, we exclude the case of upward band bending for the material BiSbTe$_3$ discussed here.

Furthermore, as long as $E^{\mathrm{top}}_F$ in BiSbTe$_3$ is positioned below the conduction band edge, as observed in ARPES \cite{Luepke_2017_2,Kellner_2015}, a \textit{strong} downward band bending which would produce a non-topological 2DEG near the surface can also be excluded. A 2DEG has been observed in recent studies \cite{Bianchi_2010,King_2011,Benia_2011}, but in all these cases $E^{\mathrm{top}}_F$ is positioned deep inside the conduction band, while the Fermi energy inside the film material is still inside the band gap.

The doping character of a thin TI film can be strongly influenced by the stoichiometric composition \cite{Kong_2011,He_2013,Weyrich_2016}. 
In some experiments on (Bi$_{1-x}$Sb$_{x}$)$_2$Te$_3$ and (Bi$_{1-x}$Sb$_{x}$)$_2$Se$_3$ with different stoichiometric parameter $x$, the surface Fermi energy $E^{\mathrm{top}}_F$ is found closer to mid-gap \cite{Zhang_2011,Satake_2018}. If such a situation was found for BiSbTe$_3$, for appropriate film thicknesses and screening lengths the thin film could possibly become fully depleted, as almost all negative mobile charge carriers flow into the TSS. This would result in a midgap position of the Fermi energy inside the film, and, in these circumstances, the film material would be completely insulating. 
However, this state of affairs can be excluded for the case of BiSbTe$_3$, since $E^{\mathrm{top}}_F$ is located very close to the conduction band edge, as measured by ARPES \cite{Luepke_2017_2,Kellner_2015}.
We can therefore conclude that for BiSbTe$_3$ from the experiment in ref. \cite{Luepke_2017_2} the band bending must be \textit{weakly} downward. 

\subsubsection{Mobile charge carrier density}
Within the Boltzmann approximation, the mobile charge carrier density can be calculated from the potential $v(z)$ by $n_e(z) = n_b\,\exp[v(z)]$ and $p_h(z) = p_b\,\exp[-v(z)]$ for electrons and holes, respectively. Integrating the sum of $n_e(z)$ and $p_h(z)$ over the thickness $d$ of the thin film results in the total mobile charge carrier density 
\begin{align}
n_{\mathrm{film}}\left(E^{\mathrm{top}}_F,E^{\mathrm{bulk}}_F\right) & = \int_{0}^{d} n_{i} \left(e^{\left[u_b\left(E^{\mathrm{bulk}}_F\right) +v\left(z,E^{\mathrm{top}}_F,E^{\mathrm{bulk}}_F\right)\right]} \right. \nonumber \\[1ex]
& \hspace{2ex} \left. +\; e^{-\left[u_b\left(E^{\mathrm{bulk}}_F\right) + v\left(z,E^{\mathrm{top}}_F,E^{\mathrm{bulk}}_F\right)\right]}\right) \,dz 
\label{eq:nfilm}
\end{align}
inside the film material, with $n_i$ denoting the intrinsic charge carrier concentration. In general, $n_{\mathrm{film}}$ is both dependent on the dopant concentration in the film, which is represented in the equation by the bulk Fermi energy $E^{\mathrm{bulk}}_{F}$ of a corresponding bulk crystal with same dopant density, and on the surface Fermi energy $E^{\mathrm{top}}_F$.

\begin{figure}[t!]
\centering
\includegraphics[width=0.45\textwidth]{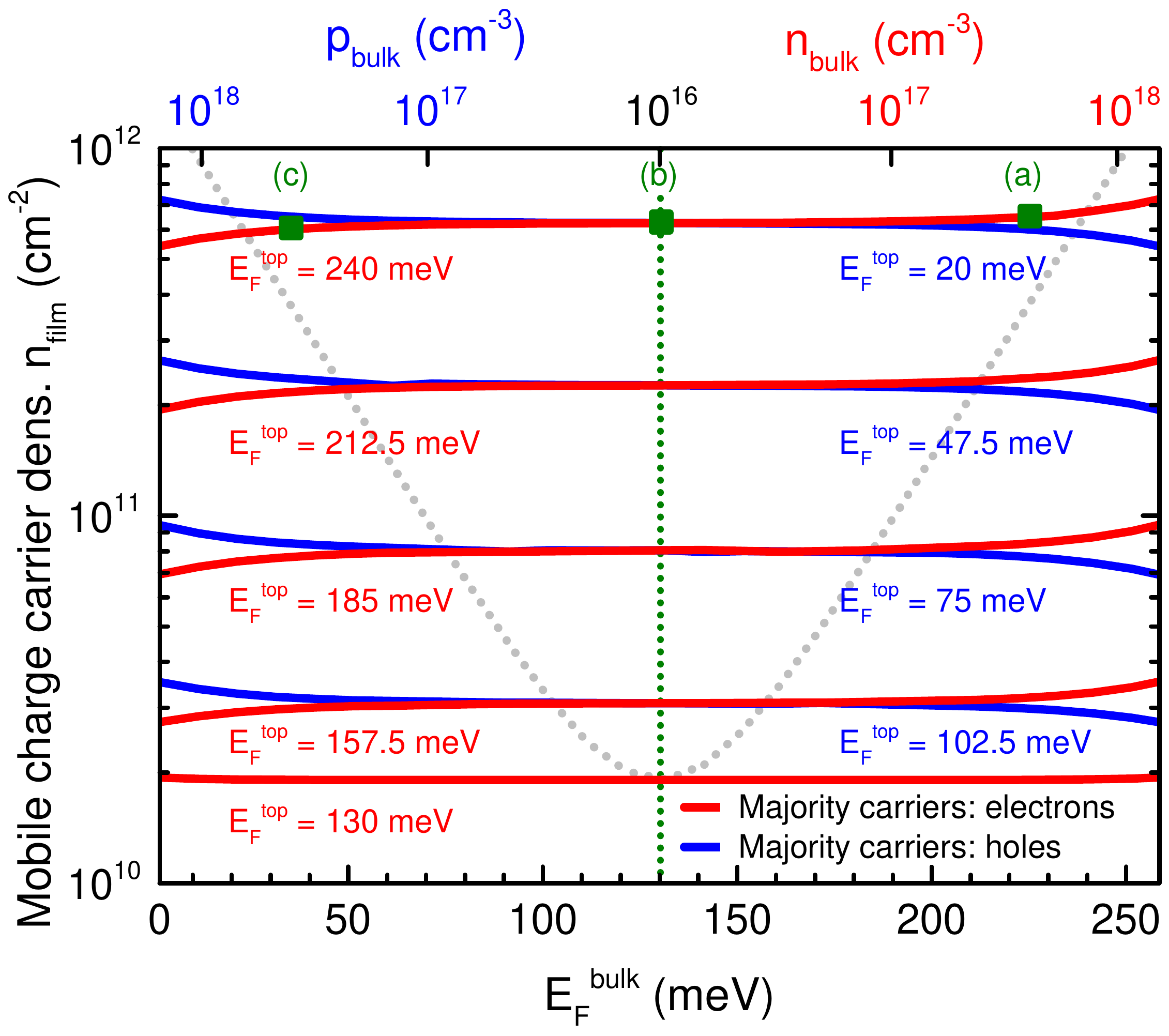}
	\caption{(Color online) Integrated total mobile charge carrier density $n_{\mathrm{film}}$ based on the symmetric approximation for a thin TI film as function of the dopant concentration represented by the bulk Fermi energy $E^{\mathrm{bulk}}_F$. The surface Fermi energy $E^{\mathrm{top}}_F$ is an additional parameter. For $E_F^{\mathrm{bulk}}$ inside the band-gap, the calculated film carrier density is approximately constant and, thus, independent of the dopant concentration. The type of majority carriers in the thin film (either electrons (red) or holes (blue)) is indicated. On the upper horizontal axis, the associated 3D majority charge carrier densities of a bulk crystal, i.e. $p_{\mathrm{bulk}}$ on the left and $n_{\mathrm{bulk}}$ on the right, are shown. The green squares correspond to the values of $E^{\mathrm{bulk}}_{F}$ and $E^{\mathrm{top}}_F$ used in the band diagrams in Fig.~\ref{fig3}(a)-(c), while the dotted green line corresponds to a vertical cut plotted in Fig.~\ref{fig8} along the red dotted diagonal. For comparison, the dotted gray line shows the strong exponential dependence of the total charge carrier density expected inside an extended bulk crystal (values converted to 2D units by integrating over a width of $10\,\mathrm{nm}$ inside the bulk).}
\label{fig4}
\vspace{-0ex}
\end{figure}

In Fig. \ref{fig4}, the total mobile charge carrier density as function of the dopant level expressed by the bulk Fermi energy $E^{\mathrm{bulk}}_{F}$ is plotted for a thin film of BiSbTe$_3$ with $d=10$\,nm \cite{Luepke_2017_2}. The surface Fermi energy $E^{\mathrm{top}}_{F}$ is an additional free parameter. As long as $E^{\mathrm{top}}_{F}$ is well within the band gap of  BiSbTe$_3$, the charge carrier concentration in the thin film is nearly independent of $E^{\mathrm{bulk}}_{F}$, i.e. $n_{\mathrm{film}}(E^{\mathrm{top}}_F,E^{\mathrm{bulk}}_F) \approx n_{\mathrm{film}}(E^{\mathrm{top}}_F)$. Only in the vicinity of the band edges, where also the Boltzmann approximation becomes less accurate, a deviation from the constant charge carrier concentration is observed. Note that the mobile charge carrier concentration is of course influenced by the surface Fermi energy $E^{\mathrm{top}}_{F}$.  
 
\begin{figure*}[t!]
\centering
\includegraphics[height=0.245\textwidth]{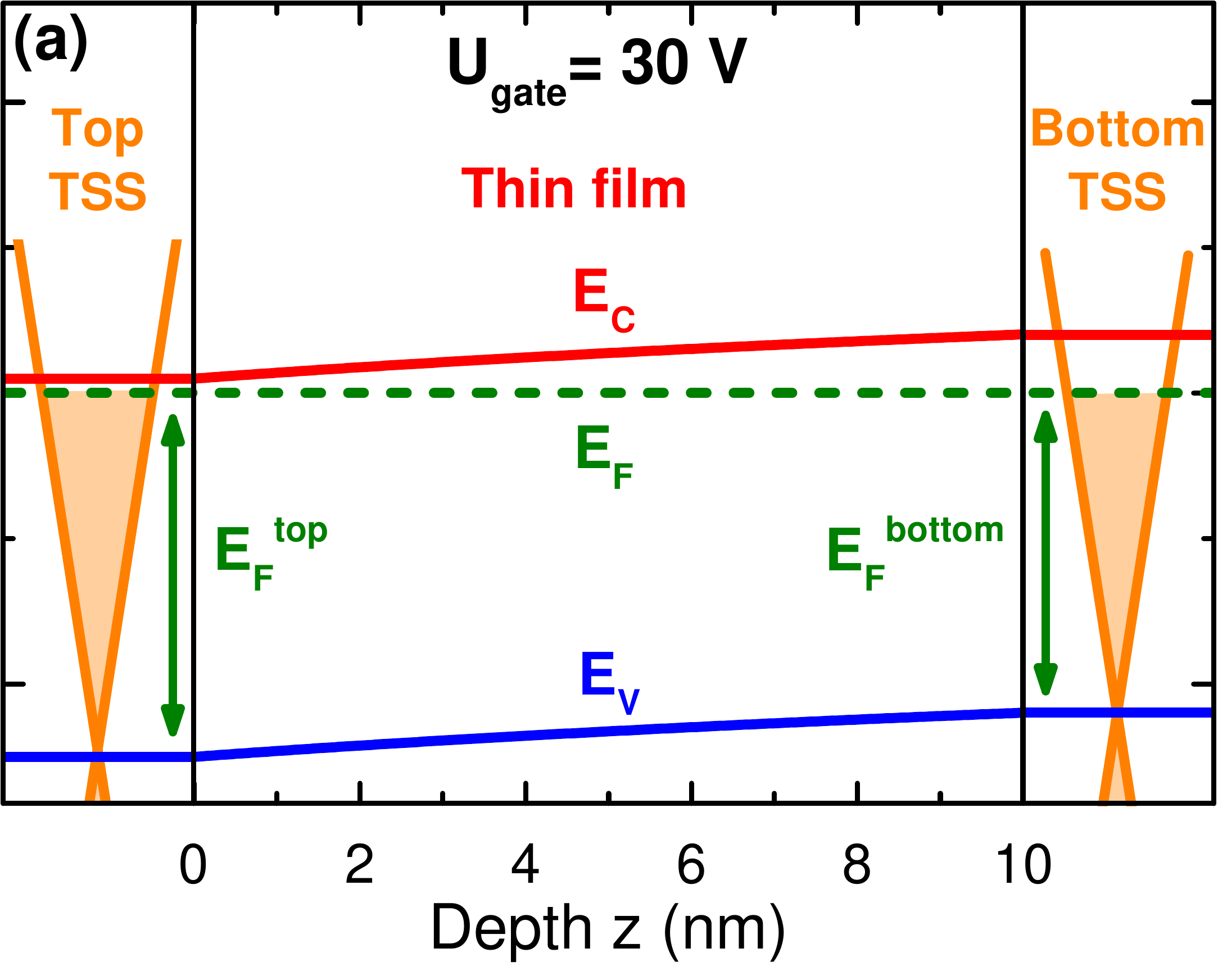}
\hspace{0.1cm}
\includegraphics[height=0.245\textwidth]{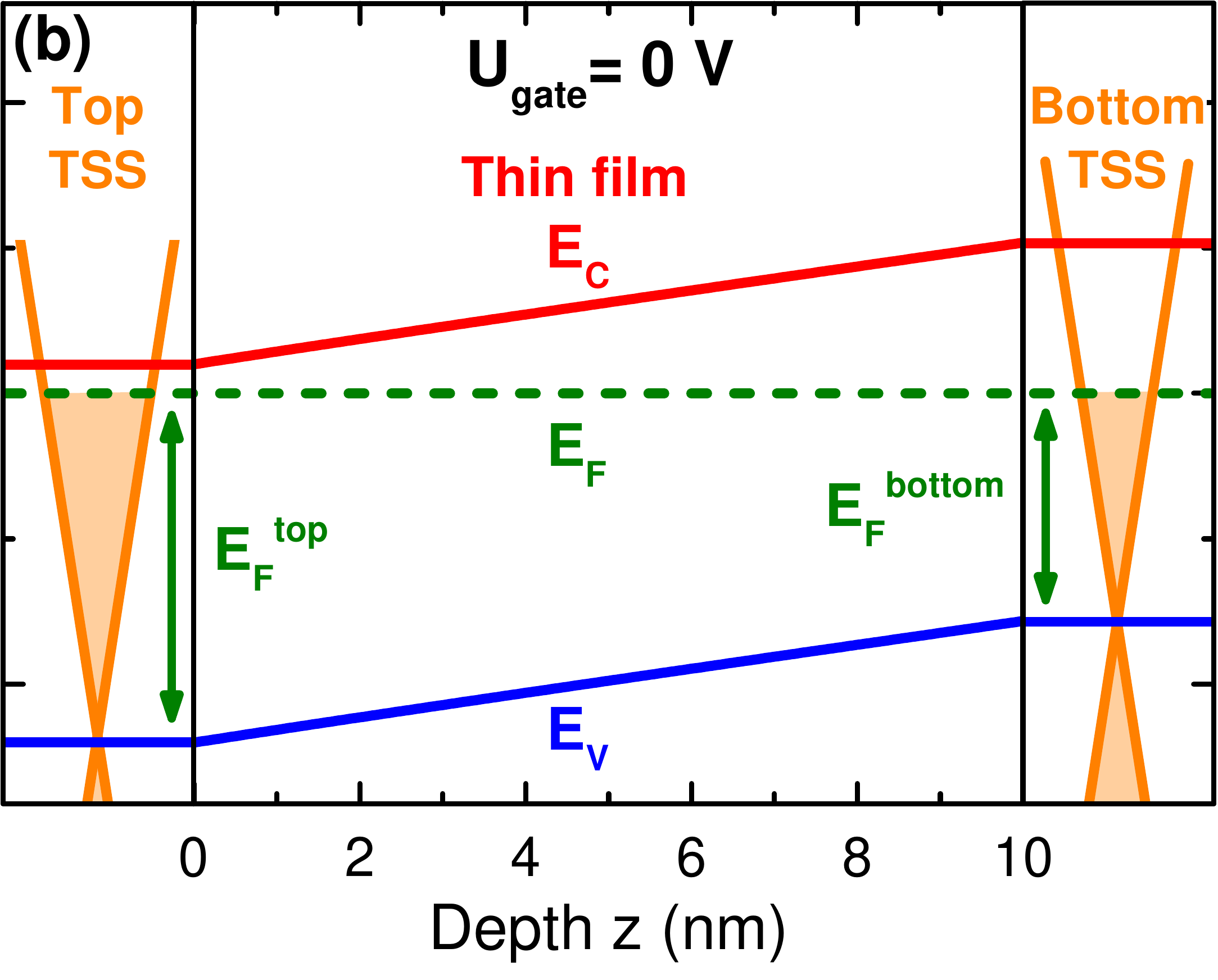}
\hspace{0.1cm}
\includegraphics[height=0.245\textwidth]{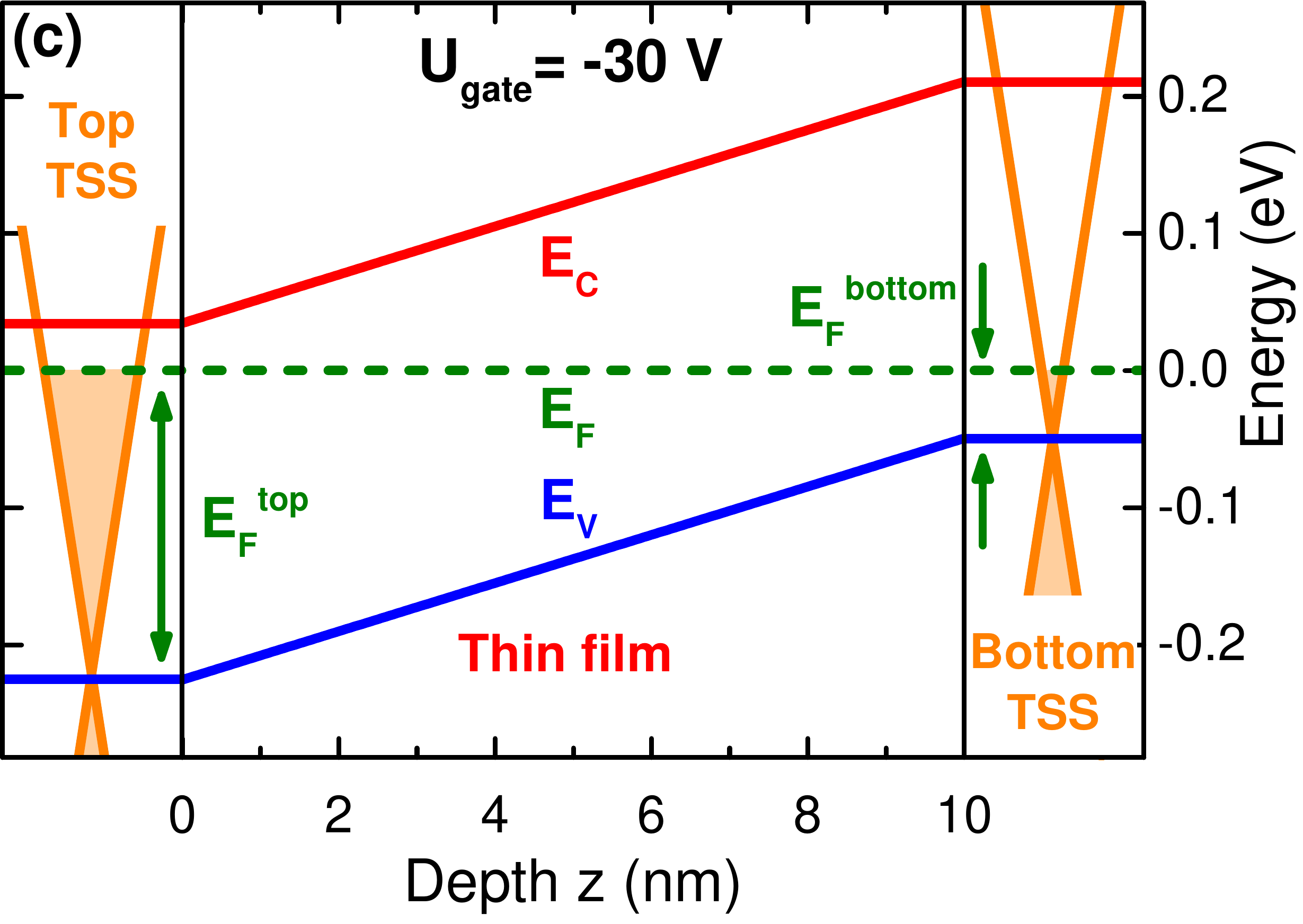}
	\caption{(Color online) Gate-dependent band-bending for a $10\,\mathrm{nm}$ thin BiSbTe$_3$ film with intrinsic dopant concentration ($E^{\mathrm{bulk}}_F = 130\,\mathrm{meV}$, $n_{\mathrm{bulk}} = p_{\mathrm{bulk}} = 1 \times 10^{16}\,\mathrm{cm}^{-3}$) as function of the depth $z$ into the film for different gate voltages, i.e. $30\,\mathrm{V}$~(a), $0\,\mathrm{V}$~(b) and $-30\,\mathrm{V}$~(c). The calculation is based on asymmetric boundary conditions with the same parameters as used in Fig. \ref{fig3} and additional information from gate-dependent transport measurements \cite{Luepke_2017_2}. 
In the central part of each diagram, the bands inside the thin film are depicted, while on the left and right side of the panels the top and bottom TSS are shown, respectively. The Fermi energy $E_{F}$ (green dashed line) and both $E^{\mathrm{top}}_{F}$ and $E^{\mathrm{bottom}}_{F}$ (green arrows) are indicated. 
Due to different surface Fermi energies $E^{\mathrm{top}}_{F}$ at $z=0$ and $E^{\mathrm{bottom}}_{F}$ at $z=10\,\mathrm{nm}$ ((a)~$250\,\mathrm{meV}$, $220\,\mathrm{meV}$; (b)~$240\,\mathrm{meV}$, $155\,\mathrm{meV}$; (c)~$225\,\mathrm{meV}$, $50\,\mathrm{meV}$) the resulting band bending in the thin film is asymmetric. 
The bottom gate voltage influences both surface Fermi levels in a different way, as described in detail in ref. \cite{Luepke_2017_2}, and leads to an increase in the strength of band bending from (a) to (c). Hence, the mobile charge carrier density inside the thin film is significantly influenced by the gate voltage.}
\label{fig6}
\vspace{-0ex}
\end{figure*}

As a quantitative example, for  $E^{\mathrm{top}}_{F}=240$\,meV, as found in ref. \cite{Luepke_2017_2}, the calculated charge carrier concentration in the BiSbTe$_3$ film is $\sim 6 \times 10^{11}\,\mathrm{cm}^{-2}$, which is close to the charge carrier density inside the Dirac cone of the top TSS of $4 \times 10^{12}\,\mathrm{cm}^{-2}$, indicating that there may be a contribution by the interior of the film to the overall charge transport in the TI system. Fortunately, in this example it turns out that the mobility of the bulk material at room temperature is very low ($<2\,\mathrm{cm}^2/\mathrm{Vs})$ compared to the TSS channels \cite{Luepke_2017_2,He_2013,Weyrich_2016,Zhang_2011,He_2012_2}, with the result that the conductivity of the film interior is negligible. However, this could be completely different for other material systems, especially at low temperatures, where the mobility may be larger by factors of 10 to 50. Thus, each case has to be considered individually. 

We can thus conclude that by measuring the surface Fermi energy $E^{\mathrm{top}}_{F}$, the charge carrier concentration inside the TI thin film can be determined, even if the dopant concentration ($E^{\mathrm{bulk}}_F$) is unknown. If also the mobility of the TI material is known, the conductivity of the parasitic conduction channel that is constituted by the film interior in the TI system can directly be calculated. We stress again that this is in strong contrast to the behavior of an extended bulk crystal, where the charge carrier density shows a strong dependence on the dopant concentration, which is indicated by the dotted gray curve in Fig. \ref{fig4}. 


\subsection{Asymmetric band bending in a thin film}
\label{sec:Asymmetric_band_bending_in_a_thin_film}

In general, different environments and thus different surface defect densities will result in distinct surface Fermi energies $E^{\mathrm{top}}_{F}$ and $E^{\mathrm{bottom}}_{F}$ at the top  and the bottom of the TI film. Therefore, the boundary conditions at the top and bottom surfaces are not identical, yielding an asymmetric band bending. In this situation, the calculation of the band bending is more complex than in the symmetric limit. An outline is given in Appendix B. 

To calculate the band bending in a thin film with asymmetric boundary conditions, both $E^{\mathrm{top}}_{F}$ and $E^{\mathrm{bottom}}_{F}$ have to be known. While $E^{\mathrm{top}}_{F}$ can be measured by ARPES (see above), $E^{\mathrm{bottom}}_{F}$ must be extracted from gate-dependent four-point transport measurements. The procedure is described in detail in ref. \cite{Luepke_2017_2}. To this end, a bottom gate electrode is integrated into the substrate of the TI film and the four-point resistance of the film is measured with a multi-tip STM as function of the applied gate voltage. Due to quantum capacitance effects that arise because the DOS in the TSS is small, the induced electric field of the gating electrode does not only have a strong influence on the filling level of the bottom TSS, but also on that of the top TSS. Thus we can conclude that $E^{\mathrm{bottom}}_{F}$ can be determined via gate-dependent measurements, if $E^{\mathrm{top}}_{F}$ is known. 

For the quantization effects, we approximate the potential in the film by a triangular well of length $d$ with infinite barriers on both sides. Its slope is determined by the difference between the top and bottom surface Fermi levels $E^{\mathrm{top}}_{F}$ and $E^{\mathrm{bottom}}_{F}$. This approximation is applicable if the band bending in the film is weak, otherwise the potential would exhibit a curvature.  Further details of the calculations are discussed in the supplemental material \cite{suppl}. 

\begin{figure*}[t!]
\centering
\includegraphics[height=0.31\textwidth]{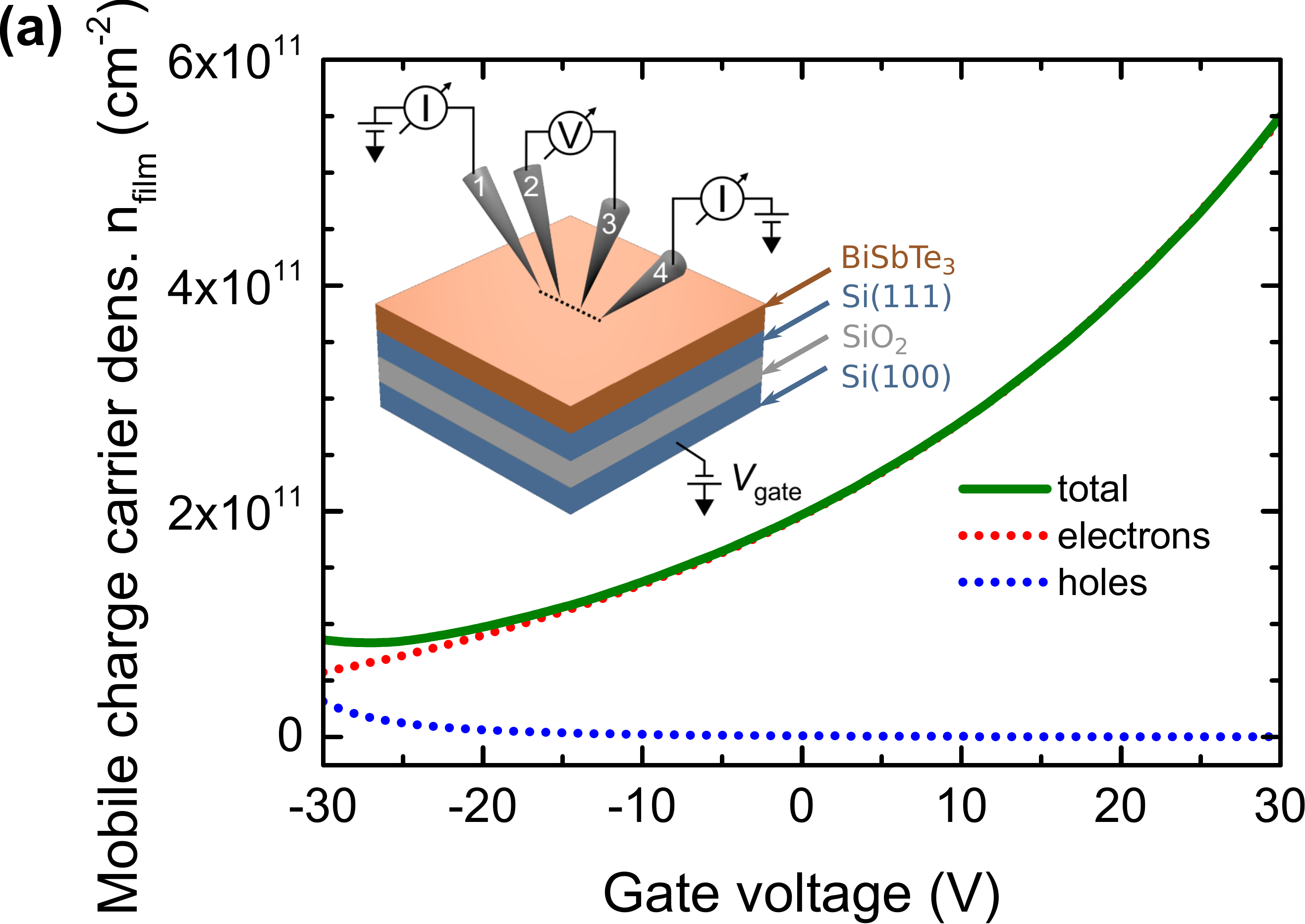}
\hspace{0.2cm}
\includegraphics[height=0.305\textwidth]{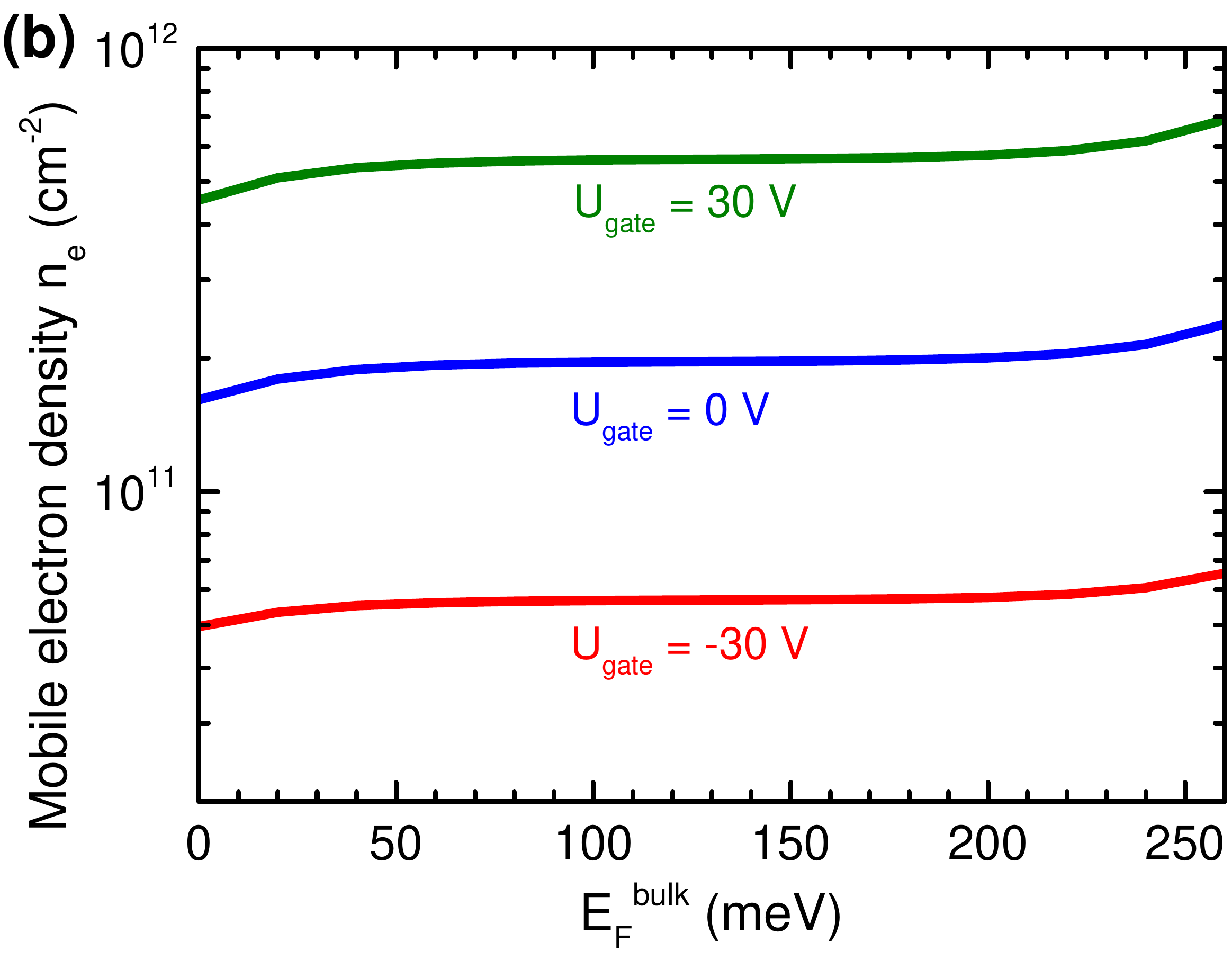}
\caption{(Color online) (a) Calculated total mobile charge carrier density $n_{\mathrm{film}}$ inside a $10\,\mathrm{nm}$ thin BiSbTe$_3$ film as function of the applied gate voltage (solid green line). The calculation is based on asymmetric boundary conditions for a thin film with intrinsic dopant concentration $n_{\mathrm{bulk}} = p_{\mathrm{bulk}} = 1 \times 10^{16}\,\mathrm{cm}^{-3}$ ($E^{\mathrm{bulk}}_F = 130\,\mathrm{meV}$), and on parameters from ref. \cite{Luepke_2017_2}. The individual contributions by electrons (red) and holes (blue) are depicted by the dotted lines. In the inset, the measurement setup and the sample configuration is shown \cite{Luepke_2017_2}. (b) Integrated total mobile charge carrier density inside the $10\,\mathrm{nm}$ BiSbTe$_3$ film as function of dopant concentration (represented by $E^{\mathrm{bulk}}_F$ of a corresponding extended crystal) and gate voltage as additional parameter.} 
\label{fig7}
\vspace{-0ex}
\end{figure*}

\begin{figure}[b!]
\centering
\includegraphics[width=0.42\textwidth]{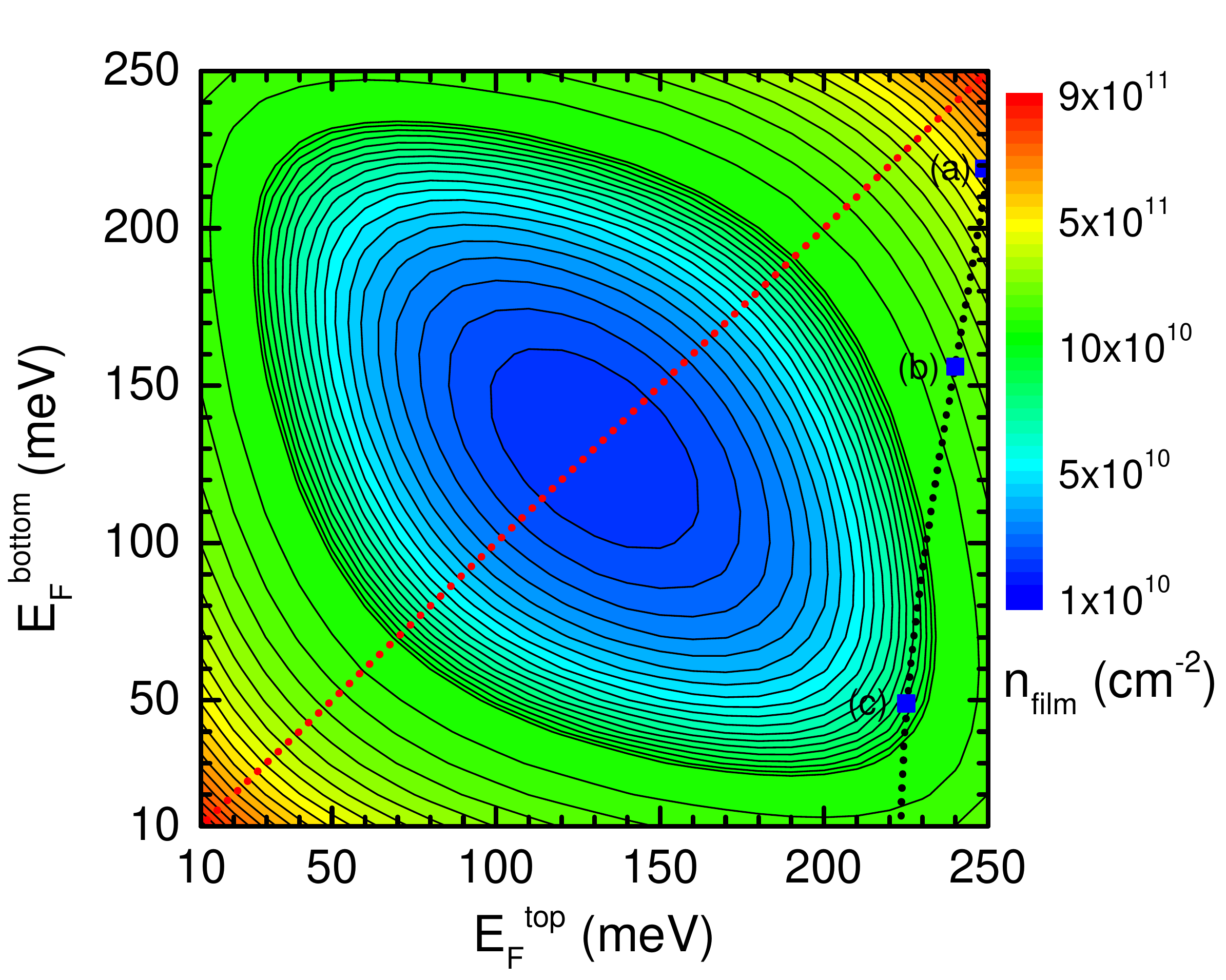}
	\caption{(Color online) Color plot of the mobile charge carrier density $n_{\mathrm{film}}$ as function of the top and bottom surface Fermi energy $E^{\mathrm{top}}_{F}$ and $E^{\mathrm{bottom}}_{F}$, respectively, for the specific case considered in ref. \cite{Luepke_2017_2}. The charge carrier density is calculated for a thin film with intrinsic dopant concentration. The diagonal dotted red line indicates the symmetric case with $E^{\mathrm{top}}_{F} = E^{\mathrm{bottom}}_{F}$ and corresponds to a cut at the position of the dotted green line in Fig. \ref{fig4}. The marked blue points correspond to the charge carrier concentrations resulting from the band diagrams depicted in Fig. \ref{fig6}(a)-(c), while the dotted black line in between shows the general gate-dependent behavior.} 
\label{fig8}
\vspace{-0ex}
\end{figure}

In Fig. \ref{fig6}, the asymmetric band bending in a BiSbTe$_3$ film with $d=10\,\mathrm{nm}$ is shown for three different gate voltages, i.e. $+30$\,V, $0$\,V, and $-30$\,V. All three band diagrams are calculated for an intrinsic film dopant concentration (i.e., $E^{\mathrm{bulk}}_F = 130\,\textrm{meV}$). Similar to the symmetric case, the band bending turns out to be approximately the same  for different dopant concentrations (not shown). However, Fig. \ref{fig6} clearly shows that the varying gate voltage causes different concentrations of induced charge carriers in the top and bottom TSS \cite{Luepke_2017_2}, and thus results in different filling levels of the Dirac cones and different surface Fermi energies $E^{\mathrm{top}}_{F}$ and $E^{\mathrm{bottom}}_{F}$. Specifically, the band bending increases if the gate voltage becomes more negative (Fig. \ref{fig6}(a)-(c)), as the charge carrier concentration in the bottom TSS is stronger influenced by the gating than in the top TSS. As a consequence, also the charge carrier density in the thin film varies strongly with gate voltage. 
This dependency of the total mobile charge carrier density on the gate voltage is plotted for the case of a BiSbTe$_3$ film with $d=10\,\mathrm{nm}$ in Fig. \ref{fig7}(a). The inset depicts the measurement setup used in ref. \cite{Luepke_2017_2}. 
However, if the dopant concentration in the thin film is changed at a fixed gate voltage, the integrated charge carrier density remains nearly constant. This is illustrated in Fig. \ref{fig7}(b), where the strongly $z$-dependent carrier density resulting from Fig. \ref{fig6} is integrated over the film (Eq. (\ref{eq:nfilm})) and plotted as function of the film dopant concentration, represented by the bulk Fermi energy of a corresponding extended crystal, and the applied gate voltage as additional parameter. We observe that nearly the same behavior results as in the symmetric approximation (Fig. \ref{fig4}), i.e. the total mobile charge carrier density $n_{\mathrm{film}}$ in the thin film is nearly independent of $E^{\mathrm{bulk}}_{F}$. It does, however, strongly depend on the applied gate voltage. 
%

Finally, we compare the results for asymmetric boundary conditions to those obtained in the symmetric case. It is to be expected that for a given parameter set $E^{\mathrm{top}}_{F}$ and $E^{\mathrm{bottom}}_{F}$ the mobile charge carrier density $n_{\mathrm{film}}$ will differ from one that is calculated in the symmetric approximation. This becomes apparent in Fig. \ref{fig8}, where we have plotted the total mobile charge carrier density of a BiSbTe$_3$ film with $d=10$\,nm (based on parameters from ref. \cite{Luepke_2017_2}) as function of the top and bottom surface Fermi levels. The diagonal dotted red line corresponds to symmetric boundary conditions with $E^{\mathrm{top}}_{F} = E^{\mathrm{bottom}}_{F}$. The plot reveals that for a specific measured $E^{\mathrm{top}}_{F}$ and unknown $E^{\mathrm{bottom}}_{F}$, i.e. along a vertical cut through the diagram, $n_{\mathrm{film}}$ can vary by up to one order of magnitude. We note that the contour lines in Fig. \ref{fig8} are symmetric, because both electrons and holes contribute to the total mobile charge carrier density. The blue points correspond to the charge carrier densities obtained for the gate-dependent band diagrams in Fig. \ref{fig6}(a)-(c), while the dotted connection line between the blue points expresses the general gate-dependency of $n_{\mathrm{film}}$ shown in Fig. \ref{fig7}(a), providing the corresponding $E^{\mathrm{top}}_{F}$ and $E^{\mathrm{bottom}}_{F}$. For example, for $E^{\mathrm{top}}_{F}=240 $\,meV and $E^{\mathrm{bottom}}_{F}=156 $\,meV, as determined in ref. \cite{Luepke_2017_2} for a vanishing gate voltage, the calculated charge carrier density in the film is $\sim 2 \times 10^{11}\,\mathrm{cm}^{-2}$, which is only one third of the value calculated in the symmetric approximation. 

We thus conclude that the charge carrier density in the thin TI film can be calculated more precisely with asymmetric boundary conditions. Because the gate-dependent measurements not only provide the band bending but also the charge carrier mobility in the film, the conductivity of the interior of the TI film can finally be determined \cite{Luepke_2017_2}.  

\section{Conclusion}

In this paper, we have shown that thin TI films usually exhibit parallel conduction channels through the interface layer and the interior of the film. These parasitic channels participate in the overall current transport and therefore significantly reduce the fraction of the total current that flows through the TSS channel. As a consequence, the desired beneficial properties of the TI, for instance spin-momentum locking, are partially lost. However, if the parasitic transport channels are understood in detail, the possibility arises to tune them towards a negligible influence compared to the auspicious TSS channel. 

In order to determine the interface conductivity of thin TI films grown by van-der-Waals epitaxy, only the initial substrate termination has to be prepared and, afterwards, the interface conductivity can directly be measured by surface-sensitive four-probe transport measurements performed with a multi-tip STM.

The conductivity of the film interior can be determined by a combination of surface-sensitive experimental methods, such as ARPES and gate-dependent four-probe transport measurements, and band bending calculations in the thin-film limit. In the latter the TI film is treated similar to semiconductors, but without the assumption of Fermi level pinning, because the DOS of the TSS is typically small. In the symmetric approximation, where the measured value of the top surface Fermi level from ARPES is also applied to the non-accessible bottom surface, the total mobile charge carrier density in the film material can be calculated. This calculation is possible even if the concentration of dopants that are unintentionally incorporated during film growth is unknown, because in the thin-film limit the carrier density is nearly independent of the film dopant concentration for the desirable case of moderate film doping, i.e. a non-degenerate doping. The band bending calculations can be refined by employing asymmetric boundary conditions, if  gate-dependent four-probe measurements on the top surface are used as additional input besides ARPES. Then, the conductivity of the interior of the thin film can be calculated unambiguously and the role of this channel for current transport in the TI system can be evaluated. 

We finally stress that the methods presented in this paper are general and can be applied to very different classes of TI materials. The information gained in this way is important for designing future electronic devices based on TI materials in such a way that the majority of current is exclusively transported by the TSS and most benefit can thus be gained from the topological properties of the material in question. 

The authors gratefully acknowledges financial support by the Deutsche Forschungsgemeinschaft (DFG, German Research Foundation) through SFB 1083 \textit{Internal Interfaces: Structure and Dynamical Properties} (project A12) and 
under Germany's Excellence Strategy–Cluster of Excellence Matter and Light for Quantum Computing (ML4Q) EXC 2004/1 – 390534769. 


\appendix 

\section{Conductivity of the interface channel}
A common substrate for the growth of thin TI films by means of van-der-Waals epitaxy is Si(111). Here, an initial passivation of the surface dangling bonds at the beginning of the growth process results in a weak van-der-Waals coupling of the TI film to the substrate \cite{Lanius_2016,Koma_1992,Borisova_2012}. However, it should be noted that the interface reconstruction of the Si(111) surface that results from the passivation can potentially have an appreciable surface conductivity, thus providing an additional parasitic channel for current transport beneath the thin TI film \cite{L_pke_2015,Just_2015,Homoth_2009,Tanikawa_2004}. 
Scanning transmission electron microscopy has revealed that thin TI films often exhibit a sharp interface to the Si(111) substrate \cite{Borisova_2012,Luepke_2017}. In combination with the predominantly weak van-der-Waals interaction between the TI and the substrate the electronic properties of the TI film and the passivated substrate are largely decoupled. Because of this decoupling, we may assume that the conductivity of the substrate's surface reconstruction without the thin TI film equals the interface conductivity of the final TI/substrate system. Evidently, it is then sufficient to prepare the pertinent passivation layer on the Si(111) surface and measure its conductivity in order to access the conductivity of the interface layer. The surface conductivity of the suitably passivated Si(111) surface can be measured by surface-sensitive four-probe measurements. In the following, we discuss the surface conductivities of some of the relevant surface reconstructions. 

\subsection{Te/Si(111)-(7$\times$7)}
\begin{figure}[t!]
\centering
\includegraphics[width=0.42\textwidth]{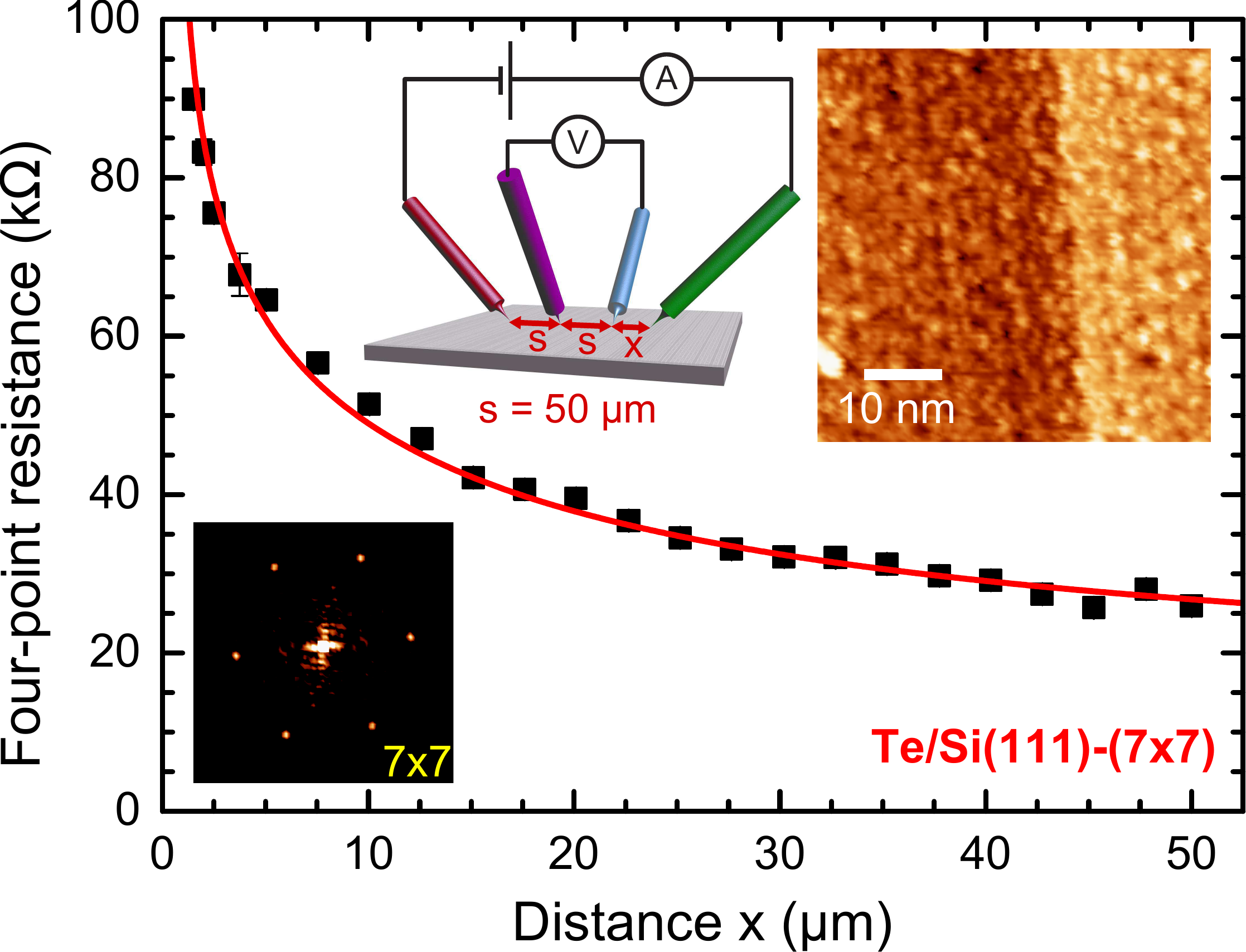}
	\caption{(Color online) Measured four-point resistance of the Te/Si(111)-(7$\times$7) surface reconstruction as function of the non-equidistant probe spacing $x$. The red line corresponds to a fit to the data using a pure 2D model according to Eq. (\ref{2D}) and results in a surface conductivity of $\sigma^{\mathrm{Te}}_{7\times7} = (8.3\pm 0.5)\times 10^{-6}\,\mathrm{S}/\square$. In the upper left inset, the probe configuration is visualized. The upper right inset shows a STM image of the Te/Si(111)-(7$\times$7) surface exhibiting traces of the characteristic corner holes of the ($7\times 7$)-reconstruction. This periodicity can also clearly be seen by the distinct spots in the Fourier transform of the STM image shown in the lower left inset.} 
\label{fig2}
\end{figure}

For Te-based van-der-Waals epitaxy, a plausible interface reconstruction is the Te/Si(111)-(7$\times$7) reconstruction, which has been reported as a template for the growth of TI films on Si(111) \cite{Zhang2_2009}. It can be prepared at room temperature by depositing Te on top of a reconstructed Si(111)-(7$\times$7) surface. 
In a first step, the (7$\times$7)-reconstruction of the Si(111) substrate (p-doped, bulk resistivity $22.5\,\mathrm{k}\Omega\mathrm{cm}$) is established by heating the substrate to 1200$^{\circ}\mathrm{C}$ and then slowly decreasing its temperature. Next, the Si(111)-(7$\times$7) surface is passivated by the deposition of one monolayer (ML) Te at 300$^{\circ}\mathrm{C}$, employing a flux of 1 ML/min from a Knudsen cell. In subsequent low energy electron diffraction (LEED) measurements, weak (7$\times$7) spots are visible (not shown here), indicating that at least some elements of the (7$\times$7) Si reconstruction are still present below the Te layer. This conclusion is corroborated by STM images in which the characteristic corner holes of the Si(111)-(7$\times$7) reconstruction can still be identified (upper right inset in Fig. \ref{fig2}). The corresponding periodicity can be discerned more clearly as distinct spots in the Fourier transform of the STM image (lower left inset in Fig. \ref{fig2}) \cite{Luepke_2019}.   

To determine the surface conductivity of the Te/Si(111)-(7$\times$7) reconstruction, we have performed distance-dependent four-probe measurements with a multi-tip STM at room temperature. The tip configuration was chosen to be linear but not equidistant (upper left inset of Fig. \ref{fig2}). In this arrangement, the distance $x$ between one of the outer current-injecting tips and the adjacent inner voltage-measuring tip is varied, while the spacing $s$ between the other tips remains constant at $s = 50\,\mu\mathrm{m}$. Under these experimental circumstances, the four-point resistance $R^{4p}_{2D}$ for a two-dimensional sheet depends on the distances $s$ and $x$ as \cite{Wells1,Luepke_2017}
\begin{align}
	R^{4p}_{2D}(s,x) & = \frac{1}{2\pi\sigma_{2D}} \left[\ln\left(\frac{2s}{x}\right) - \ln\left(\frac{s}{x+s}\right)\right] \label{2D}\,\mathrm{.}
\end{align}
In Fig. \ref{fig2} the measured four-point resistance of the Te/Si(111)-(7$\times$7) surface is plotted as function of the spacing $x$. The solid red line corresponds to a fit according to Eq. (\ref{2D}).  The fit corresponds well to the data and results in a surface conductivity of $\sigma^{\mathrm{Te}}_{7\times 7} = (8.3\pm0.5) \times 10^{-6}\,\mathrm{S}/\square$, which is slightly larger than the surface conductivity of the bare Si(111)-(7$\times$7) surface, measured as $\sigma^{\mathrm{Si}}_{7\times7} = (5.1 \pm 0.7) \times 10^{-6}\,\mathrm{S}/\square$ \cite{Just_2015}. 
In combination with the results from STM and LEED the surface conductivity measurements suggest that the Si(111)-(7$\times$7) reconstruction is still partly intact underneath the deposited Te, while its conductivity is moderately increased by doping from Te. But if $\sigma^{\mathrm{Te}}_{7\times 7}$ is compared to typical TSS conductivities of $\sigma_{\mathrm{TSS}}\approx 4\,\, \mathrm{to} \,\,8 \times 10^{-4}\,\mathrm{S}/\square$ \cite{Luepke_2017_3,Bauer_2016}, only 1 to 2\% of the total current would flow through the interface channel, indicating that it does not play a significant role in the overall current transport.

\subsection{Te/Si(111)-(1$\times$1)}
A further plausible interface termination for Te-based van-der-Waals epitaxy is the Te/Si(111)-(1$\times$1) surface.  In fact, this reconstruction forms the most common template for Te-based van-der-Waals epitaxy. For example, it has been used for the growth of Bi$_2$Te$_3$ \cite{Lanius_2016} and BiSbTe$_3$ \cite{Luepke_2017_2}. The reconstruction exhibits a surface conductivity of $\sigma^{\mathrm{Te}}_{1\times 1} = (2.6 \pm 0.5) \times 10^{-7}\,\mathrm{S}/\square$ \cite{Luepke_2017}, which is substantially lower than the conductivity of the Te/Si(111)-(7$\times$7) surface. If this value is compared to typical TSS conductivities of $\sigma_{\mathrm{TSS}} \approx 4\,\, \mathrm{to} \,\,8 \times 10^{-4}\,\mathrm{S}/\square$, it turns out to be much smaller, such that the interface channel would contribute less than 1\% to the total current transport in the TI/substrate system. Evidently, this contribution is negligible. 

\subsection{Bi/Si(111)-($\sqrt{3}\times\sqrt{3}$)}
For Bi-based van-der-Waals epitaxy of TI films, a common growth template is the Bi/Si(111)-($\sqrt{3}\times\sqrt{3}$) surface reconstruction with one monolayer Bi coverage (also named $\beta$-phase), which remains stable under Te flux \cite{Bauer_2016,Zhang2_2009,Li_2010,Sakamoto_2010,Hirahara_2010}. The conductivity of this surface reconstruction is reported to be $\sigma^{\mathrm{Bi}}_{\sqrt{3}\times\sqrt{3}} = (1.4\pm0.1) \times 10^{-4}\,\mathrm{S}/\square$ \cite{Just_2015}. This value is in the range of typical TSS conductivities, such that in this case a substantial fraction (20 to 35\%) of the total current through the TI/substrate system would be transmitted by the interface channel. Because of this high parasitic conductance a Bi-terminated interface of the Si(111) substrate exhibiting the Bi/Si(111)-($\sqrt{3}\times\sqrt{3}$) reconstruction is not favorable for designing TI devices. 

\begin{figure*}[t!]
\centering
\includegraphics[width=0.31\textwidth]{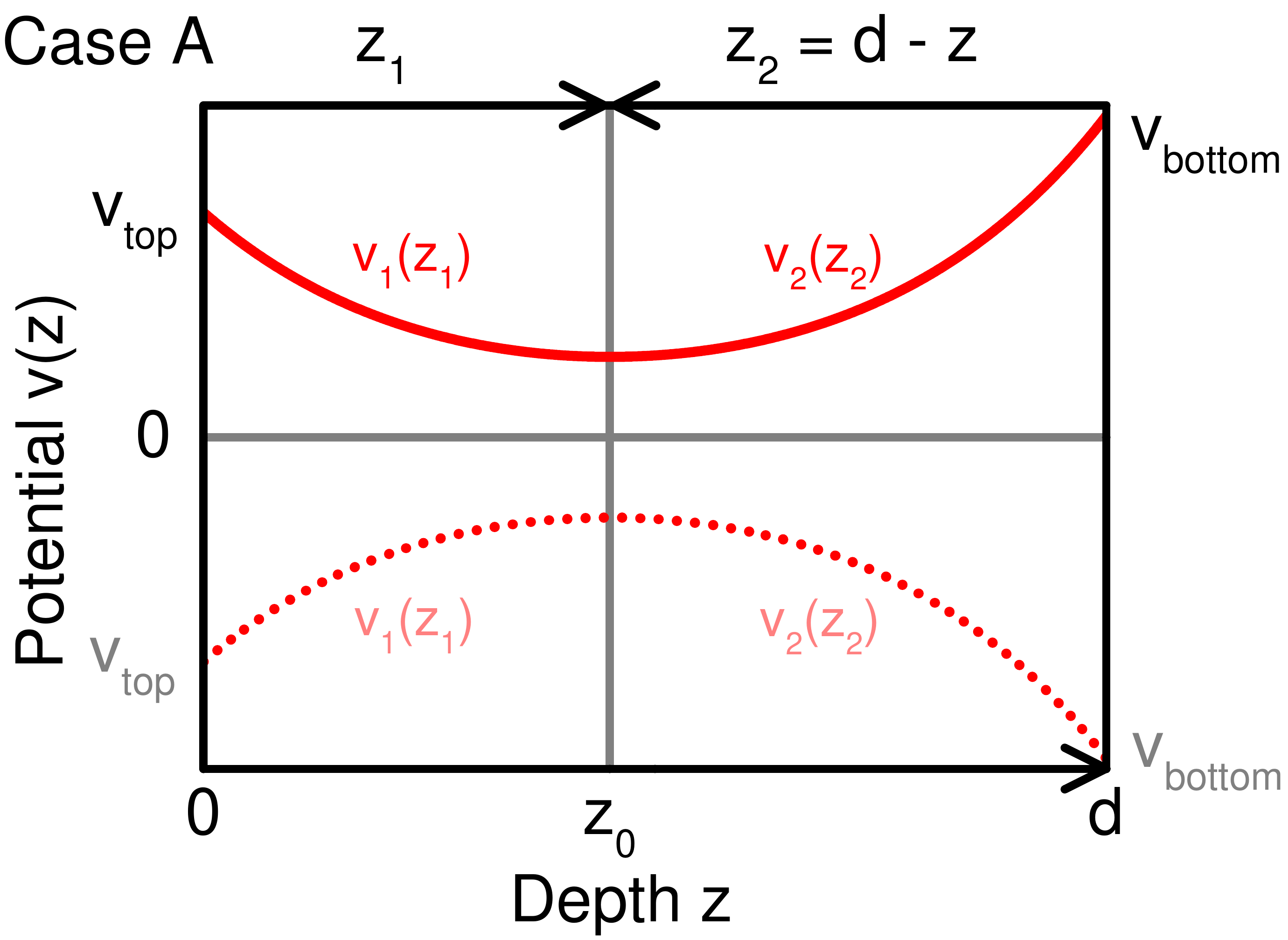}
\hspace{0.2cm}
\includegraphics[width=0.31\textwidth]{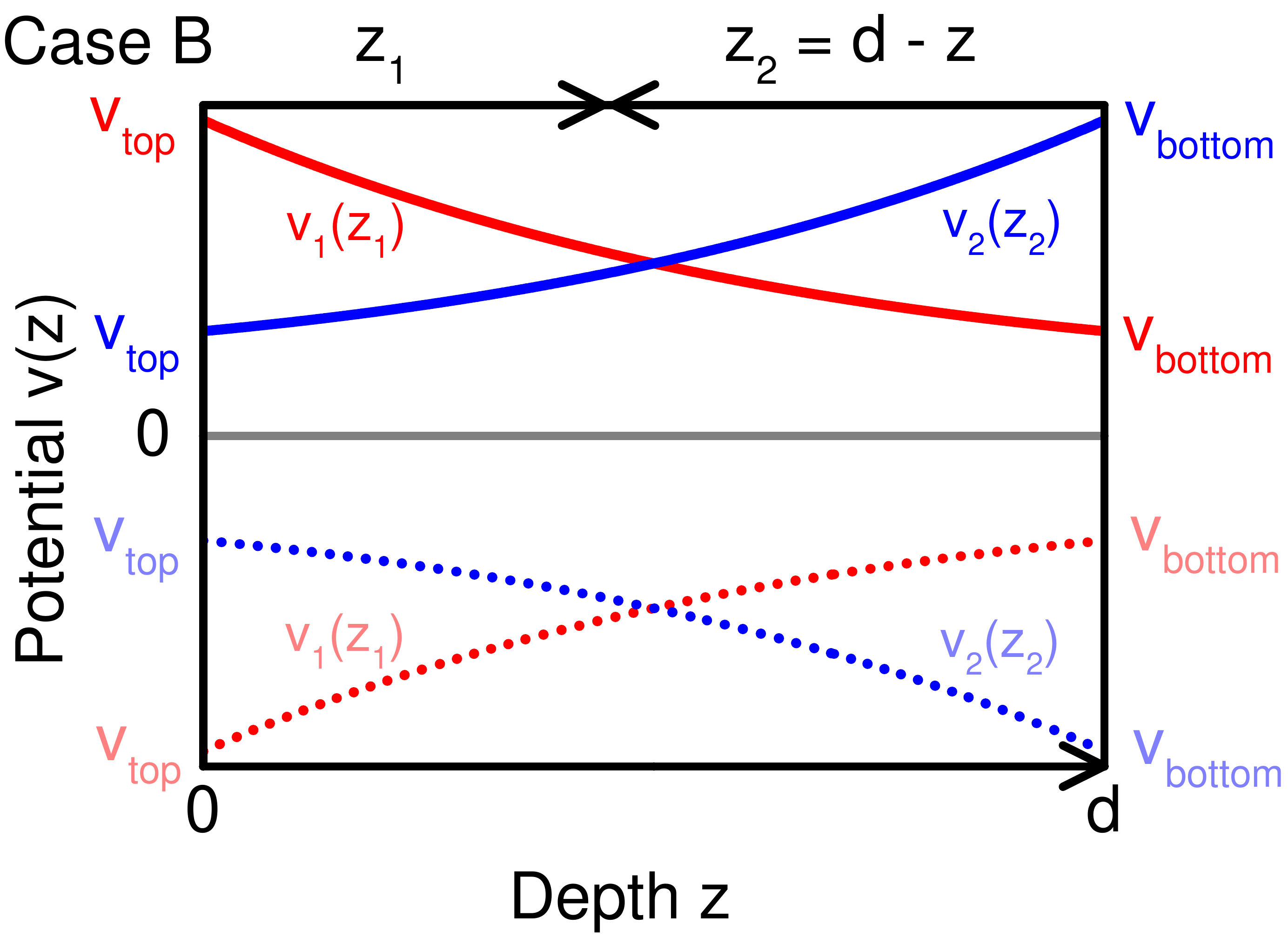}
\hspace{0.2cm}
\includegraphics[width=0.31\textwidth]{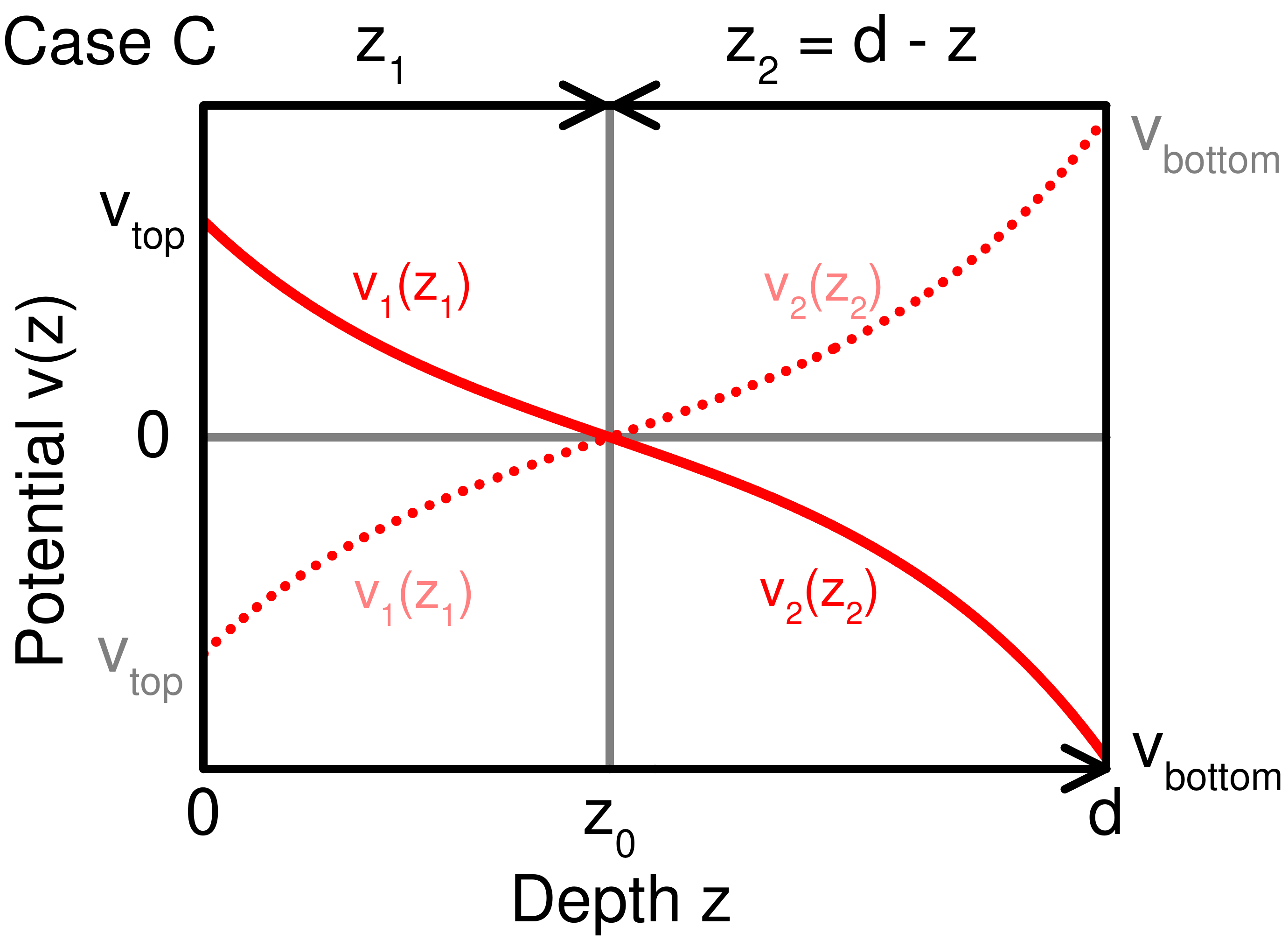}
	\caption{(Color online) Three different cases A - C for the asymmetric potential $v(z)$ of a $10\,\mathrm{nm}$ thin film with different surface potentials $v_{\mathrm{top}}$ and $v_{\mathrm{bottom}}$ on the top and bottom surface of the TI film, respectively, plotted as function of depth $z$ from the top surface. The coordinates $z_1$ and $z_2$ are indicated. In order to visualize that the shape of potential depends strongly on the boundary conditions, several distinct potential curves are plotted: solid and dotted lines in each panel exhibit boundary conditions with inverted signs. For case B, the blue curves display the potentials obtained by interchanging the values of $v_{\mathrm{top}}$ and $v_{\mathrm{bottom}}$.}
\label{fig5}
\end{figure*}  

\subsection{Se/Si(111)}
Similar results as for the Te/Si(111) interface terminations can also be expected for Se-based van-der-Waals epitaxy on Si(111) \cite{Bansal_2011,Bringans_1989,Wu_2011}. However, exact values for the surface conductivities of the respective (1$\times 1$) and (7$\times$7) reconstructions have not yet been reported in the literature. Furthermore, in the case of initial Se termination, depending on the preparation parameters additional amorphous interface layers up to several nm in thickness can occur \cite{Li_2010,He_2011}. Such extended interface regions can also have a significant effect on the total interface conductivity, which then involves a superposition of the conductivities of the Se/Si(111) substrate surface reconstruction and the additional amorphous layer(s). 

\subsection{Conclusion}
 The interface conductivity can have a significant influence on the overall current transport in a thin TI film on a substrate. It strongly depends both on the material system and the preparation parameters. Reported literature values for reconstructed interfaces are summarized in Tab. \ref{tab1}. For the example of TI  films grown by Te-based van-der-Waals epitaxy, such as Bi$_2$Te$_3$, Sb$_2$Te$_3$ and corresponding ternary and quaternary compounds, the Te/Si(111)-(1$\times$1) interface reconstruction is the best choice, since it exhibits a very low parasitic conductivity that hardly influences the current transport through the TSS channel. 
We finally note that potentially high interface conductivities are a general problem not only for TI films but also for other van-der-Waals thin films. They must therefore be determined individually for each material system at hand.   

\section{Formalism for asymmetric band bending calculations}

In this appendix the equations for the calculation of the asymmetric band bending, based on the solution of Eq. (\ref{ersteAbl}), are presented in detail.

Concerning the boundary conditions, three different cases, labeled A to C, can be distinguished. They are shown schematically in Fig. \ref{fig5}: (A) $\frac{dv}{dz}\big |_{z=z_0} = 0$ for $0\le z_0\le d$, (B) $\frac{dv}{dz}\big |_{z=z_0} \ne 0$ for $ 0 \le z_0 \le d$, and (C) $v(z=z_0) = 0 $ for $ 0 \le z_0 \le d $. For cases A and B the signs of the surface potentials $v_{\mathrm{top}}$ and $v_{\mathrm{bottom}}$ are equal, while for case C the signs are opposite. 
To simplify the calculation of the band bending for cases A and C, the function $v(z)$ is split into two branches, namely $v_1(z_1)$ with $z_1\equiv z$ for $0\le z \le z_0$ and $v_2(z_2)$ with $z_2\equiv d-z$ for $z_0\le z \le d$ (or equivalently $d-z_0\ge z_2 \ge 0)$. Note that the origin of $z_2$ is located at the bottom surface of the thin film and that $z_2$ increases towards the top surface, whereas $z_1$ (and $z$) have their origin at the top surface and increase towards the bottom one. The transition between $v_1$ and $v_2$ at $z_0$ must be continuous and differentiable, leading to the following matching conditions: (A) $v_1(z_0) = v_2(d-z_0):=v_0$ and $\frac{dv_1}{dz_1}\big|_{z_1=z_0} = -\frac{dv_2}{dz_2}\big|_{z_2=d-z_0}=0$,  (C) $v_1(z_0)=v_2(d-z_0)=0$ and $\frac{dv_1}{dz_1}\big|_{z_1=z_0} = -\frac{dv_2}{dz_2}\big|_{z_2=d-z_0}$.  Two additional boundary conditions are determined by the initial values of the potential $v_1(0)=v_{\mathrm{top}}$ and $v_2(0)=v_{\mathrm{bottom}}$ at the surfaces of the film. 
For case B only one function, either $v_1(z_1)$ or $v_2(z_2)$, needs to be calculated and the two boundary conditions are $v_{1,2}(0)=v_{\mathrm{top}}$ and $v_{1,2}(d)=v_{\mathrm{bottom}}$. 
In detail, the band bending $v_i(z_i)$ for cases A to C is obtained by inserting the appropriate boundary conditions in Eq. (\ref{ersteAbl}). Hereby, the index $i = 1,2$ denotes the two branches $v_1(z_1)$ and $v_2(z_2)$ of the potential $v(z)$, and $v_{\mathrm{top},\mathrm{bottom}}$ represents the corresponding surface potential, i.e. for $i = 1$ the first index ($_{\mathrm{top}}$) and for $i=2$ the second index ($_{\mathrm{bottom}}$) has to be used. 

\subsection{Case A}  

In the left panel of Fig. \ref{fig5}, the band bending corresponding to case A is shown. The surface potentials $v_{\mathrm{top}}$ and $v_{\mathrm{bottom}}$ are either both positive (red solid line) or both negative (red dotted line) and exhibit only a small difference. If we choose $z_0$ at the extremum (minimum or maximum) of $v(z)$, the two branches $v_1(z_1)$ and $v_2(z_2)$ can be calculated in analogy to Eq. (\ref{eq:symmetric}) by 
\begin{widetext}
\begin{align}
	z_{i}(v_{i}) & = \sgn(-v_{i}) \frac{L}{\sqrt{2}} \int_{v_{\mathrm{top},\mathrm{bottom}}}^{v_{i}(z_{i})} \underbrace{\sqrt{\frac{e^{u_b} + e^{-u_b}}{e^{u_b}\left(e^{v_i} - e^{v_0} - v_{i} + v_0\right) + e^{-u_b}\left(e^{-v_{i}} - e^{-v_0} + v_{i} - v_0\right)}}}_{:=F_A(u_b,v_i,v_0)}\, dv_{i} 
\end{align}
\end{widetext}
for $0 \le z_1 \le z_0$ and $ 0 \le z_2 \le d-z_0$, with the additional auxiliary condition 

\begin{align}
	& \sgn(-v_{\mathrm{top}}) \frac{L}{\sqrt{2}} \left[\int_{v_{\mathrm{top}}}^{v_0} F_A(u_b,v_1,v_0) \, dv_1 \right. \nonumber \\[1ex]
	&  \hspace{11.5ex} \left. + \int_{v_{\mathrm{bottom}}}^{v_0} F_A(u_b,v_2,v_0) \,dv_2 \right] - d = 0 \; 
\end{align}
which determines the constant $v_0$.

\subsection{Case B} 

In case B (center panel of Fig. \ref{fig5}) there is no local extremum in the potential $v(z)$ because the difference between the boundary values $v_{\mathrm{top}}$ and $v_{\mathrm{bottom}}$ is larger than in case A, resulting in a more pronounced slope of the band potential. The calculation of the gate-dependent band bending in section \ref{sec:Asymmetric_band_bending_in_a_thin_film} assumed case B. Again, both $v_{\mathrm{top}}$ and $v_{\mathrm{bottom}}$ have the same sign, either positive (solid lines) or negative (dotted lines). Although not imperative, we also define two branches of $v(z)$ in the present case: Specifically, we place $z_0$ at the surface which has the smaller surface potential, i.e. $z_0 \equiv 0$ if $|v_{\mathrm{top}}| < |v_{\mathrm{bottom}}|$ and $z_0 \equiv d$ if $|v_{\mathrm{bottom}}| < |v_{\mathrm{top}}|$. In the former case  $v(z)$ is given by $v_2(z_2)$, in the latter by $v_1(z_1)$. If the values of $v_{\mathrm{top}}$ and $v_{\mathrm{bottom}}$ are interchanged, $v_1(z_1)$ has to be replaced by $v_2(z_2)$ or vice versa. Both potential functions can be calculated by
\begin{align}
	& z_i(v_i) = \sgn(-v_i) \frac{L}{\sqrt{2}} \;\times \nonumber \\[1ex]
	& \int_{v_{\mathrm{top},\mathrm{bottom}}}^{v_i(z_i)} \underbrace{\sqrt{\frac{e^{u_b} + e^{-u_b}}{e^{u_b}\left(e^{v_i} - v_i + c\right) + e^{-u_b}\left(e^{-v_i} + v_i + c\right)}}}_{:=F_B(u_b,v_i,c)}\, dv_i, \label{caseB}  
\end{align}
with $(z_i=z_1 \land 0 \le z_1 \le d) \Leftrightarrow |v_{\mathrm{top}}| > |v_{\mathrm{bottom}}| $ and $(z_i=z_2 \land 0 \le z_2 \le d) \Leftrightarrow |v_{\mathrm{top}}| < |v_{\mathrm{bottom}}|$. The auxiliary condition which determines the constant $c$ reads
\begin{align}
	B \int_{v_{\mathrm{top}}}^{v_{\mathrm{bottom}}} F_B(u_b,v_i,c) \, dv_i - d = 0 
\end{align}
with prefactor $B=\sgn(|v_{\mathrm{top}}|-|v_{\mathrm{bottom}}|)\sgn(-v_{\mathrm{top}}) \frac{L}{\sqrt{2}} $. 

\subsection{Case C}

Case C is visualized in the right panel of Fig. \ref{fig5}. As $v_{\mathrm{top}}$ and $v_{\mathrm{bottom}}$ have opposite signs, there must appear a root at a finite  $z$ between $0$ and $d$. We identify this $z$ with the $z_0$ that separates the potential into two branches $v_1$ and $v_2$. In the thin TI film, this root corresponds to a change of the type of band bending, from a depletion to an accumulation zone or vice versa. Both branches can be calculated by applying Eq. (\ref{caseB}) 
separately for $0 \le z_1 \le z_0$ and for $ 0 \le z_2 \le d-z_0$, with the modified auxiliary condition  
\begin{align}
	\sgn(-v_{\mathrm{top}}) \frac{L}{\sqrt{2}} \int_{v_{\mathrm{top}}}^{v_{\mathrm{bottom}}} F_B(u_b,v_1,c) \, dv_1 - d = 0 
\end{align}
that determines the constant $c$.

\subsection{Transition between the cases A - C}
Starting from the symmetric case of band bending with $v_{\mathrm{top}} = v_{\mathrm{bottom}}$, case A has to be applied if there is only a slight imbalance between $v_{\mathrm{top}}$ and $v_{\mathrm{bottom}}$ that breaks the symmetry between the top and bottom surfaces of the film. The transition to case B occurs once the local minimum (maximum) of the potential function $v(z)$ reaches one of the surfaces of the film ($z_0=0$ or $z_0=d$) because the difference between $|v_{\mathrm{top}}|$ and $|v_{\mathrm{bottom}}|$ becomes sufficiently large. Within the coordinates of the two branches $v_{1,2}(z_{1,2})$ (see above) the potential minimum (maximum) is then located at $z_{1,2}=d$. So, the condition for the transition point is $\frac{dv_{1,2}}{dz_{1,2}}\big|_{z_{1,2}=d} = 0$. This condition is fulfilled for a potential value $v_{1,2}(d)$ which can be expressed relative to the surface potential $v_{\mathrm{top},\mathrm{bottom}}$ by introducing a threshold $\Delta$ as $v_{1,2}(d) = v_{\mathrm{top},\mathrm{bottom}} - \Delta$. This threshold then determines the maximum difference between $v_{\mathrm{top}}$ and $v_{\mathrm{bottom}}$ up to which no transition between case A and B occurs. 
From Eq. (\ref{eq:symmetric}) the threshold $\Delta$ follows as 
\begin{widetext}
\begin{align}
	C \int_{v_{\begin{subarray}{l}\scriptscriptstyle{\mathrm{top},}\\\scriptscriptstyle{\mathrm{bottom}}\end{subarray}}}^{v_{\begin{subarray}{l}\scriptscriptstyle{\mathrm{top},}\\\scriptscriptstyle{\mathrm{bottom}}\end{subarray}} - \Delta} \sqrt{\frac{e^{u_b} + e^{-u_b}}{e^{u_b}\left(e^{v_i} - e^{v_{\mathrm{top},\mathrm{bottom}}-\Delta} - v_i + v_{\begin{subarray}{l}\scriptscriptstyle{\mathrm{top},}\\\scriptscriptstyle{\mathrm{bottom}}\end{subarray}}-\Delta\right) + e^{-u_b}\left(e^{-v_i} - e^{-v_{\mathrm{top},\mathrm{bottom}} + \Delta} + v_i - v_{\begin{subarray}{l}\scriptscriptstyle{\mathrm{top},}\\\scriptscriptstyle{\mathrm{bottom}}\end{subarray}} + \Delta\right)}}\, dv_i = d \;\textrm{.}
\end{align}
\end{widetext}
with the prefactor $C = \sgn(-v_i) \frac{L}{\sqrt{2}}$. 
When the absolute value of the difference between $v_{\mathrm{top}}$ and $v_{\mathrm{bottom}}$ is smaller than $\Delta$, i.e. $\big||v_{\mathrm{top}}|-|v_{\mathrm{bottom}}|\big| \le \Delta$, case A is used. When the difference is larger than $\Delta$, i.e. $\big||v_{\mathrm{top}}|- |v_{\mathrm{bottom}}|\big| > \Delta$, case B has to be applied. 
If the signs of $v_{\mathrm{top}}$ and $v_{\mathrm{bottom}}$ become different, the transition to case C occurs.  

\bibliography{./bandbending}

\end{document}


\renewcommand{\theequation}{S\arabic{equation}}


\title{Parasitic conduction channels in topological insulator thin films \\ - Supplemental Material - 
} 
\author{Sven Just}
\altaffiliation[Present address: ]{Leibniz-Institut f\"{u}r Festk\"{o}rper und Werkstoffforschung Dresden, 01069 Dresden, Germany}
\affiliation{Peter Gr\"{u}nberg Institut (PGI-3), Forschungszentrum J\"{u}lich, 52425 J\"{u}lich, Germany}
\affiliation{J\"ulich Aachen Research Alliance (JARA), Fundamentals of Future Information Technology, 52425 J\"{u}lich, Germany}
\affiliation{Experimental Physics VI A, RWTH Aachen University, Otto-Blumenthal-Stra\ss e, 52074 Aachen, Germany}

\author{Felix L\"upke}
\altaffiliation[Present address: ]{Oak Ridge National Laboratory, TN, USA}
\affiliation{Peter Gr\"{u}nberg Institut (PGI-3), Forschungszentrum J\"{u}lich, 52425 J\"{u}lich, Germany}
\affiliation{J\"ulich Aachen Research Alliance (JARA), Fundamentals of Future Information Technology, 52425 J\"{u}lich, Germany}

\author{Vasily Cherepanov}
\affiliation{Peter Gr\"{u}nberg Institut (PGI-3), Forschungszentrum J\"{u}lich, 52425 J\"{u}lich, Germany}
\affiliation{J\"ulich Aachen Research Alliance (JARA), Fundamentals of Future Information Technology, 52425 J\"{u}lich, Germany}

\author{F. Stefan Tautz}
\affiliation{Peter Gr\"{u}nberg Institut (PGI-3), Forschungszentrum J\"{u}lich, 52425 J\"{u}lich, Germany}
\affiliation{J\"ulich Aachen Research Alliance (JARA), Fundamentals of Future Information Technology, 52425 J\"{u}lich, Germany}
\affiliation{Experimental Physics VI A, RWTH Aachen University, Otto-Blumenthal-Stra\ss e, 52074 Aachen, Germany}

\author{Bert Voigtl\"{a}nder}
\email[Corresponding author: ]{b.voigtlaender@fz-juelich.de}
\affiliation{Peter Gr\"{u}nberg Institut (PGI-3), Forschungszentrum J\"{u}lich, 52425 J\"{u}lich, Germany}
\affiliation{J\"ulich Aachen Research Alliance (JARA), Fundamentals of Future Information Technology, 52425 J\"{u}lich, Germany}
\affiliation{Experimental Physics VI A, RWTH Aachen University, Otto-Blumenthal-Stra\ss e, 52074 Aachen, Germany}

\date{\today}

\pacs{Valid PACS appear here}%

\begin{abstract}
In part A of this supplemental material, further examples of the near-surface band bending in topological insulators are presented. This reveals the mechanism of band bending as well as the influence of the various parameters, i.e. the film thickness, the position of CNL, the surface Fermi energy $E_{F}^{\mathrm{top}}$ and the dopant level $E_F^{\mathrm{bulk}}$, on the resulting shape of the bands in more detail.  

In part B, the Schr\"{o}dinger-Poisson approach to band bending calculations in thin TI films is discussed in more detail. First, the effective density of states that is used in the purely classical Poisson approach  is derived. Then, modifications to the effective density of states that arise from the confinement of charge carriers in the thin film are discussed. 

Finally, in part C plots of the calculated mobile charge carrier densities in a thin TI film are shown separately for electrons and holes. 
\end{abstract}

\maketitle

\section{A. Near-surface band bending in topological insulators}

\begin{figure*}[b!]
	\centering
	\includegraphics[height=0.23\textwidth]{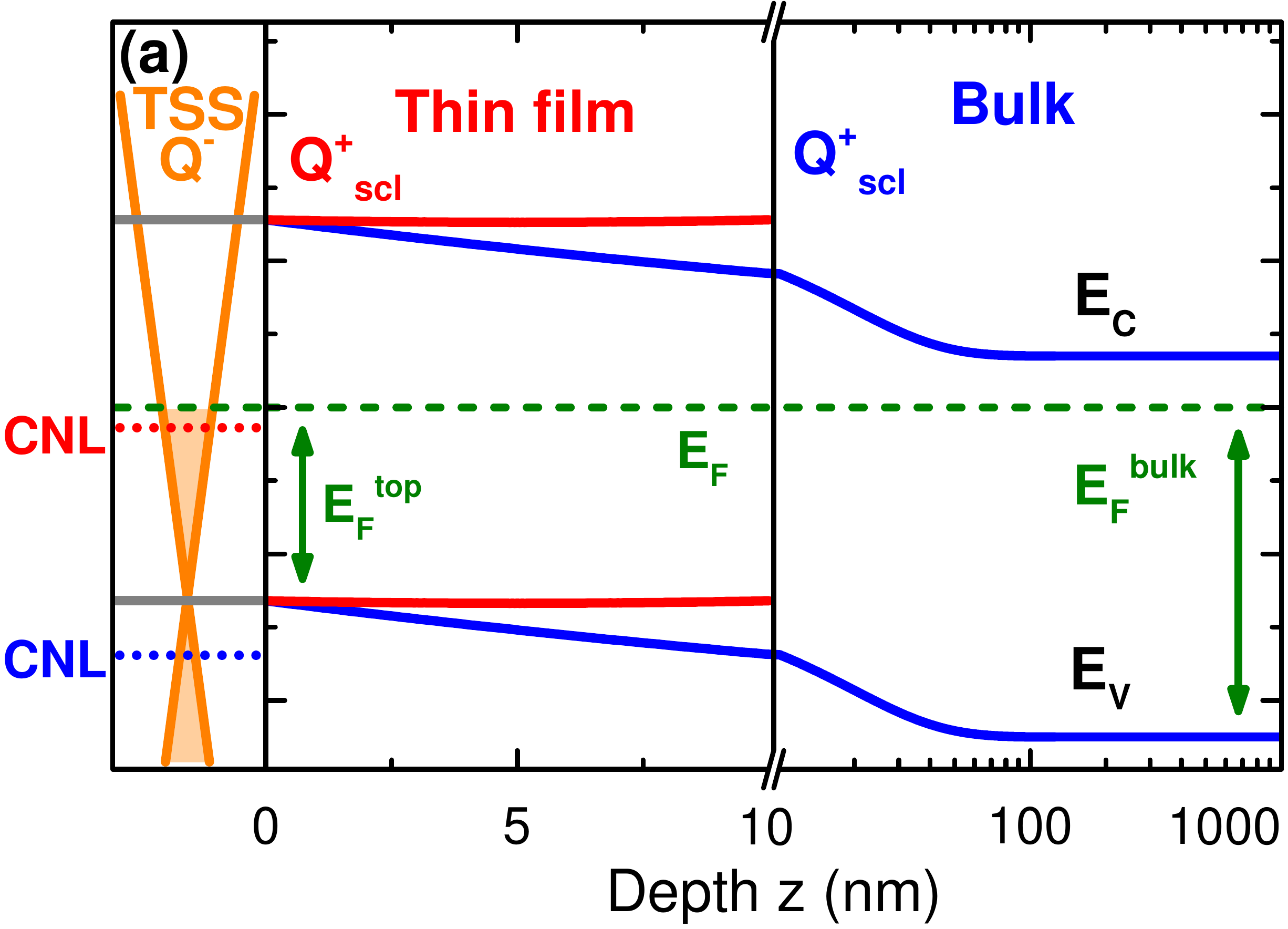}
	\hspace{-0.15cm}
	\includegraphics[height=0.23\textwidth]{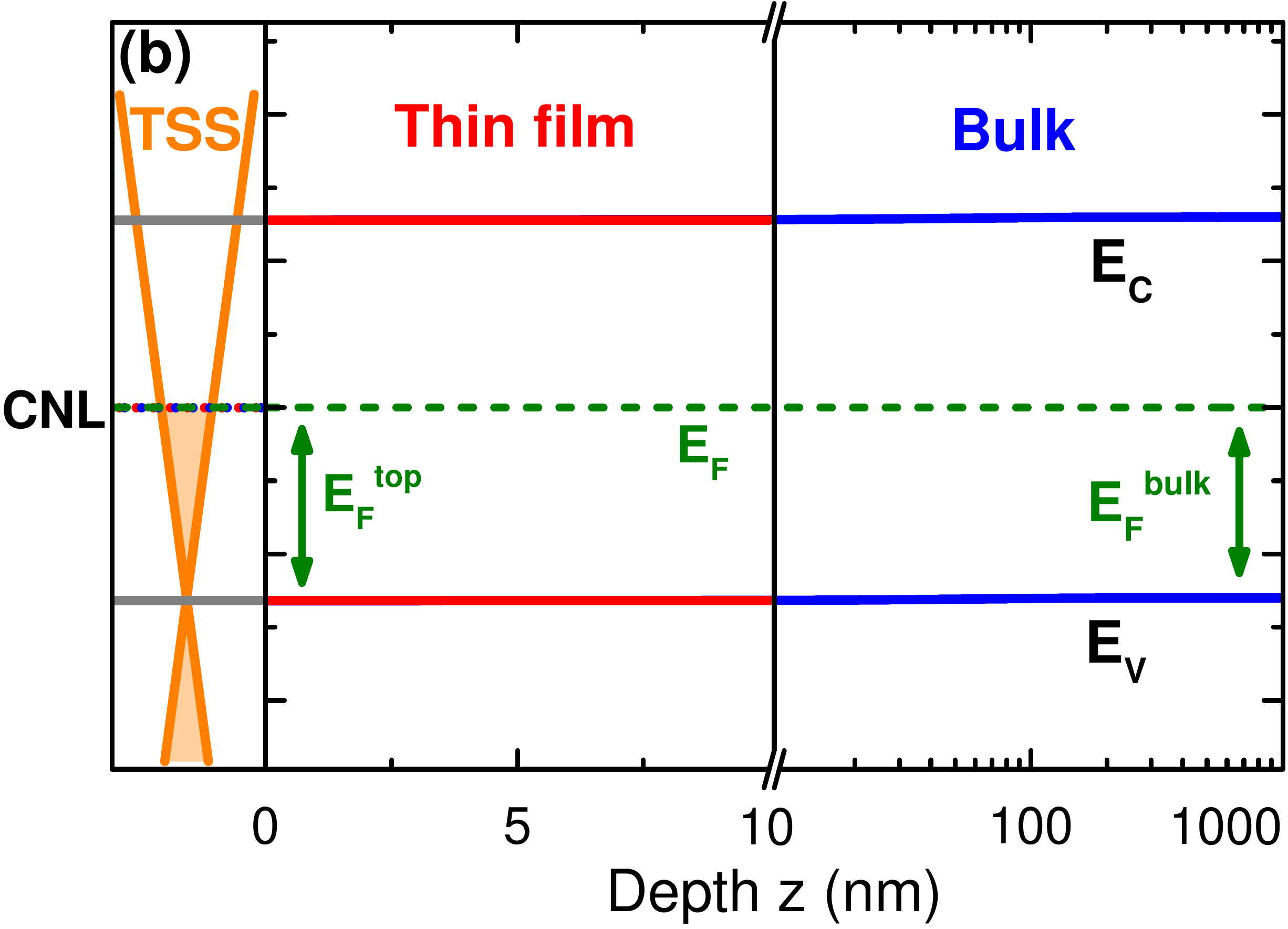}
	\hspace{-0.15cm}
	\includegraphics[height=0.23\textwidth]{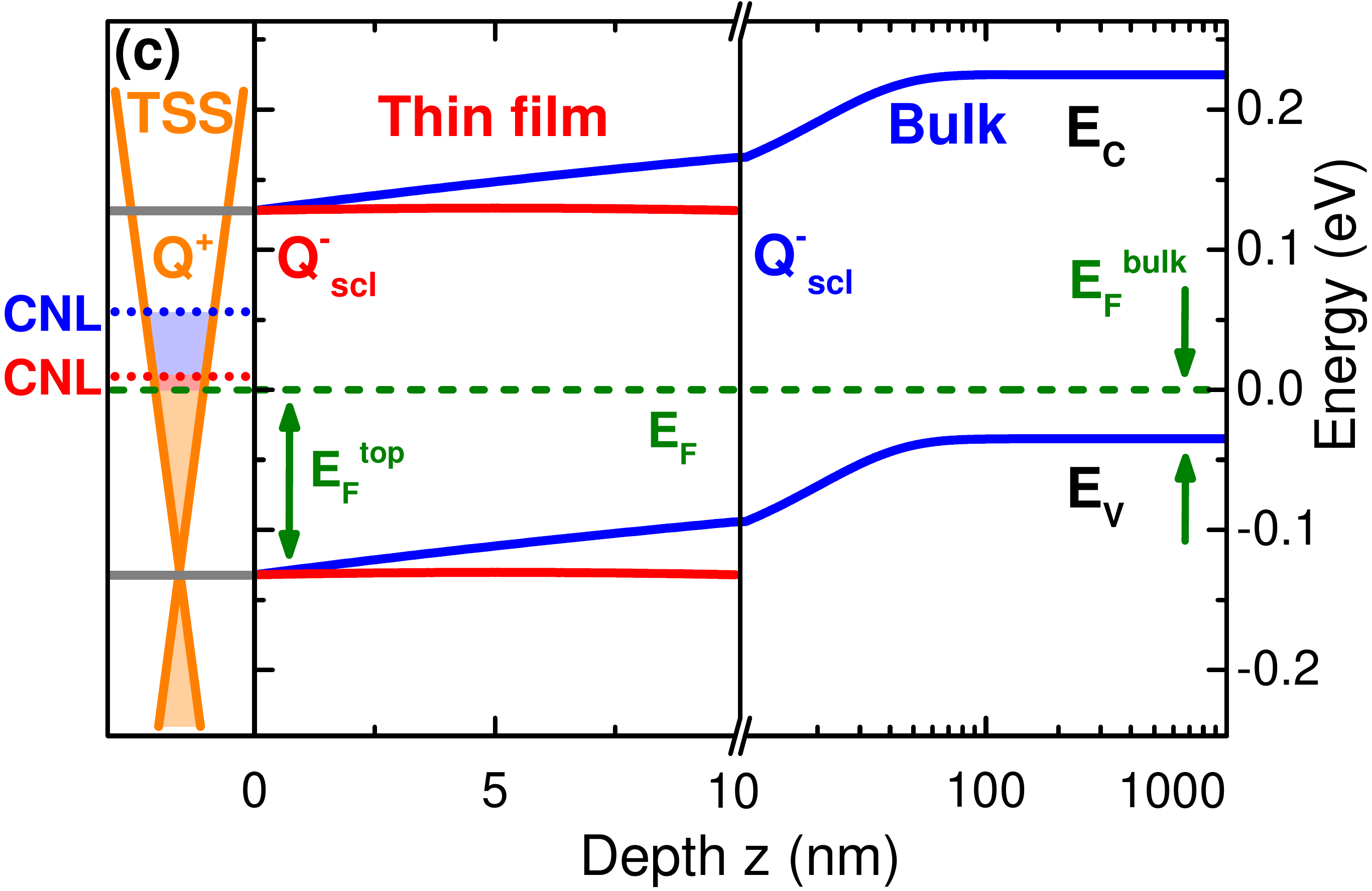}
	\caption{(Color online) Calculated band bending for a $10\,\mathrm{nm}$ thin film (red) and equivalent bulk crystal (blue) of the TI BiSbTe$_3$ as presented in Fig. 4 of the paper, but now for a fixed mid-gap position of the surface Fermi energy $E_F^{\mathrm{top}} = 130\,\mathrm{meV}$. From (a) to (c) the dopant concentration is varied, which is represented by different values for the bulk Fermi energy $E_F^{\mathrm{bulk}}$: (a)~$225\, \mathrm{meV}$, (b) $130\, \mathrm{meV}$, and (c) $35\, \mathrm{meV}$. Because the parameter $E_F^{\mathrm{top}}$ differs from the case displayed in Fig. 4 of the paper, the resulting band bendings and CNL are also different.}
	\label{supfig1}
	\vspace{1ex}
\end{figure*}

As discussed in detail in the paper, the charge transfer between the TSS and the bulk material and thus the resulting near-surface band bending is determined by four parameters: the charge neutrality level (CNL) of the combined TSS/DS system, the surface Fermi energy $E_F^{\mathrm{top}}$, the bulk Fermi energy $E_F^{\mathrm{bulk}}$ (given by the bulk dopant concentration), and the film thickness $d$. If three of these parameters are given, the fourth parameter is fixed by the condition of charge neutrality.

\begin{figure*}[t!]
	\centering
	\includegraphics[height=0.206\textwidth]{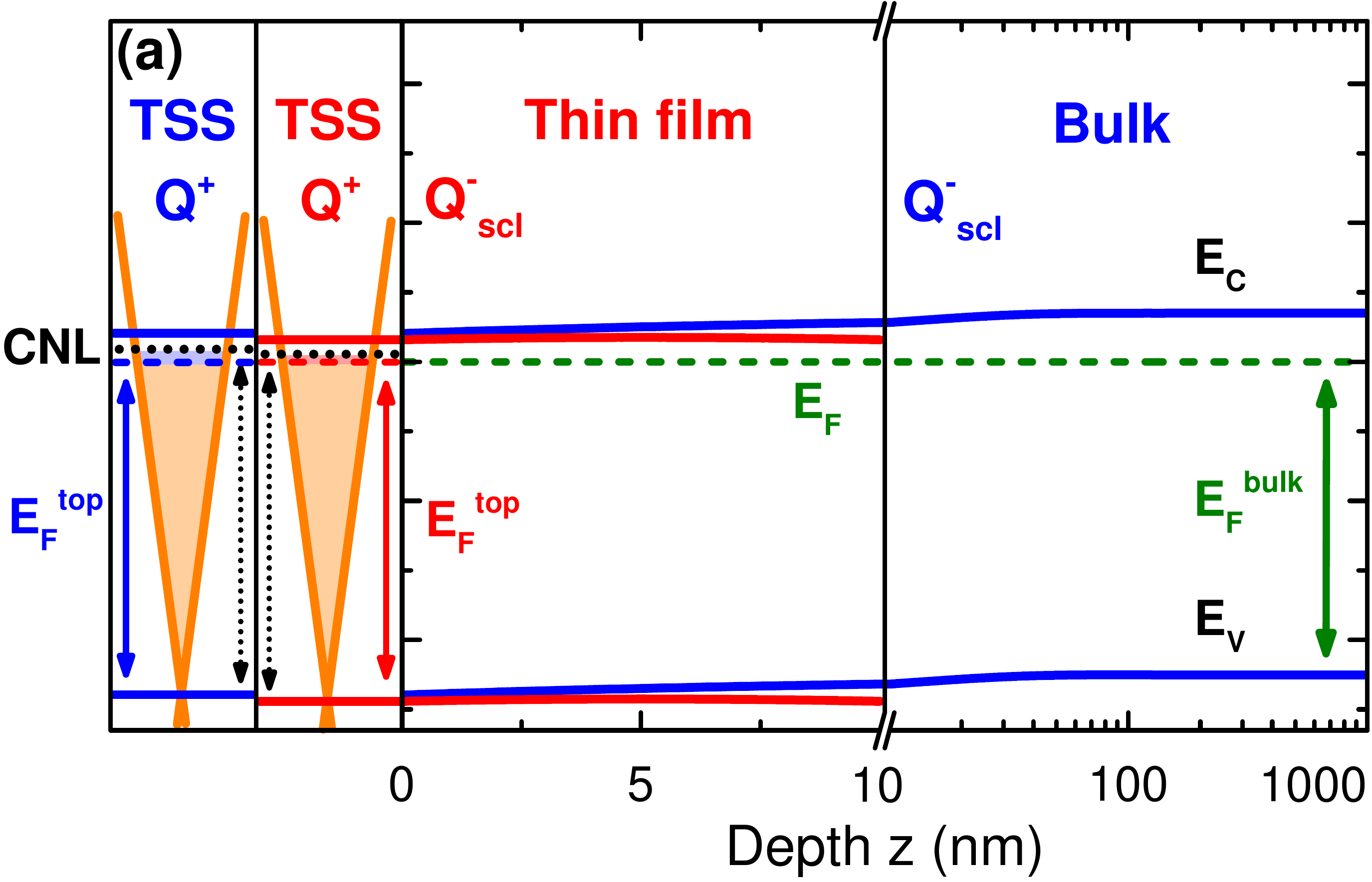}
	\hspace{-0.15cm}
	\includegraphics[height=0.206\textwidth]{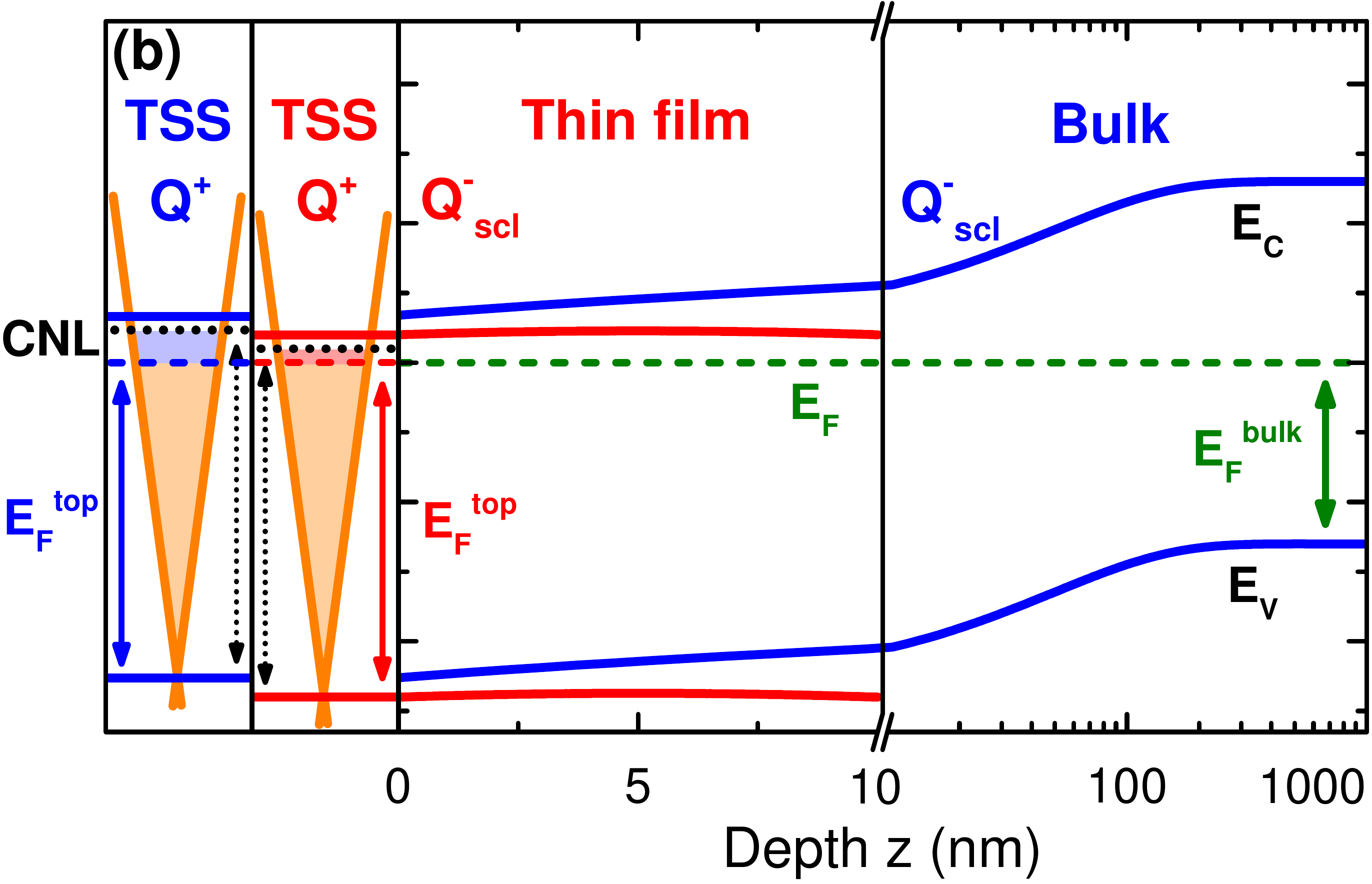}
	\hspace{-0.15cm}
	\includegraphics[height=0.206\textwidth]{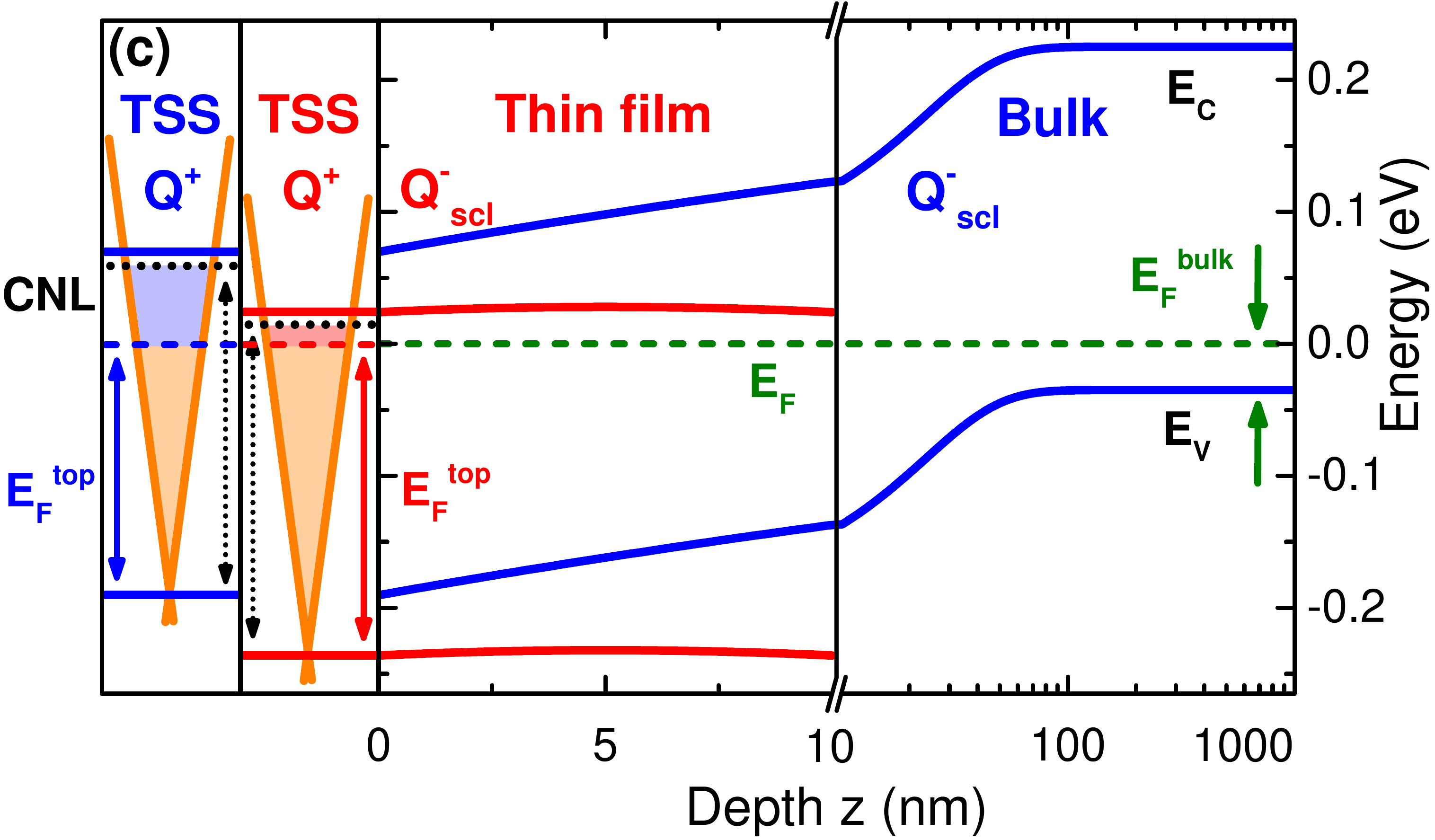}
	\vspace{0cm}	
	
	\includegraphics[height=0.206\textwidth]{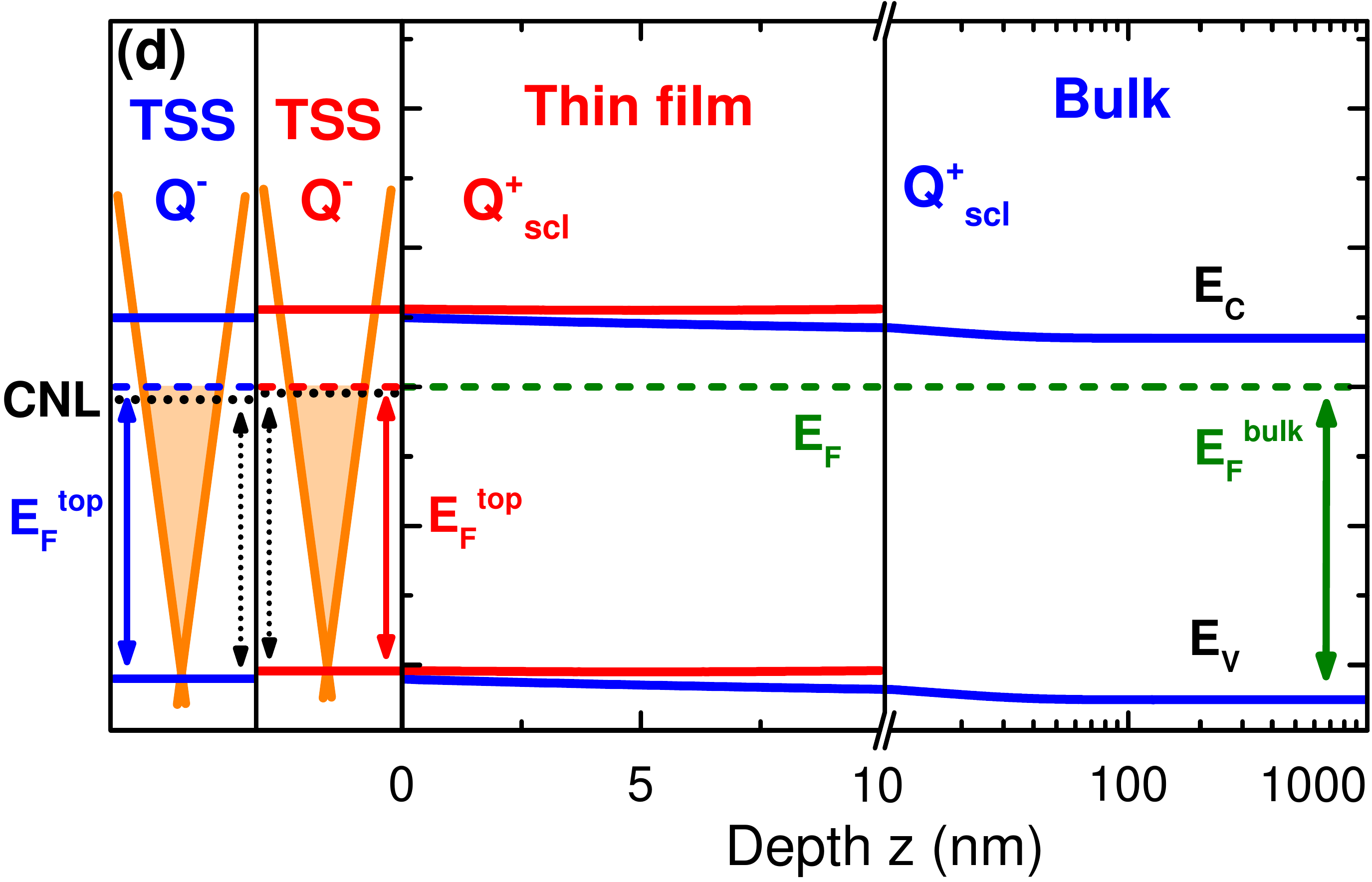}
	\hspace{-0.15cm}
	\includegraphics[height=0.206\textwidth]{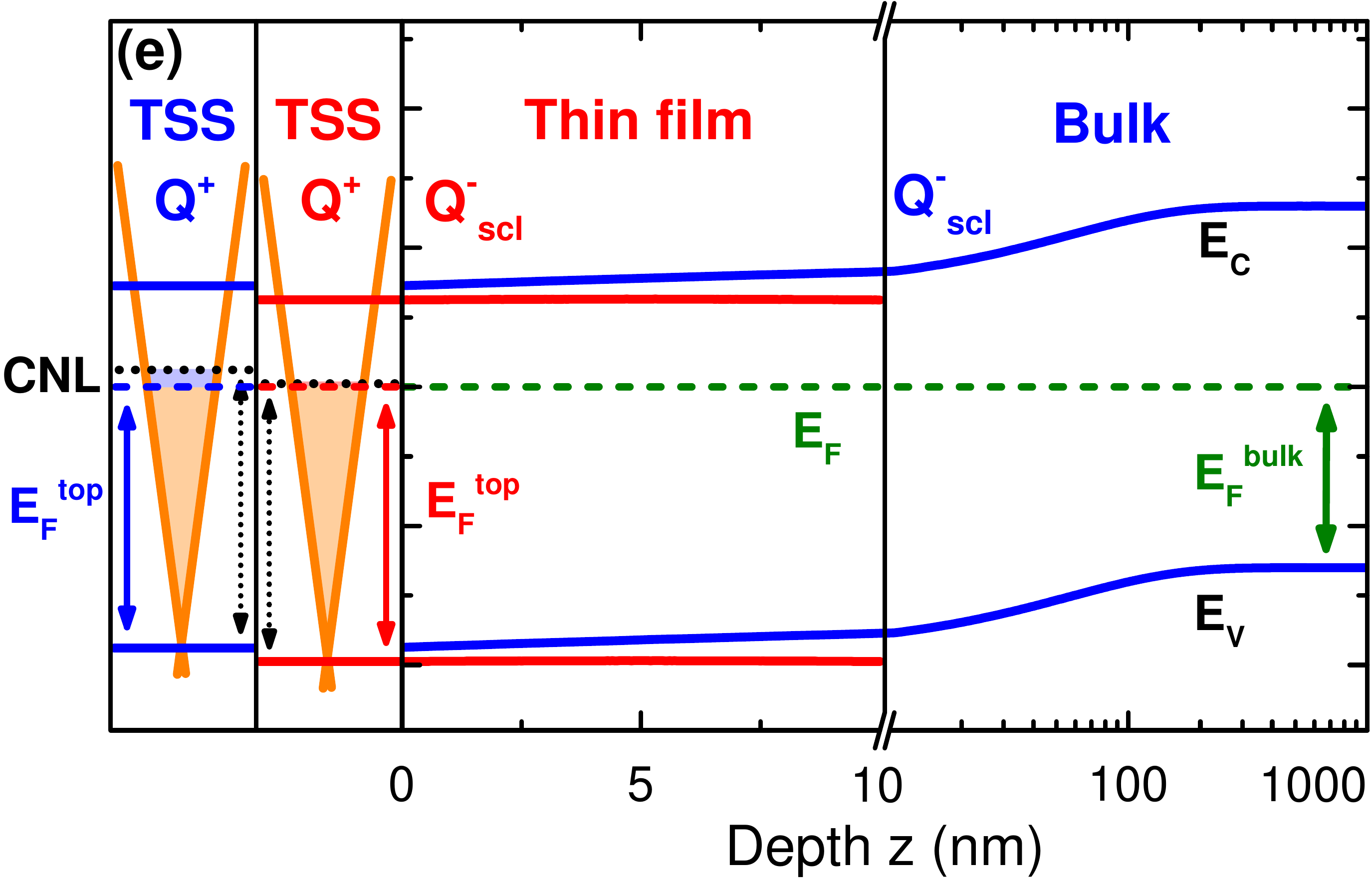}
	\hspace{-0.15cm}
	\includegraphics[height=0.206\textwidth]{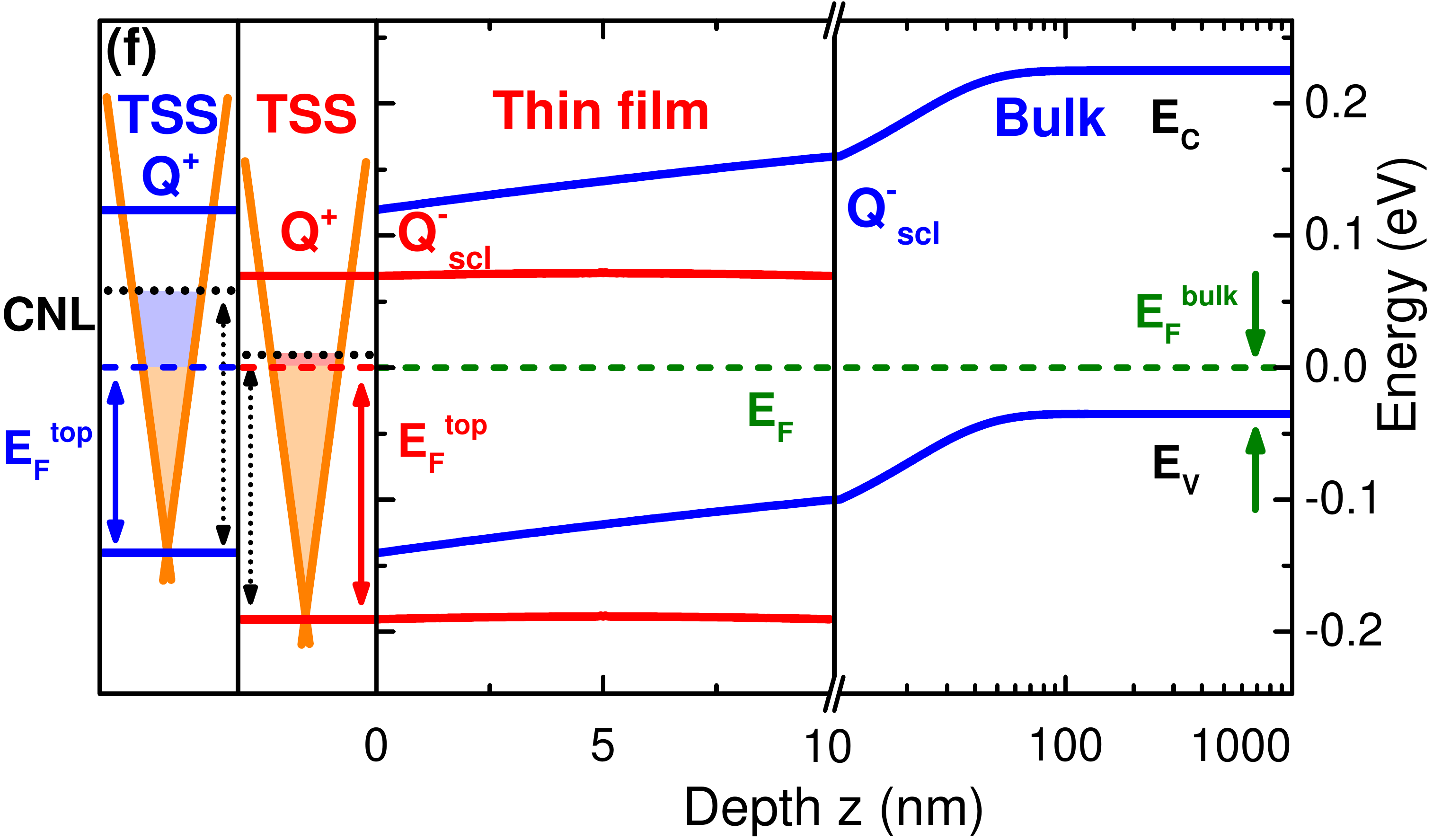}
	\caption{(Color online) Calculated band bending for a $10\,\mathrm{nm}$ thin film (red) and equivalent bulk crystal (blue) of the TI BiSbTe$_3$ as presented in Fig. 4 of the paper, but now for a fixed value of the charge neutrality level (black dotted line), i.e. for a fixed surface defect density, with $\mathrm{CNL} = 250\,\mathrm{meV}$ for (a) to (c)  and $\mathrm{CNL} = 200\,\mathrm{meV}$ for (d) to (f). From (a) to (c) and (d) to (f) the dopant concentration is varied, as represented by the bulk Fermi energy $E_F^{\mathrm{bulk}}$: (a),(d) 225 meV, (b),(e) 130 meV, and (c),(f) 35 meV. In contrast to the data shown in Fig. 4 of the paper, the parameter CNL is constant among the top (bottom) panels, such that the calculated values for $E_F^{\mathrm{top}}$ come out different for both thin film and extended bulk crystal on the one hand, and for different dopant concentrations on the other hand. Due to the alignment of the Fermi energy $E_F$ (green dashed line) throughout all plots and the different values of $E_F^{\mathrm{top}}$, the Dirac cones for the thin film and the extended bulk are vertically displaced with respect to each other and plotted separately on the left of each panel.}
	\label{supfig2}
	\vspace{3ex}
\end{figure*}

In order to demonstrate the interplay of these four parameters and their effect on the band bending, several cases with different initial conditions are considered and plotted as band diagrams in Fig. \ref{supfig1} and Fig. \ref{supfig2}. Each band diagram is calculated for a $10\,\mathrm{nm}$ thin film (red) and the equivalent bulk crystal (blue) for the TI material BiSbTe$_3$, which was also considered in the paper. The difference to Fig. 4 in the paper is that now the parameters of the calculation are not based on measurements, but chosen with the aim to exemplify as clearly as possible their strong influence on the resulting band bending. In each set of band diagrams, i.e. Fig. \ref{supfig1}(a)-(c), Fig. \ref{supfig2}(a)-(c) and Fig. \ref{supfig2}(d)-(f), the dopant concentration, represented by $E_F^{\mathrm{bulk}}$, is varied from n- to p-type.

In Fig. \ref{supfig1}, the surface Fermi energy is set to a constant mid-gap position of $E_F^{\mathrm{top}} = 130\,\mathrm{meV}$. This results in distinct CNL values for the thin film and the bulk crystal, and also for the three different dopant concentrations. While for the extended bulk crystal the change of CNL and band bending with dopant concentration is significant, it is rather small for the thin film: For all dopant concentrations in Fig. \ref{supfig1}, the Fermi level in the thin film remains very close to mid-gap, because mobile charges flow almost completely into the TSS. The interior of the film is thus intrinsically doped (fully depleted), irrespective of the dopant concentration in the film material. If moreover the TI material is intrinsic (i.e. $E_F^{\mathrm{bulk}}$ in mid-gap position), the CNL is independent of the film thickness $d$ and therefore coincides for the thin film and the bulk cases (Fig. \ref{supfig1}(b)), resulting in flat bands in both cases.

In Fig. \ref{supfig2}, the charge neutrality level (black dotted line) is kept constant and fixed at $250\,\mathrm{meV}$ for Fig. \ref{supfig2}(a)-(c) and at $200\,\mathrm{meV}$ for Fig. \ref{supfig2}(d)-(f). Because all band diagrams in Fig. \ref{supfig2} are aligned to a common Fermi level (green), the CNL and the Dirac cones for the thin film and for the extended bulk appear vertically displaced, but the position of the CNL with respect to the conduction band is the same in both cases (black dotted arrows). A fixed CNL corresponds to the same specific concentration of surface defects in all cases. As a result, the calculated values for $E_F^{\mathrm{top}}$ are now different for both thin film and extended bulk crystal on the one hand, and for different dopant concentrations on the other hand. Also the shape of band bending varies between all cases (even upward band bending is observed in Fig.~\ref{supfig2}(d)). 

The calculations in Fig. \ref{supfig2} show that the CNL (resulting from a certain concentration of surface defects) directly influences the position of the surface Fermi energy $E_F^{\mathrm{top}}$, and correspondingly the overall band bending. As $E_F^{\mathrm{top}}$ is accessible by ARPES, this effect is directly visible in measurements. For example, a variation of $E_F^{\mathrm{top}}$ with time due to an increasing surface contamination has been experimentally observed \cite{Bianchi_2010,Kong_2011,Analytis_2010,Benia_2011}. Concomitantly, the filling level of the TSS changes, and thus the concentration of charge carriers inside. This in turn may lead to different properties of the TSS, e.g., a different conductivity. For this reason, it is difficult to characterize the TI properly after exhibiting it to ambient conditions, as it is for example necessary for the lithographic fabrication of electrical contacts for subsequent conductivity measurements. Only as long as ultra-high vacuum conditions are maintained, a change of the TSS properties due to adsorbates on the surface (and corresponding change of CNL) can be excluded. This highlights the great advantage of \textit{in situ}  transport measurements by means of multi-probe STM.    

\section{B. Schr\"{o}dinger-Poisson approach for band bending in thin films}

Below, we present in some detail the concept of the effective density of states and the modifications to the classical calculation of the band bending that are necessitated by the Schr\"{o}dinger-Poisson approach. 

\subsection{Effective density of states for the classical Poisson approach} 

The charge carrier density in the conduction band (electrons) is given by 
\begin{align}
	n_e(z) &= \int_{E_C(z)}^{\infty} n^{\mathrm{dos}}_C \left(E-E_C(z)\right)\,\frac{1}{e^\frac{E-E_F}{k_B T}+ 1}\, dE \label{ndens}
\end{align}
with the Fermi energy $E_F$, the conduction band edge $E_C(z)$ and the density of states $n^{\mathrm{dos}}_{C,\mathrm{2D}}(E) = \frac{m^*_e}{\pi \hbar^2}$ and $n^{\mathrm{dos}}_{C,\mathrm{3D}}(E) = \frac{\left(2 m^*_e\right)^{3/2}}{2 \pi^2 \hbar^3} \sqrt{E}$ for the 2D and 3D cases, respectively, assuming parabolic bands. For non-degenerate semiconductors with the Fermi energy positioned well in between the band edges the Boltzmann approximation for the Fermi-Dirac distribution can be used. Eq. (\ref{ndens}) can then be approximated as \cite{Lueth}
\begin{align}
	n_e(z) &= N^{\mathrm{eff}}_{C}\, e^{-\frac{\left(E_C(z) - E_F\right)}{k_B T}} \label{ndensboltz}
\end{align}
with the effective density of states $N^{\mathrm{eff}}_{C,\mathrm{2D}} = \frac{m^*_e k_B T}{\pi \hbar^2}$ and $N^{\mathrm{eff}}_{C,\mathrm{3D}} = 2 \left(\frac{2 m^*_e \pi k_B T}{h^2}\right)^{3/2}$ for the 2D and 3D cases, respectively. Conceptually, the effective density of states $N^{\mathrm{eff}}$ corresponds to a theoretical density of states that would be present, if all empty states of the conduction band were concentrated at the lower conduction band edge at $E_C$. The charge carrier density in the valence band (holes) can be described in the same way by using the effective hole mass $m^*_h$ and the valence band edge $E_V(z)$ in Eq. (\ref{ndensboltz}), resulting in 
\begin{align}
	p_h(z) &= N^{\mathrm{eff}}_{V}\, e^{\frac{E_V(z)-E_F}{k_B T}} \label{pdensboltz}
\end{align}
with $N^{\mathrm{eff}}_{V,\mathrm{2D}} = \frac{m^*_h k_B T}{\pi \hbar^2}$ and $N^{\mathrm{eff}}_{V,\mathrm{3D}} = 2 \left(\frac{2 m^*_h \pi k_B T}{h^2}\right)^{3/2}$ for the 2D and 3D cases, respectively. 

For the calculation of the near-surface band bending $E_C(z)$ and $E_V(z)$ within the classical approach, the Poisson equation (Eq. (1) in the manuscript)   
\begin{align}
	\frac{d^2v}{dz^2} & = -\frac{q^2}{\epsilon_0 \epsilon_r k_B T}  \left(n_b - p_b + p_b e^{-v(z)} - n_b e^{v(z)}\right) \label{poisdiff}
\end{align}
for the dimensionless potential $v(z)$ has to be solved. In this equation, the two terms on the left correspond to the static charge density $\rho_{\mathrm{static}} = q [N_D^+-N_A^-]$ caused by the homogeneous concentration of charged donors and acceptors in the material, which can be expressed by the mobile bulk charge carriers $n_b$ and $p_b$ by using the condition of charge neutrality $n_b + N_A^- = p_b + N_D^+$. The two terms on the right of Eq. (\ref{poisdiff}) correspond to the position-dependent mobile charge density $\rho_{\mathrm{mobile}}(z) = q [p(z)-n(z)]$ of electrons and holes, which can be expressed by the equations (\ref{n_end}) and (\ref{p_end}) (see below).    
With the definition given in the paper
\begin{align}
q \Phi(z) &= k_BT u(z) = E_F - E_i(z) \label{u_pot} 
\end{align}
the potential $v(z)$ can be written as
\begin{align}
	v(z) &= \frac{q}{k_BT} \Big[\Phi(z) - \Phi_b \Big] \,=\, \frac{1}{k_BT} \Big[E_{i}^b - E_i(z)\Big] \label{potential},
\end{align}
where the indices $i$ and $b$ denote the intrinsic level and the bulk limit (for $z \rightarrow \infty$), respectively. The intrinsic level $E_i^b$ in the bulk material depends on the effective densities of states as
\begin{align}
	E_i^b\left(N^{\mathrm{eff}}_C,N^{\mathrm{eff}}_V\right) &= \frac{E_C^b + E_V^b}{2} + \frac{k_BT}{2}\,\ln\left(\frac{N^{\mathrm{eff}}_V}{N^{\mathrm{eff}}_C}\right)\mbox{.} \label{intrinsbulk}
\end{align}
Eq. (\ref{intrinsbulk}) is also valid for all other values of $z$, such that 
\begin{align}
	E_i(z) &= \frac{E_C(z) + E_V(z)}{2} + \frac{k_BT}{2}\,\ln\left(\frac{N^{\mathrm{eff}}_V}{N^{\mathrm{eff}}_C}\right)\mbox{.} \label{intrins}
\end{align}
If the equations (\ref{intrinsbulk}) and (\ref{intrins}) are inserted into Eq. (\ref{potential}) and additionally the relation between the band edges $E_{\mathrm{gap}} = E_C(z) - E_V(z) = E_C^b - E_V^b$ is used, the potential $v(z)$ can be expressed as
\begin{align}
v(z) & = \frac{1}{k_BT} \Big[E_V^b - E_V(z)\Big]\,\mbox{.}
\end{align}
Thus, from a solution $v(z)$ of the Poisson equation (Eq. (\ref{poisdiff})) the final shape of band bending $E_V(z)$ with respect to the bulk valence band edge $E_V^b$ can be calculated as
\begin{align}
E_V(z) & = E_V^b - k_BT v(z) \,\mbox{.} \label{band}
\end{align}
Eq. (\ref{band}) does not directly depend on the effective densities of states. 
However, in order to obtain the potential $v(z)$ by solving Poisson's equation (Eq. (\ref{poisdiff})), the knowledge of the effective densities of states $N^{\mathrm{eff}}_C$ and $N^{\mathrm{eff}}_V$ is required, because the charge carrier densities inside the bulk $n_b$ and $p_b$ depend on them, which will be shown in the following. 


By using Eq. (\ref{ndensboltz}) and additionally inserting the relation for the intrinsic charge carrier density (resulting from $n_i^2 = n_e(z) \cdot p_h(z)$)
\begin{align}
	n_i\left(N^{\mathrm{eff}}_C, N^{\mathrm{eff}}_V\right) &= \sqrt{N^{\mathrm{eff}}_C \cdot N^{\mathrm{eff}}_V} \, e^{-\frac{E_{\mathrm{gap}}}{2k_BT}} \label{depintrinsdens}
\end{align}
with the band gap $E_{\mathrm{gap}} = E_C(z) - E_V(z)$, the mobile electron charge carrier density $n_e(z)$ can be rewritten as
\begin{align}
	n_e(z) &\,\overset{\mathrm{(\ref{ndensboltz})}}{=}\, N^{\mathrm{eff}}_{C}\, e^{-\frac{\left(E_C(z)-E_F\right)}{k_B T}} \, = \,\sqrt{N^{\mathrm{eff}}_{C} N^{\mathrm{eff}}_{V}}\, e^{\frac{-E_C(z) + E_V(z)}{2 k_B T}} \sqrt{\frac{N^{\mathrm{eff}}_{C}}{N^{\mathrm{eff}}_{V}}} \, e^{\frac{- E_C(z) - E_V(z)}{2 k_B T}} \, e^{\frac{E_F}{k_B T}} \nonumber \\
	& \,\overset{\mathrm{(\ref{depintrinsdens})}}{=}\, n_i\left(N^{\mathrm{eff}}_C, N^{\mathrm{eff}}_V\right) \, e^{\frac{E_F}{k_BT}} \sqrt{\frac{N^{\mathrm{eff}}_C}{N^{\mathrm{eff}}_V}} \, e^{\frac{-E_V(z)-E_C(z)}{2k_BT}} .\label{ndensdep}
\end{align}
From Eq. (\ref{intrins}) it follows that 
\begin{align}
	e^{-\frac{E_i(z)}{k_B T}} & = e^{\frac{-E_C(z) - E_V(z)}{2 k_B T} - \frac{1}{2}\,\ln\left(\frac{N^{\mathrm{eff}}_V}{N^{\mathrm{eff}}_C}\right)} \, = \,  e^{\frac{-E_C(z) - E_V(z)}{2 k_B T}} \, \left(e^{\ln\left(\frac{N^{\mathrm{eff}}_V}{N^{\mathrm{eff}}_C}\right)}\right)^{-\frac{1}{2}} \, = \, e^{\frac{-E_C(z) - E_V(z)}{2 k_B T}} \, \sqrt{\frac{N^{\mathrm{eff}}_C}{N^{\mathrm{eff}}_V}} \mbox{,} \label{efunceq}
\end{align}
and by inserting Eq. (\ref{efunceq}) into Eq. (\ref{ndensdep}), we obtain
\begin{align}	
n_e(z) & \,\overset{\mathrm{(\ref{efunceq})}}{=}\, n_i\left(N^{\mathrm{eff}}_C, N^{\mathrm{eff}}_V\right) \, e^{\frac{E_F-E_i(z)}{k_BT}} \,\overset{\mathrm{(\ref{u_pot})}}{=}\, n_i\left(N^{\mathrm{eff}}_C, N^{\mathrm{eff}}_V\right) \, e^{u(z)}   \mbox{.} \label{n_e}
\end{align}
Then, the electron charge carrier density inside the bulk $n_b$ can be expressed as
\begin{align}
	n_b &= n_i\left(N^{\mathrm{eff}}_C, N^{\mathrm{eff}}_V\right) \, e^{\frac{E_F-E_i^b\left(N^{\mathrm{eff}}_C, N^{\mathrm{eff}}_V\right)}{k_BT}} \,\overset{\mathrm{(\ref{u_pot})}}{=}\, n_i\left(N^{\mathrm{eff}}_C, N^{\mathrm{eff}}_V\right) \, e^{u_b\left(N^{\mathrm{eff}}_C, N^{\mathrm{eff}}_V\right)} \,\mbox{.} \label{nbulk}
\end{align}
If the exponential term in Eq. (\ref{n_e}) is expanded by $u_b$ and the definition for the potential $v(z)$ in Eq. (\ref{potential}) is used, the electron charge carrier density can be rearranged as 
\begin{align}
n_e(z) & \,\overset{\mathrm{(\ref{potential})}}{=}\, n_i\left(N^{\mathrm{eff}}_C, N^{\mathrm{eff}}_V\right) \, e^{u_b\left(N^{\mathrm{eff}}_C, N^{\mathrm{eff}}_V\right) + v(z)} \,\overset{\mathrm{(\ref{nbulk})}}{=}\, n_b\left(N^{\mathrm{eff}}_C, N^{\mathrm{eff}}_V\right) \, e^{v(z)}   \mbox{.} \label{n_end}
\end{align}

Following similar considerations as shown in Eqs. (\ref{ndensdep}) - (\ref{n_end}), but starting with Eq. (\ref{pdensboltz}), the hole charge carrier density results in
\begin{align}
p_h(z) & \,=\, n_i\left(N^{\mathrm{eff}}_C, N^{\mathrm{eff}}_V\right) \, e^{- \frac{E_F-E_i(z)}{k_BT}} \,\overset{\mathrm{(\ref{u_pot})}}{=}\, n_i\left(N^{\mathrm{eff}}_C, N^{\mathrm{eff}}_V\right) \, e^{-u(z)} \,\mbox{,} \label{p_e}
\end{align}
inside the bulk it is given by 
\begin{align}
	p_b &= n_i\left(N^{\mathrm{eff}}_C, N^{\mathrm{eff}}_V\right) \, e^{- \frac{E_F-E_i^b\left(N^{\mathrm{eff}}_C, N^{\mathrm{eff}}_V\right)}{k_BT}} \,\overset{\mathrm{(\ref{u_pot})}}{=}\, n_i\left(N^{\mathrm{eff}}_C, N^{\mathrm{eff}}_V\right) \, e^{-u_b\left(N^{\mathrm{eff}}_C, N^{\mathrm{eff}}_V\right)} \label{pbulk}
\end{align}
and by introducing the potential $v(z)$ it can be expressed as 
\begin{align}
p_h(z) & \,\overset{\mathrm{(\ref{potential})}}{=}\, n_i\left(N^{\mathrm{eff}}_C, N^{\mathrm{eff}}_V\right) \, e^{-u_b\left(N^{\mathrm{eff}}_C, N^{\mathrm{eff}}_V\right) - v(z)} \,\overset{\mathrm{(\ref{pbulk})}}{=}\, p_b\left(N^{\mathrm{eff}}_C, N^{\mathrm{eff}}_V\right) \, e^{-v(z)} \, \mbox{.} \label{p_end}
\end{align}

Thus, the effective densities of states $N^{\mathrm{eff}}_C$ and $N^{\mathrm{eff}}_V$ can have a crucial influence on the shape of band bending. They are necessary for the calculation of the bulk charge carrier densities $n_b$ and $p_b$ (equations (\ref{nbulk}) and (\ref{pbulk})), which in turn are needed for the solution $v(z)$ of the Poisson equation (Eq. (\ref{poisdiff})) and for the determination of the charge carrier densities $n_e(z)$ and $p_h(z)$ inside the space-charge region (equations (\ref{n_end}) and (\ref{p_end})). For an extended 3D crystal or a pure 2D sheet, in conjunction with the assumption of parabolic bands, the effective densities of states that were presented above can be used. However, for a system in which charge carriers are confined in one direction to within a length of the order of the wave length of the electrons, quantization effects have to be taken into account. This is outlined in the following section. The above considerations for semiconductors can be transferred straightforwardly to topological insulators, employing the arguments presented in section (II) of the paper.

\subsection{Quantum mechanical modifications to the effective density of states}

In order to take quantization effects on the occupation of the valence and conduction bands into account, which arise from an electron confinement on the length scale of the wave function, e.g. in thin films with thicknesses in the range of a few nm, a quantum mechanical approach must be used and the Schr\"{o}dinger equation has to be solved in addition to the Poisson equation. For thin TI films we use a model system which is extended in the $x$- and $y$-directions, but confined in the $z$-dimension. The quantized band states resulting from the solution of the Schr\"{o}dinger equation can then be expressed by modified effective densities of states and used as input for the band bending calculation. 

Specifically, the solution preceeds in three steps: In the first step, Poisson's equation is solved for a classical 3D system with a finite width in the $z$-direction, assuming a parabolic shape of the bands (as shown in detail in section (II) in the paper). In the next step, the resulting band bending potential $v(z)$ is used in the Schr\"{o}dinger equation to calculate quantization levels. To this end, the top and bottom surfaces of the thin film are treated as potential barriers which are very high compared to the potential inside the film. This is justified by the large work function of the TI in comparison to its band gap \cite{Haneman_1959,Ryu_2018,Qian_2016}. Moreover, it turns out that the band bending in all considered cases has a rather weak curvature, such that the problem can be framed in terms of generic potential wells. In other words, the exact solutions of the Schr\"{o}dinger equation for the given $v(z)$ are approximated very well by the solutions for potential wells of either square or triangular shapes. Their quantized eigenenergies result in a modification of the densities of states within the conduction and valence bands. In particular, the bands are not continuous any more, but instead multiple discrete two-dimensional subbands arise for each of both bands due to the confinement in one dimension. The correspondingly modified effective densities of states of this quasi-two-dimensional system are used in the final calculation step, in which Poisson's equation is solved once again to determine the band bending $v(z)$ in the thin film. 

\subsubsection{Square well}

Symmetric boundary conditions for the calculation of the band bending in thin films, which are presented in section (II.B) in the paper, result in a potential without a slope and only with very weak curvature. This potential can be approximated by a square well with infinite barriers on both sides, resulting in a confinement in the $z$-direction, while the $x$- and $y$- directions are assumed to be infinitely extended. The eigenenergies are \cite{Davies_1997}
\begin{align}
	E_j(k_x,k_y) & = \epsilon_j + \frac{\hbar^2}{2m^*}\left(k_x^2 + k_y^2  \right)\qquad \mathrm{with}\; \epsilon_j = \frac{\hbar^2}{2m^*} \left(\frac{j\pi}{d} \right)^2 \label{square_E}
\end{align}
with the integer number $j = 1,2,\dots$ and the film thickness $d$. The electron charge carrier density in the conduction band then results from the sum over all subbands $j$ as
\begin{align}
	n_{e,\mathrm{square}} & = \sum_{j\ge 1} n^{\mathrm{2D}}_j \,\overset{\mathrm{(\ref{ndensboltz})}}{=}\, \sum_{j\ge 1} \frac{m^*_ek_B T}{\pi \hbar^2} \, e^{-\frac{\left(E_C + \epsilon_j - E_F \right)}{k_B T}} = \underbrace{\left[\frac{m^*_e k_B T}{\pi \hbar^2} \sum_{j \ge 1}\, e^{-\frac{\epsilon_j}{k_B T}}\right]}_{N_{\mathrm{C,square}}^{\mathrm{eff}}} e^{-\frac{(E_C - E_F)}{k_BT}} \,\mbox{,} \label{ne_square}
\end{align}
where the two-dimensional charge carrier density $n^{\mathrm{2D}}_j$ of the subband $j$ is given by Eq. (\ref{ndens}) with the integration starting from $E_C + \epsilon_j$. According to Eq. (\ref{ndensboltz}) it can be expressed by using the Boltzmann approximation. From a comparison of Eq. (\ref{ne_square}) with Eq. (\ref{ndensboltz}) the modified effective density of states $N^{\mathrm{eff}}_{C,\mathrm{square}}$ can be extracted as 
\begin{align}
	N^{\mathrm{eff}}_{C,\mathrm{square}} & = \frac{m^*_e k_B T}{\pi \hbar^2} \, \sum_{j \ge 1} e^{-\frac{\epsilon_j}{k_B T}} \, \overset{\mathrm{(\ref{square_E})}}{=}\, \frac{m^*_e k_B T}{\pi \hbar^2} \,\sum_{j \ge 1} e^{-\frac{\hbar^2 \pi^2 j^2}{2m^*_e k_B T d^2}}\,\mathrm{.} \label{effdos}
\end{align}  
The modified effective density of states $N^{\mathrm{eff}}_{V,\mathrm{square}}$ of the valence band is given by the same expression, albeit replacing the effective mass $m^*_e$ with $m^*_h$. Note that $N^{\mathrm{eff}}_{C,\mathrm{square}}$ and $N^{\mathrm{eff}}_{V,\mathrm{square}}$ only depend on the width of the square well, i.e. the thickness $d$ of the thin TI film. In particular, the solutions are independent of the value of the surface Fermi energy $E_F^{\mathrm{top}}$, since the potential barriers (work function) at the top and bottom surfaces of the film are high compared to the potential in the film. Because of the relatively high work function (e.g. $\approx 5\,\mathrm{eV}$ for Bi$_2$Te$_3$ \cite{Haneman_1959,Ryu_2018}), any corrections resulting from the principally more accurate solution of a square well with finite barrier height remain rather small, such that the approximation of the barriers as being infinitely high is well justified in this case.

\subsubsection{Triangular well}

If the band bending in thin films is calculated with asymmetric boundary conditions (section (II.C) of the paper), the square well potential is not a suitable approximation any more. While the band bending potential still shows a very weak curvature, it now exhibits a constant slope that is determined by the difference between the top and bottom surface Fermi energies $E_F^{\mathrm{top}}$ and $E_F^{\mathrm{bottom}}$ (c.f. Fig. 6 of the paper). Hence, a triangular well of thickness $d$ with infinitely high barriers at either surface of the thin film provides a better description of the band bending potential. We express the potential in a triangular well as
\begin{align}
	V_{\mathrm{triangle}}(z) & = q F z
\end{align}
where the slope $F$ is determined by the difference between the top and bottom surface Fermi energies
\begin{align}
F\left(E_F^{\mathrm{top}},E_F^{\mathrm{bottom}}\right) & = \frac{E_F^{\mathrm{top}}-E_F^{\mathrm{bottom}}}{q\, d} \,. 
\end{align}
With this potential the Schr\"{o}dinger equation reads
\begin{align}
	\left[-\frac{\hbar^2}{2 m^*} \, \frac{\mathrm{d}^2}{\mathrm{d} z^2} + q F z \right] \Psi(z)  & = E \Psi(z)	
\end{align}
which after the variable substitution  
\begin{align}
	\xi(z) & = \underbrace{\left(\frac{2 m^* q F}{\hbar^2} \right)^{\frac{1}{3}}}_{:= \xi_0} \left(z - \frac{E}{q F}\right) 
\end{align}
becomes
\begin{align}
	\frac{\mathrm{d}^2}{\mathrm{d}\xi^2} \Psi(\xi) & = \xi \Psi(\xi) .\label{psidiff2}
\end{align}
A solution of Eq. (\ref{psidiff2}) is given by \cite{Davies_1997}
\begin{align}
	\Psi(\xi) & = A_i(\xi) + C B_i(\xi)
\end{align}
where $A_i(\xi)$ and $B_i(\xi)$ are the Airy functions and $C$ is a constant. With the boundary conditions $\Psi(\xi(z=0)) = 0$ and $\Psi(\xi(z=d)) = 0$ at the top and bottom surfaces of the thin film we obtain 
\begin{align}
	A_i(\xi(0)) + C B_i(\xi(0)) & = 0 \nonumber \\[0.5ex]
	A_i(\xi(d)) + C B_i(\xi(d)) & = 0.    
\end{align}
Eliminating the constant $C$ and resubstitution finally results in 
\begin{align}
	A_i\left(\xi_0 \left(d - \frac{E}{q F}\right)\right) - \frac{A_i\left(-\xi_0 \frac{E}{q F} \right)}{B_i\left(-\xi_0 \frac{E}{q F} \right)} \; B_i\left(\xi_o \left(d - \frac{E}{q F} \right)  \right) = 0. \label{psisolution}
\end{align}
Eq. (\ref{psisolution}) has to be solved numerically to determine the quantized eigenenergies $E_j$ that satisfy the equation. If these eigenenergies are calculated for a specific potential slope $F$, they can be inserted for $\epsilon_j$ in Eq. (\ref{effdos}) to determine the modified effective densities of states $N^{\mathrm{eff}}_{C,\mathrm{triangle}}$ and $N^{\mathrm{eff}}_{V,\mathrm{triangle}}$ for the conduction and valence bands, respectively. These effective densities of states can then be used as input for the final step of the band bending calculation. However, it has to be noted that they now depend not only on the width $d$ of the triangular well (the film thickness), but also on the slope $F$ of the potential, which in turn depends on the difference between the top and bottom surface Fermi energies. Thus, $N^{\mathrm{eff}}_{C,\mathrm{triangle}}$ and $N^{\mathrm{eff}}_{V,\mathrm{triangle}}$ have to be calculated for each pair $E^{\mathrm{top}}_F$, $E^{\mathrm{bottom}}_F$ separately. 

\vspace{3ex} 
Within the above approximation of the Schr\"{o}dinger-Poisson approach, only the eigenenergies resulting from the solution of the Schr\"{o}dinger equation are used, while the simultaneously determined wave functions $\Psi(z)$ are not considered. However, in principle the probability density $|\Psi(z)|^2$ resulting from the electron and hole wave functions must be inserted as an additional factor in front of the corresponding mobile charge carrier densities in the differential equation (Eq. (\ref{poisdiff})) before solving the Poisson equation in the final step. Evidently, this complicates the solution due to the additional $z$-dependence. Since the effect of the $|\Psi(z)|^2$-induced modification of the mobile charge carrier densities on the band bending is expected to be small, we neglect the wave functions and assume a uniform distribution of the mobile carriers in each subband across the thin film, as discussed above. This may lead to a slight overestimation of the band bending in the immediate vicinity of the surfaces, where the effect of the probability density on the mobile charge carrier distribution is strongest, because (1) the weak curvature in conjunction with the large barrier height implies that the subband wave functions tend to vanish towards the film surfaces, and (2) at room temperature nearly only the first subband is occupied ($\approx 95\%$) such that the probability density of the mobile carriers in the center of the thin film is at maximum and decreases towards the surfaces (e.g., as for the case of a square well the wave function of the lowest energetic state is the first mode of sine). 

\section{C. Mobile electron and hole charge carrier densities}

\begin{figure*}[b!]
	\vspace{-0ex}
	\centering
	\includegraphics[height=0.335\textwidth]{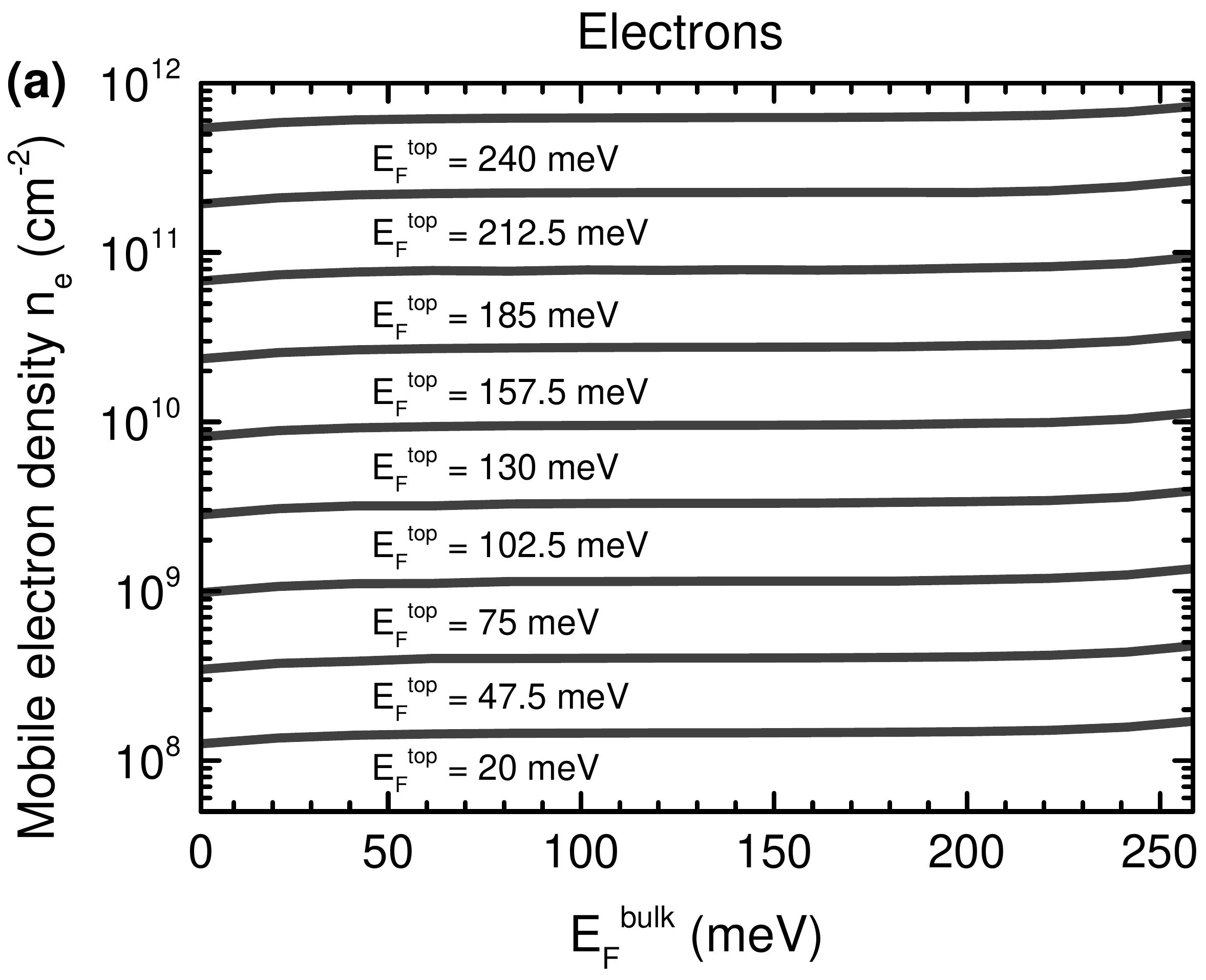}
	\hspace{0.4cm}
	\includegraphics[height=0.335\textwidth]{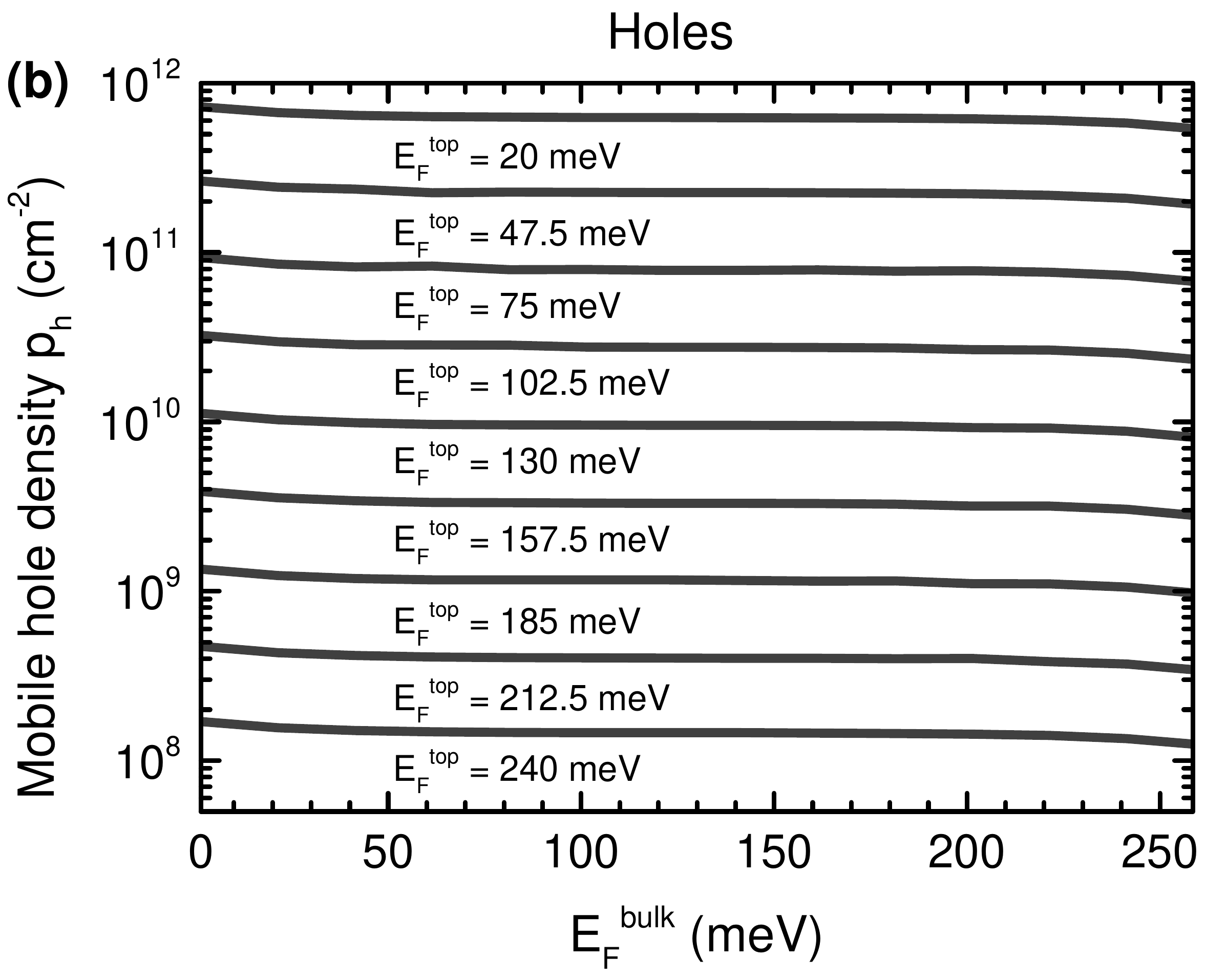}
	\vspace{-2ex}
	\caption{(Color online) Integrated mobile charge carrier densities for (a) electrons and (b) holes in a thin TI film as function of the dopant concentration (represented by  $E_F^{\mathrm{bulk}}$) and with $E_F^{\mathrm{top}}$ as parameter. All  parameters of the calculation are the same as in Fig. 5 of the paper.}
	\label{supfig5}
	\vspace{-0ex}
\end{figure*}

\begin{figure*}[t!]
	\vspace{-0ex}
	\centering
	\includegraphics[height=0.35\textwidth]{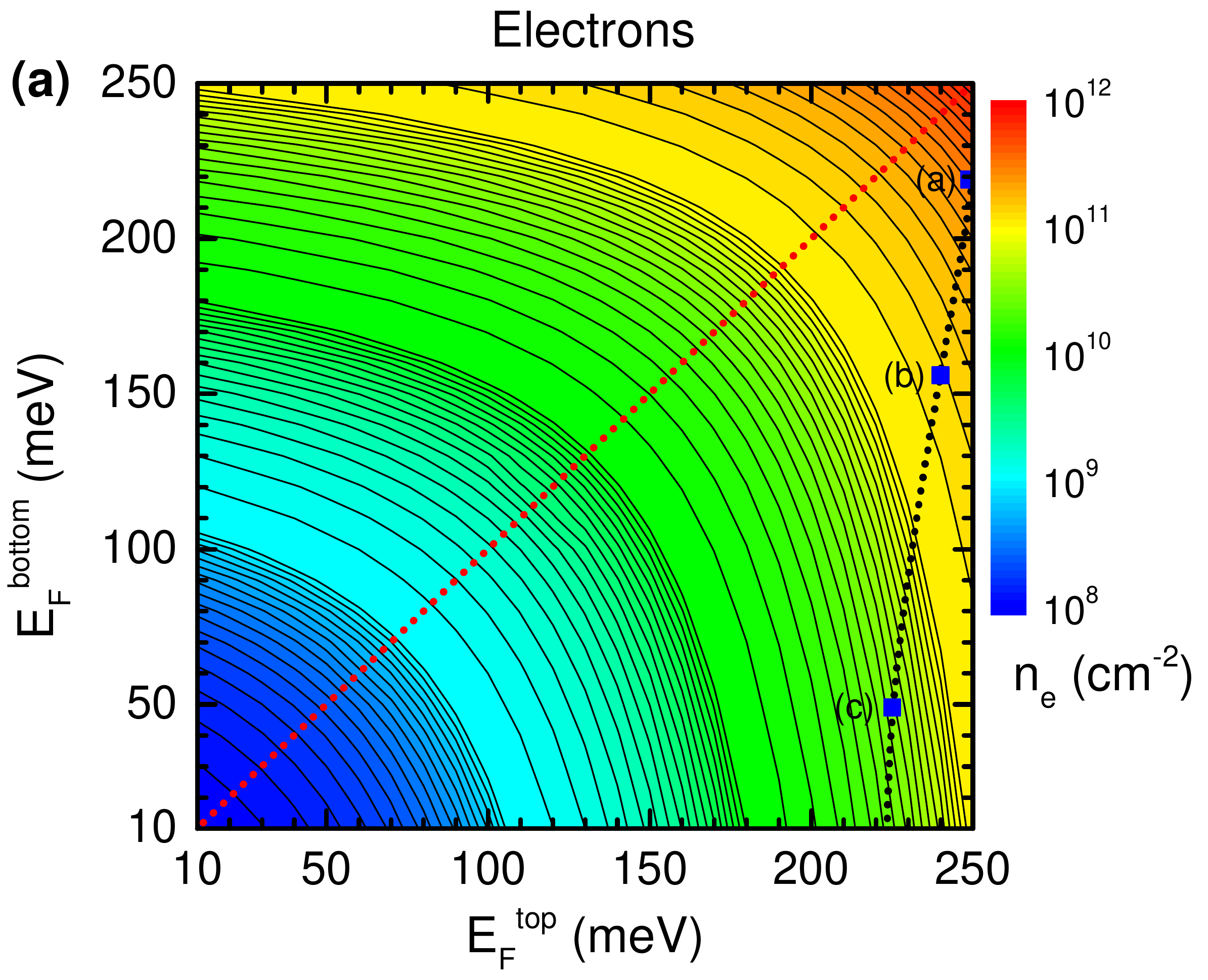}
	\hspace{0.2cm}
	\includegraphics[height=0.35\textwidth]{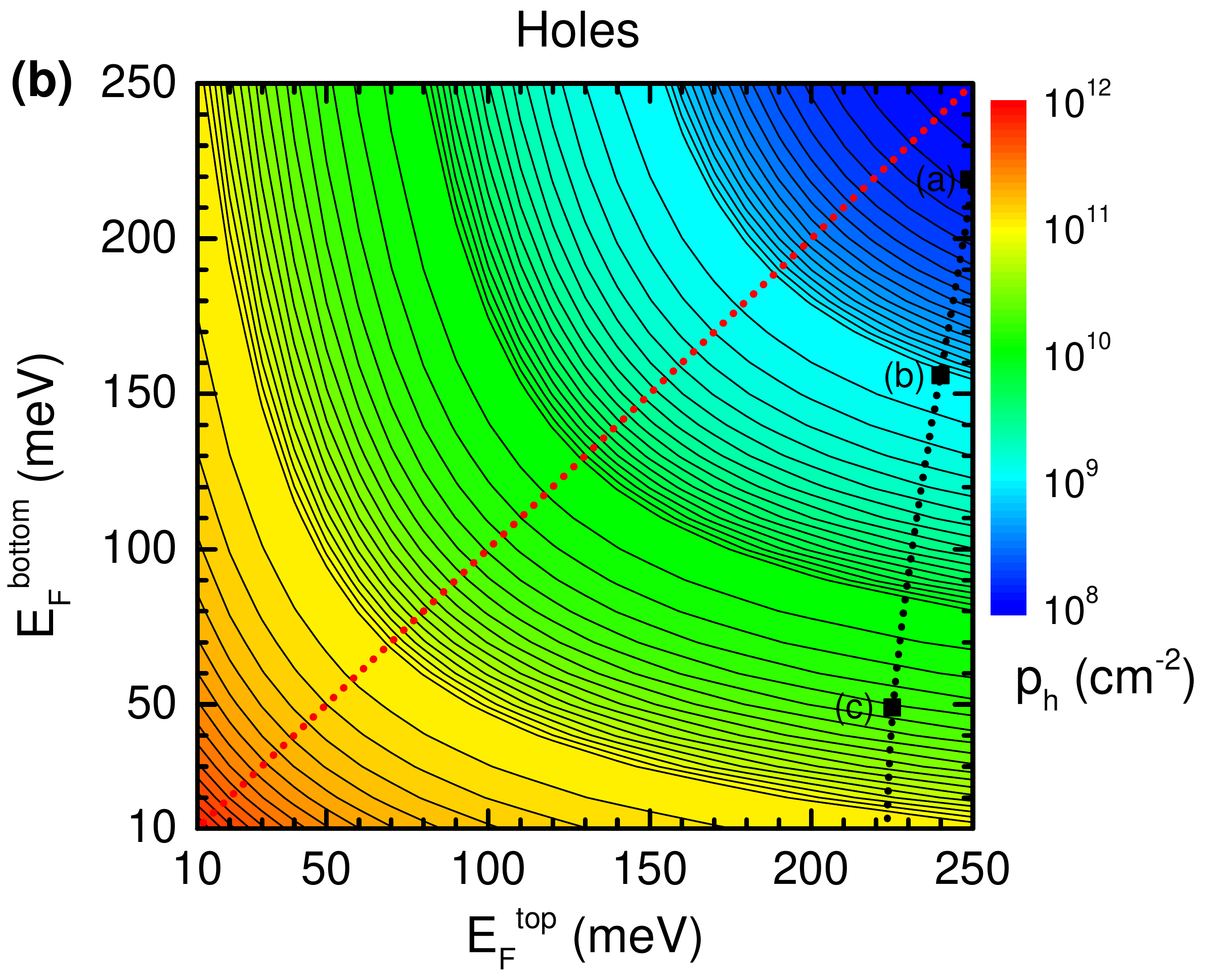}
	\vspace{-2ex}
	\caption{(Color online) Color plot of the integrated mobile charge carrier densities for (a) electrons and (b) holes in a thin TI film as function of $E_F^{\mathrm{top}}$ and $E_F^{\mathrm{bottom}}$. All parameters of the calculation are the same as in Fig. 8 of the paper.}
	\label{supfig6}
	\vspace{-0ex}
\end{figure*}

Fig. \ref{supfig5} and \ref{supfig6} correspond to Fig. 5 and Fig. 8 in the paper, with the only difference that here the individual contributions of electrons and holes to the integrated mobile charge carrier density in the thin TI film are displayed separately. 

If not the integrated density, but the position-dependent distribution of the mobile charge carrier density $n_{\mathrm{film}}(z) = n_e(z) + p_h(z)$ in the thin film needs to be determined, one may use the following approximation for obtaining a more exact solution by additionally including the probability density of the corresponding wave functions: In Eq. (8) in the paper, the probability densities $|\Psi(z)|^2$ could be used as prefactor to the corresponding charge carrier densities $n_e(z)$ and $p_h(z)$. With this approximation, the probability density has no effect on the band bending potential $v(z)$ itself (see section B above), but simply modifies the distribution of the mobile charge carriers in the film. Evidently, this approach is only useful if the $z$-dependence of the carrier density $n_{\mathrm{film}}(z)$ in the thin film is of interest. If only the total mobile charge carrier density of the film is important, as it results from the integration of Eq. (8), the additional factor $|\Psi(z)|^2$  makes no difference.

\bibliography{./suppl_bandbending}